\newcommand{\vf}{\boldsymbol{f}}
\newcommand{\vg}{\boldsymbol{g}}

\newcommand{\vx}{\boldsymbol{x}}
\newcommand{\vy}{\boldsymbol{y}}
\newcommand{\vw}{\boldsymbol{w}}

\documentclass[times,twocolumn,final]{elsarticle}
\usepackage{medima}
\usepackage{framed,multirow}

\usepackage{amssymb}
\usepackage{url}
\usepackage{xcolor}
\newcommand{\modified}[1]{\textcolor{black}{#1}}

\usepackage{amsfonts}
\usepackage{amsmath}
\usepackage{graphicx}
\usepackage[tight]{subfigure}
\definecolor{newcolor}{rgb}{.8,.349,.1}
\usepackage{hyperref}
\hypersetup{
    colorlinks=true,
    linkcolor=cyan,
    filecolor=cyan,  
    citecolor=cyan,   
    urlcolor=cyan,
}

\journal{Medical Image Analysis}

\begin{document}

\verso{Yixing Huang \textit{et~al.}}

\begin{frontmatter}

\title{Cephalogram Synthesis and Landmark Detection in Dental Cone-Beam CT Systems}%
%\tnotetext[tnote1]{This is an example for title footnote coding.}

\author[1]{Yixing Huang\corref{cor1}}
%\author[1]{Yixing Huang}
\cortext[cor1]{Corresponding author.}
\ead{yixing.yh.huang@fau.de}
\author[1]{Fuxin Fan}
\author[1]{Christopher Syben}
\author[1,2]{Philipp Roser}
\author[1]{Leonid Mill}
\author[1,2]{Andreas Maier}

\address[1]{Pattern Recognition Lab, Friedrich-Alexander-University Erlangen-Nuremberg, 91058 Erlangen, Germany}
\address[2]{Erlangen Graduate School in Advanced Optical Technologies (SAOT), 91052 Erlangen, Germany}

\received{9 September 2020}
\accepted{26 February 2021}
\availableonline{5 March 2021}

\begin{abstract}
%%%
Due to the lack of a standardized 3D cephalometric analysis methodology, 2D cephalograms synthesized from 3D cone-beam computed tomography (CBCT) volumes are widely used for cephalometric analysis in dental CBCT systems. However, compared with conventional X-ray film based cephalograms, such synthetic cephalograms lack image contrast and resolution, which impairs cephalometric landmark identification. In addition, the increased radiation dose applied to acquire the scan for 3D reconstruction causes potential health risks. In this work, we propose a sigmoid-based intensity transform that uses the nonlinear optical property of X-ray films to increase image contrast of synthetic cephalograms from 3D volumes. To improve image resolution, super resolution deep learning techniques are investigated. For low dose purpose, the pixel-to-pixel generative adversarial network (pix2pixGAN) is proposed for 2D cephalogram synthesis directly from two cone-beam projections. For landmark detection in the synthetic cephalograms, an efficient automatic landmark detection method using the combination of LeNet-5 and ResNet50 is proposed. Our experiments demonstrate the efficacy of pix2pixGAN in 2D cephalogram synthesis, achieving an average peak signal-to-noise ratio (PSNR) value of 33.8 with reference to the cephalograms synthesized from 3D CBCT volumes. Pix2pixGAN also achieves the best performance in super resolution, achieving an average PSNR value of 32.5 without the introduction of checkerboard or jagging artifacts. Our proposed automatic landmark detection method achieves 86.7\% successful detection rate in the 2\,mm clinical acceptable range on the ISBI Test1 data, which is comparable to the state-of-the-art methods. The method trained on conventional cephalograms can be directly applied to landmark detection in the synthetic cephalograms, achieving 93.0\% and 80.7\% successful detection rate in 4\,mm precision range for synthetic cephalograms from 3D volumes and 2D projections, respectively.
%%%%
\end{abstract}

\begin{keyword}
%% MSC codes here, in the form: \MSC code \sep code
%% or \MSC[2008] code \sep code (2000 is the default)
%\MSC 41A05\sep 41A10\sep 65D05\sep 65D17
%% Keywords
\KWD Deep learning\sep cephalogram synthesis \sep super resolution \sep landmark detection
\end{keyword}

\end{frontmatter}

%\linenumbers

%% main text
\section{Introduction}
\label{Sect:Introduction}

\begin{figure*}
\centering
\includegraphics[width = 0.72\linewidth]{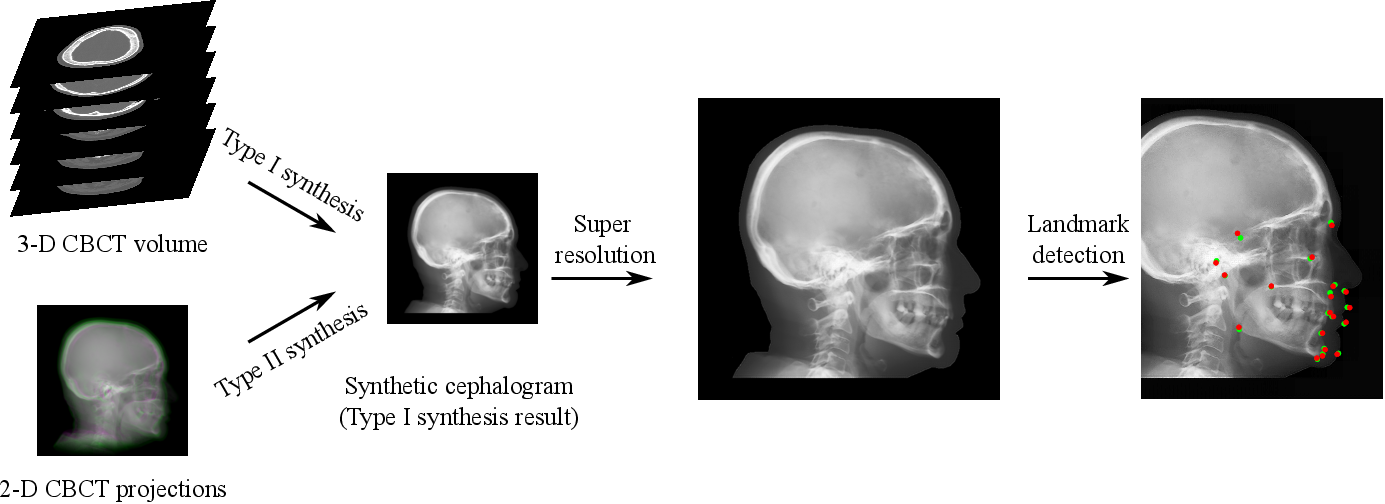}
\caption{\modified{A graphical abstract of our contributions to cephalometric analysis in dental CBCT systems.}}
\label{Fig:workflow}
\end{figure*}

Since its introduction in 1931 \citep{broadbent1931new}, cephalometric examination using two-dimensional (2D) lateral cephalograms is a standard procedure for orthodontic diagnostics and treatment planning. In cephalograms, various landmarks are sketched to form lines and angles, which are essential to assess patients' skeletal and dental relationships. Such cephalograms are acquired in specialized radiographic cephalometer systems.
Cone-beam computed tomography (CBCT) was introduced to dentistry at the end of last century \citep{mozzo1998new}. Since then it has been playing an important role in various dental applications \citep{scarfe2006clinical}, including oral surgery, orthodontics, endodontics, and implant treatment planning. In dentistry, a single system capable of multiple functions is a trend to empower dentists, facilitate management and save cost. For this purpose, systems combining CBCT and cephalograms (as well as panoramic radiographs typically) emerged. One approach to achieve such multi-functions is to equip CBCT systems with additional physical modules. However, such systems require additional acquisitions for cephalograms, causing extra dose exposure to patients. Therefore, achieving multi-functions on a standalone CBCT system with one single acquisition, where cephalometric analysis is performed based on CBCT data, is preferable for the sake of low dose and further cost reduction.

CBCT reconstructs a three-dimensional (3D) volume of anatomical structures. For the application to orthodontics, large field of view (FOV) volumes are typically reconstructed to cover the whole maxillofacial or craniofacial region. In the 3D volume, landmarks can be visualized directly without superimposition and perspective distortion. Therefore, a lot of research work has been conducted to evaluate the measurement accuracy, reliability and reproducibility of 3D landmark identification, in order to justify whether 3D cephalometric analysis is more beneficial than the standard 2D analysis. For example, \citep{park2006proposal} proposed to use 19 landmarks to examine the zygoma, maxilla, mandible and facial convexity and a 3D chart is provided to record measurements. \citep{kamiishi2007development} proposed a 3D analysis method using two types of surface rendering. Some systematic reviews on 3D cephalometric analysis are presented in \citep{pittayapat2014three,lisboa2015reliability,scarfe2018maxillofacial}. \modified{Recently, deep learning algorithms have been proposed to detect 3D cephalometric landmarks directly and they have achieved promising results \citep{zhang2017joint,o2018attaining,torosdagli2018deep,payer2019integrating}.}
 However, such 3D cephalometric analysis methods are still under development and mostly limited to research, due to the requirement of special softwares, the lack of standardized analytical methodology and insufficient evidence for diagnostic efficacy \citep{pittayapat2014three}. In addition, as practitioners are used to \modified{working} with conventional 2D cephalograms, it usually takes time for them to gain proficient skills on 3D cephalometric analysis.
 %In addition, it takes time for practitioners to command proficient skills on 3D cephalometric analysis, since most of them are accustomed to working with 2D conventional cephalograms. 
 Therefore, 3D cephalometric analysis is not yet widely used in practice.

Instead, synthesizing 2D cephalograms from 3D CBCT volumes is a widely used practical way for cephalometric analysis in dental CBCT systems \citep{farman2005dentomaxillofacial,farman2006development,moshiri2007accuracy,kumar2007comparison,cattaneo2008comparison}. In such a way, additional physical 2D cephalometer modules are not necessary, while the existing 2D cephalometric databases and standardized methodologies are inherited. Many studies have reported that CBCT synthetic cephalograms are equivalent or even superior to conventional cephalograms in terms of landmark identification error and reproducibility \citep{van2009comparison,hwang2013use,chen2014intraobserver}. However, CBCT synthetic cephalograms typically have different appearance from conventional cephalograms in terms of image contrast (see Fig.\,\ref{Fig:sigmoidCurve}) and resolution, since X-ray films used in conventional cephalograms have nonlinear optical properties \citep{ritenour1996physics} and higher image resolution than digital detectors in CBCT systems \citep{hatvani2018tensor,hatvani2018deep}. Such differences require practitioners to have further training to get familiar with synthetic cephalograms. In addition, although CBCT has lower radiation dose than \modified{multi-slice CT (MSCT)}, it still requires considerably more projections than conventional 2D cephalograms. Hence, the potential health \modified{risk} caused by radiation dose is still a concern considering the as-low-as-reasonably-achievable principle.

For cephalometric analysis in synthetic cephalograms, landmark detection is necessary. Manual cephalometric landmark identification is tedious and time-consuming. And intra- and inter-observer variability may lead to unreproducible measurements. Therefore, computer aided automatic landmark detection is highly desired \citep{ibragimov2014automatic,lindner2015fully,arik2017fully,qian2019cephanet,chen2019cephalometric}. 

In order to address the above mentioned aspects in cephalometric analysis, the following contributions, displayed in Fig.\,\ref{Fig:workflow} as an overview, are made in this work:
%We propose a nonlinear sigmoid-based intensity transform according to the optical property of X-ray films to improve image contrast for synthetic cephalograms;

1. Image contrast: a nonlinear sigmoid-based intensity transform according to the optical property of X-ray films is proposed for Type I cephalogram synthesis;

%Super resolution (SR) techniques using deep learning are applied to improve \todo{image resolution};
2. Low dose: direct cephalogram synthesis from dual CBCT projections is proposed, where the advantage of using dual projections over one projection, the selection of patches, and the feasibility of one model for multi-quadrant patches are elaborated;

3. Image resolution: super resolution (SR) techniques using different adversarial generative networks (GANs) are investigated;

4. Landmark detection: an efficient automatic landmark detection method is proposed, which is applicable to real and synthetic cephalograms.

\section{Related Work}
\subsection{Cephalogram Synthesis}
Various methods for cephalogram synthesis from 3D CBCT volumes have been proposed. Ray-sum multi-planar reformatting (MPR), also called ray casting (RayCast), using orthogonal projection was the first reported method \citep{farman2005dentomaxillofacial,farman2006development,moshiri2007accuracy}. Since real cephalometer systems use cone-beam X-rays, which cause perspective deformation, 
%As a standard, in a Wehmer cephalostat the distance between the patient midsaggital plane to the detector distance 11.5\,cm and the distance between the midsaggital plane to the X-ray source is 5 feet (152.4\,cm). 
\citep{kumar2007comparison} proposed to use perspective projection based on the Wehmer cephalostat geometry into the RayCast method to reproduce conventional cephalometric geometry with similar accuracy. However, they concluded that synthetic cephalograms with orthogonal projection provide greater accuracy of measurement for midsagittal plane dimensions than those with perspective projection. Other than RayCast methods, maximum intensity projection (MIP) \citep{cattaneo2008comparison} is also used for cephalogram synthesis from 3D CBCT volumes. Since only the largest intensity pixels are projected, low intensity structures are omitted. As a consequence, MIP is proven to produce less reproducible measurements than RayCast.

Synthesizing cephalograms from 2D cone-beam projections is an image-to-image translation problem. Due to the severe perspective deformation in cone-beam projections, it is very challenging to restore such deformation with conventional methods. Recently, deep learning methods, particularly using generative adversarial networks (GANs) \citep{yi2019generative}, have achieved promising results in image synthesis in various medical applications such as 3T MRI images to 7T MRI images \citep{qu2020synthesized}, PET images to CT images \citep{armanious2020medgan}, and MRI cone-beam projections to X-ray cone-beam projections \citep{stimpel2019projection}. However, to the best of our knowledge, such projection-to-cephalogram synthesis using GANs \modified{has} not been investigated yet. For parallel-beam projection to cone-beam projection conversion, \citep{syben2020known} have proposed a novel rebinning algorithm using known operator learning \citep{maier2019learning}. It reconstructs an intermediate volume with learnt filters from parallel-beam MRI projections in a specialized trajectory and afterwards reprojects the volume with the desired cone-beam geometry to generate CBCT projections. Due to the requirement of the special trajectory and the large number of projections, the method in \citep{syben2020known} cannot be applied in our application where direct synthesis of parallel-beam cephalogram from a few number of CBCT projections is desired.

\subsection{Image Super Resolution}
Image SR aims at recovering high resolution (HR) images from low resolution (LR) images. Benefiting from the strong capacity of extracting effective high level features between LR and HR images, deep learning has achieved the state-of-the-art performance for various SR applications \citep{yang2019deep}. One of the first neural networks in this field is called super resolution convolutional neural network (SRCNN) proposed by \citep{dong2015image}. It learns the mapping between interpolated low resolution (ILR) images and HR images based on conventional sparse-coding approaches. Follow-up researchers proposed to use deeper neural networks such as the VGG network \citep{kim2016accurate}, deep Laplacian pyramid networks \citep{lai2017deep}, and deep residual networks \citep{lim2017enhanced,kim2016deeply}. Although such deep learning methods achieve high peak signal-to-noise ratio (PSNR), generated images still lack high frequency details. That is why adversarial learning is introduced, where a generator network is trained to generate realistic HR images and a discriminator network is trained to tell the difference between generated HR images and target HR images. Super resolution generative adversarial network (SRGAN) \citep{ledig2017photo} is the first GAN-based deep learning method introduced for SR, which became the benchmark method in SR. The generator of SRGAN uses 5 residual blocks. The replacement of these residual blocks by residual dense blocks (RDBs) or residual-in-residual dense blocks (RRDBs) results in two enhanced super resolution generative adversarial networks (ESRGANs) \citep{zhang2018residual,wang2018esrgan}. Both ESGANs further adjust the architecture design, perceptual loss and adversarial loss of SRGAN to avoid the introduction of different artifacts. 

\subsection{Landmark Detection}
 Many efforts have been devoted to automatic cephalometric landmark detection. In particular, several benchmark methods have been proposed in the challenges organized by the International Symposium on Biomedical Imaging (ISBI) in 2014 \citep{wang2015evaluation} and 2015 \citep{wang2016benchmark}. The method proposed by \citep{ibragimov2014automatic} applies game theory and random forests, which won the ISBI Challenge 2014 with 72.7\% successful detection rate (SDR) within the clinical acceptable 2\,mm precision range. The random forest regression-voting method proposed by \citep{lindner2015fully} won the ISBI 2015 challenge with 74.8\% 2\,mm-SDR. \citep{arik2017fully} introduced a convolutional neural network (CNN) for landmark detection, achieving 75.3\% 2\,mm-SDR. In 2019, the CephaNet \citep{qian2019cephanet} using the faster R-CNN architecture as a backbone obtains 82.5\% 2\,mm-SDR on ISBI Test1 data. \citep{chen2019cephalometric} proposed a method combining a VGG-19 feature extraction module, an attentive feature pyramid fusion module and a regression-voting module, which achieves 86.7\% 2\,mm-SDR on ISBI Test1 data. The latest method proposed by \citep{song2020automatic} applies the ResNet50 to detect landmarks on region-of-interest (ROI) patches extracted by a registration step, achieving 86.74\% 2\,mm-SDR on ISBI Test1 data.
 
\section{Materials And Methods}

%\todo{Our contributions to cephalometric analysis in dental CBCT systems mainly lie in cephalogram synthesis and landmark detection, as displayed in Fig.\,\ref{Fig:workflow}.} When reconstructing 3D volumes is necessary, e.g., in craniofacial surgeries, a cephalogram synthesis method from 3D volumes is proposed. In some applications, the patients may only need 2D lateral cephalograms without the necessity of 3D volumes, which require hundreds of X-ray projections for each reconstruction. In such applications, we propose a method to synthesize cephalograms directly from a few 2D X-ray projections for low-dose purpose. In the following, we refer to the above mentioned synthesis methods from 3D volumes and 2D projections as Type I and Type II, respectively. Since digital X-ray detectors typically have lower resolution than X-ray films used in conventional cephalograms, the synthetic cephalograms have blurry structures. Therefore, an SR step is proposed to generate HR cephalograms. Afterwards, the synthetic cephalograms are evaluated by an efficient and effective automatic landmark detection method. Using the detected landmarks, further cephalometric classification methods \citep{wang2016benchmark,lindner2016fully,yu2020automated} can be applied.
In this section, we introduce the contents of Fig.\,\ref{Fig:workflow} in detail.

\subsection{Type I: Cephalogram Synthesis from 3D CBCT Volumes}
For Type I synthesis, our method includes the steps of skeleton enhancement, ray casting, and sigmoid-based transform. For the sigmoid-based transform, the modification from the original sigmoid transform is explained.

%a rigid transform $\vec{T} \in \text{SE}3$
\subsubsection{Skeleton enhancement}
We denote the intensity distribution of a patient head by $\vf(x,y,z)$. We further denote a reconstructed 3D volume of the head by $\tilde{\vf}(x, y, z)$. In dental CBCT systems, the patient head is typically well aligned by a fixation device. If not, a rigid transform $\vec{T} \in \text{SE}3$ can be applied to $\tilde{\vf}$ to adjust the orientation of the head facing to the positive $Y$ direction. In cephalograms, the projection of skeletal structures and airways plays an important role. In order to enhance such structures, we choose two thresholds -500\,HU and 1000\,HU to preprocess $\tilde{\vf}$ in the following way,
\begin{equation}
\vf^\ast(x,y,z) =
\left\lbrace \begin{array}{ll}
a \cdot \tilde{\vf}(x,y,z), & \text{if } \tilde{\vf}(x,y,z) > 1000\,\text{HU},\\
-1000\,\text{HU}, & \text{if } \tilde{\vf}(x,y,z) < -500\,\text{HU},\\
\tilde{\vf}(x,y,z), & \text{otherwise. } \\
\end{array}
\right.
\label{eqn:skeletonEnhancement}
\end{equation}
The threshold 1000\,HU is used to segment skeletal structures and $a$ is a weight to slightly highlight them. To preserve soft tissue visualization, we empirically choose $a=1.3$ in this work. 
%The threshold -500\,HU is used to segment airways including air background. In CBCT reconstruction, the airway areas do not have an ideal value of -1000\,HU due to noise and artifacts caused by various effects such as scattering and beam hardening. 
By resetting the values below -500\,HU to -1000\,HU, the noise and artifacts (e.g., scattering and beam hardening artifacts) in the airway areas are suppressed. Any other structures between these two thresholds are mainly soft tissues. Their values are preserved.

\subsubsection{Ray casting}
According to \citep{farman2005dentomaxillofacial,farman2006development,moshiri2007accuracy,kumar2007comparison}, RayCast is applied to synthesize preliminary 2D cephalograms from 3D volumes,
\begin{equation}
\vg(y, z) = \mathcal{P}\vf^\ast(x,y,z),
\end{equation}
where $\vg(y, z)$ is an (enhanced) attenuation integral image, and $\mathcal{P}$ is a projection operator. The pixel intensity values of $\vg$ are typically in the range of [0, 6] for human heads. With parallel-beam X-rays, $\mathcal{P}$ is an orthogonal projection along the $X$ direction; with cone-beam X-rays, $\mathcal{P}$ is a perspective projection using the geometry of a standard Wehmer cephalostat \citep{kumar2007comparison}, i.e., with the isocenter-to-detector distance of 11.5\,cm and the source-to-isocenter distance of 152.4\,cm. Since synthetic cephalograms with orthogonal projection provide better measurement accuracy than those with perspective projection \citep{kumar2007comparison}, orthogonal projection is mainly used in this work.

% To generate a 2D cephalogram from a 3D CT volume, ray casting (RayCast), also called ray-sum, using orthogonal projection \citep{farman2005dentomaxillofacial,moshiri2007accuracy} or perspective projection \citep{kumar2007comparison} is typically used. In cephalometer systems, the patient head is placed close to the detector while the cone-beam X-ray source is placed far away to get as parallel X-rays as possible. As a standard, the distance between the patient midsaggital plane to the detector distance 11.5\,cm and the distance between the midsaggital plane to the X-ray source is 5 feet (152.4\,cm) in a Wehmer cephalostat. To consider the magnification factor in such a system, RayCast using perspective projection is used \citep{kumar2007comparison}. In virtual cephalogram synthesis from 3D CT volumes, no physical constraints exist for the X-ray source. As an alternative, ideal orthogonal projection using parallel-beam X-rays is utilized in this paper, 
%\begin{equation}
%\vg(y, z) = \mathcal{}\int_{x_{\min}}^{x_{\max}} \vf^\ast(x,y,z)\text{d}x,
%\label{eqn:projection}
%\end{equation}
%where $x_{\min}$ and $x_{\max}$ are the minimum and maximum index values in the $X$ dimension. Here we use the continuous form for simplicity. For discrete implementation, the integral operation can be simply replaced by the sum operation.

\subsubsection{Original sigmoid transform}
\begin{figure}
\centering
\begin{minipage}[b]{0.38\linewidth}
\subfigure[Real cephalogram]{
\includegraphics[width = \linewidth]{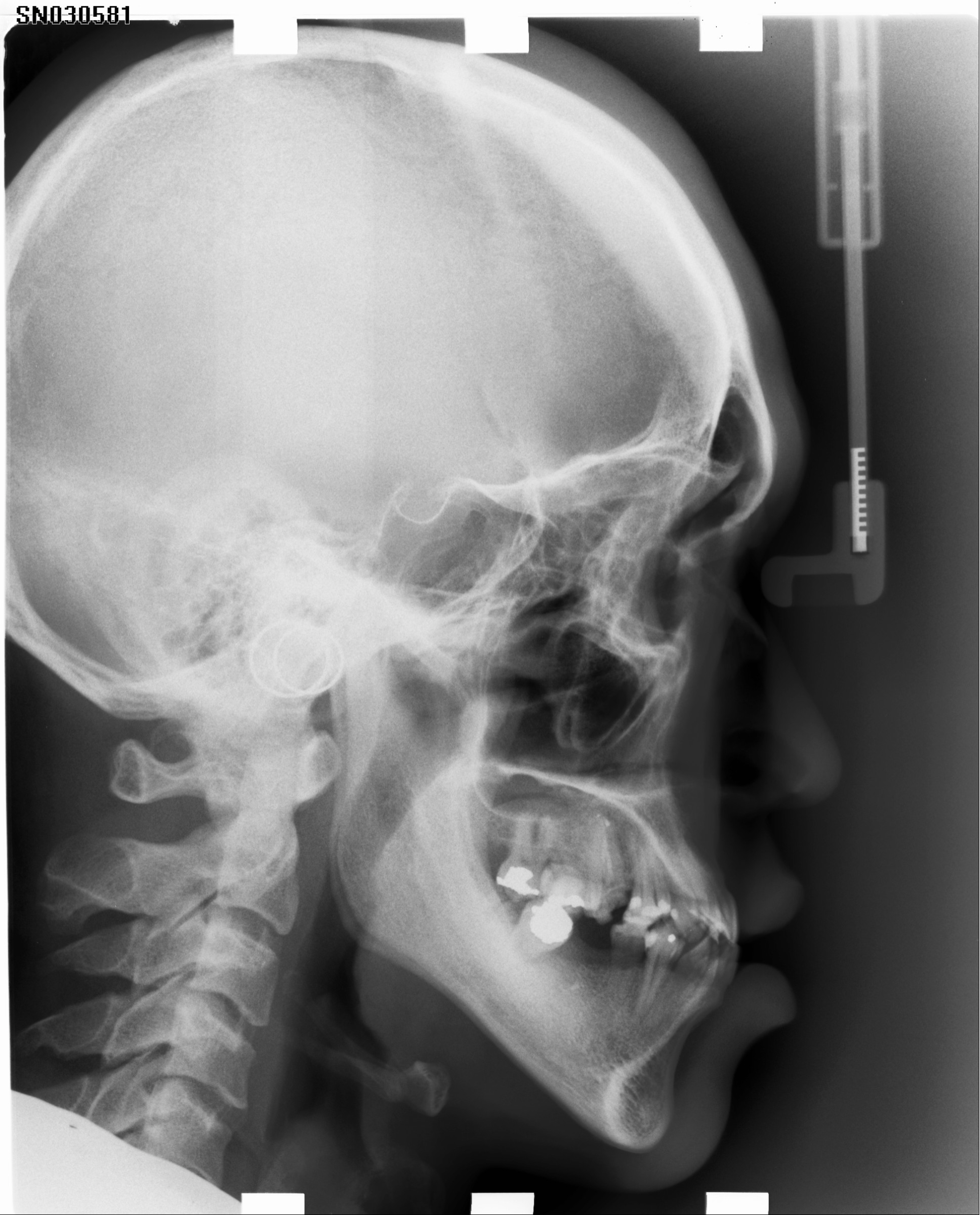}
}
\end{minipage}
\begin{minipage}[b]{0.368\linewidth}
\subfigure[Synthetic cephalogram]{
\includegraphics[width = \linewidth]{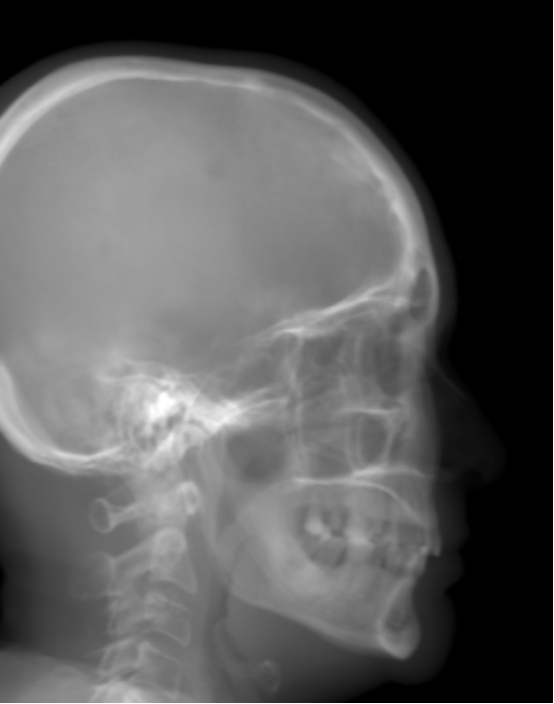}
}
\end{minipage}

\begin{minipage}[b]{0.7\linewidth}
\subfigure[Plot of samples and sigmoid curves]{
\includegraphics[width = 1\linewidth]{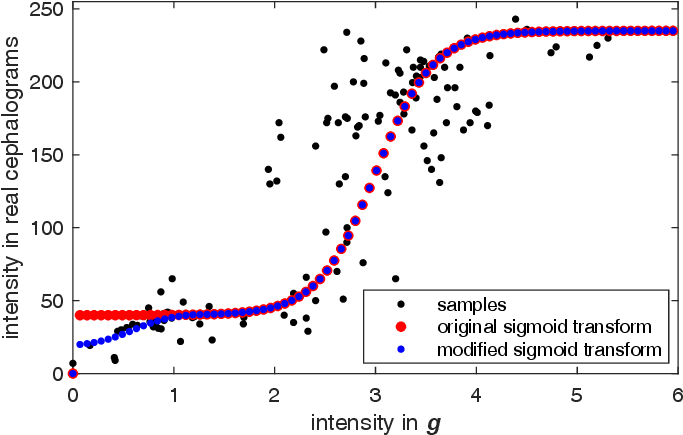}
}
\end{minipage}
\caption{Image contrast difference between real conventional cephalograms and RayCast synthetic cephalograms: (a) a real cephalogram example; (b) a RayCast synthetic cephalogram example; (c) the plot of samples between RayCast synthetic cephalograms and real cephalograms with \modified{an} original sigmoid curve (red) and our proposed modified sigmoid curve (blue). }
\label{Fig:sigmoidCurve}
\end{figure}
In Fig.\,\ref{Fig:sigmoidCurve}, one real conventional cephalogram example and one example of $\vg$ are displayed, where evident image contrast difference between these two images is observed. It is because the X-ray films used in conventional cephalograms have the special nonlinear optical property that the characteristic curve between optical density and logarithmized X-ray exposure has a sigmoid-like shape \citep{ritenour1996physics}. According to the Lambert-Beer law, the logarithmized X-ray exposure is equivalent to the attenuation integral. It indicates that the intensity relation between the desired cephalogram and the attenuation integral image $\vg$ should also exhibit a sigmoid-like curve. Therefore, to make the image contrast of synthetic cephalograms similar to conventional cephalograms, a sigmoid intensity transform is necessary. 

The standard sigmoid function is defined as the following,
\begin{equation}
\sigma(x) = 1/(1 + e^{-x}).
\end{equation}
 Considering shift, scaling and the intensity range [0, 255] in 8-bit gray scale images, the following general sigmoid function is proposed to transform the intensities of $\vg$,
\begin{equation}
\tilde{\vg}(y,z) = c_1 + (255 - c_1 - c_2) /\left(1 + e^{-s\cdot \left(\vg(y,z) - t\right)}\right), 
\label{eqn:sigmoidTransform}
\end{equation}
where $c_1$ is a base intensity value caused by film base attenuation and fog or unwanted exposure during storage and handling \citep{ritenour1996physics}, $c_2$ is a parameter to decide the intensity saturation value, $t$ is an intensity shift parameter, and $s$ is a scaling factor for the slope of the curve. Since the standard sigmoid function has a value between 0 and 1, with the above sigmoid transform, $\tilde{\vg}$ has an intensity range of [$c_1$, $255 - c_2$].

%In X-ray imaging, the attenuation of X-ray intensity follows the Lambert-Beer's law,
%\begin{equation}
%I = I_0 \cdot e^{-p},
%\end{equation}
%where $I_0$ is the incident X-ray intensity without attenuation, $p$ is the attenuation value of an imaged object, and $I$ is the transmitted X-ray intensity. In cephalometers, cephalgram images are obtained by X-ray films, where the relationship between the optical density and the logarithmized X-ray exposure follows the shape of a sigmoid function \citep{ritenour1996physics}, i.e.,
%\begin{equation}
%v \propto \sigma\left(\log(I) - \tau\right)= \sigma\left( \right),
%\label{eqn:tanh}
%\end{equation} 
%where $\sigma(x) = 1/(1 + e^{-x})$, $v$ is the pixel intensity in a cephalogram and $\tau$ is a shift parameter.
To find the parameters for the above general sigmoid transform, the mean intensity values of samples chosen in certain regions of $\vg$ together with the corresponding values sampled from real conventional cephalograms are plotted as black dots in Fig.\,\ref{Fig:sigmoidCurve}(c). The sigmoid transform of Eqn.\,(\ref{eqn:sigmoidTransform}) is plotted as the red-dotted curve, whose parameters are determined by least squares curve fitting. As displayed, most samples are located near the sigmoid curve, which is consistent with the characteristic curve in X-ray films. Note that the positions of the samples are approximated very coarsely due to the absence of dental CBCT volumes and their corresponding cephalograms. With matching pairs, a more accurate intensity transform can be learned.
%
%\subsubsection{Air background recovery}
%After the above sigmoid transform, the air background is not easy to distinguish from low intensity soft tissue areas, since they both have values close to the base value $c_1$. To detect the air background, a thresholding operation with a parameter $\tau$ is applied to the attenuation integral image $\vg$ to get a binary mask $\vm$ for the non-background area, i.e., $\vm(y,z)=0$ if $\vg(y,z) < \tau$; otherwise $\vm(y,z)=1$.
%%\begin{equation}
%%\vm(y,z)=
%%\left\lbrace \begin{array}{ll}
%%0, & \text{if } \vg(y,z) < \tau,\\
%%1, & \text{otherwise, } \\
%%\end{array}
%%\right.
%%\label{eqn:thresholding}
%%\end{equation}
%%where $\vm$ is a mask for the non-background area. 
%With the mask $\vm$, the final cephalogram $\vg^\ast$ is obtained as the point-wise multiplication of $\vm$ and $\vg'$,
%\begin{equation}
%\vg^\ast(y,z)=\vm(y,z) \cdot \vg'(y,z).
%\end{equation}

\subsubsection{Modified sigmoid transform}
With the original sigmoid transform, the air background and low intensity soft tissues both have values close to $c_1$. To recover air background, pixel values smaller than a threshold $\tau_1$ in $\vg$ are set to 0. In addition, to recover the contrast in soft tissues, for the low intensity range [$\tau_1$, $\tau_2$], another sigmoid function is used,
\begin{equation}
\hat{\vg}(y,z) = c_3 + c_4 /\left(1 + e^{- \left(\vg(y,z) - (\tau_1+\tau_2)/2\right)}\right),
\end{equation}
where $c_3$ is a modified base intensity value and $c_4$ is an intensity parameter determined by setting $\hat{\vg}(y,z) = \tilde{\vg}(y,z)$ at $\vg(y,z) = \tau_2$ for continuity. Here we choose a second sigmoid function instead of a linear function to make the curve smoother at the transition point $\vg(y,z) = \tau_2$.

In summary, the final cephalogram $\vg^\ast$ is obtained as
\begin{equation}
\vg^\ast(y,z)= \left\lbrace
\begin{array}{ll}
0, & \vg(y,z) < \tau_1, \\
\hat{\vg}(y,z), & \tau_1 \leq \vg(y,z) \leq \tau_2, \\
\tilde{\vg}(y,z), & \vg(y,z) > \tau_2,
\end{array}
\right.
\label{eqn:twoStepSigmoidTransform}
\end{equation}
where the intensity transform is a modified sigmoid function, illustrated as the blue-dotted curve in Fig.\,\ref{Fig:sigmoidCurve}.

\subsection{Type II: Cephalogram Synthesis from 2D CBCT Projections}
In this subsection, we propose a deep learning method to synthesize 2D cephalograms directly from 2D logarithmized projections for low-dose purpose.
\modified{The effective dose of dental CBCT systems, varying from 50\,$\mu$Sv to 600\,$\mu$Sv depending on systems, is noticeably lower than that of MSCT systems which is typically more than 1000\,$\mu$Sv \citep{lorenzoni2012cone}. For example, the effective dose of a 3D scan for a modern NewTom9000 system (23\,cm field-of-view) is 56.2\,$\mu$Sv \citep{silva2008cone}. In comparison, the effective dose for a cephalogram acquired from sephlometer systems is typically between 1\,$\mu$Sv and 6\,$\mu$Sv \citep{lorenzoni2012cone}. For example, it is 2.3\,$\mu$Sv for a film-based Siemens Orthophos C (Sirona Dental) system \citep{visser2001dose} and 1\,$\mu$Sv for the digital cephalometer module in the NewTom9000 system \citep{silva2008cone}. For a 3D scan, the number of projections is typically more than 300 for a CBCT system. Our Type II synthesis uses two projections only and hence the effective dose is lower than that of a conventional cephalometer system. For example, it is below 0.38$\mu$Sv for the NewTom9000 system.}

 In order to train deep learning models \modified{to synthesize 2D cephalograms}, it is beneficial to have clinical dental CBCT projections and their corresponding conventional cephalograms as pairs. However, in practice, it is infeasible to obtain a sufficient number of such pairs due to ethic considerations, privacy concerns, and clinical regulations. For a proof of concept, we choose to use synthetic projection images from publicly available CBCT head data \citep{chilamkurthy2018development} as a surrogate  in this work. The projections simulated based on a regular dental CBCT system configuration and the cephalograms synthesized by the above Type I method with orthogonal projection are used as pairs. 

Dental CBCT systems typically have a shorter source-to-isocenter distance and a longer isocenter-to-detector distance than cephalometer systems. As a result, dental CBCT projections have more severe perspective deformation than conventional cephalograms, in addition to the image contrast difference. Therefore, the neural network needs to learn both the perspective deformation and the image contrast transform. 

\subsubsection{Neural network}
For image-to-image translation, GANs are the state-of-the-art. Therefore, in this work, we propose to apply a pixel-to-pixel generative adversarial network (pix2pixGAN) \citep{isola2017image} for cephalogram synthesis. The U-Net is used as the generator $G$ while a 5-layer CNN is used as the discriminator $D$ \citep{isola2017image}. $G$ learns to convert a cone-beam projection to a cephalogram. $D$ learns to distinguish the synthetic cephalogram from the target cephalogram. The objective of the conditional GAN is,
\begin{equation}
\mathcal{L}_{\text{cGAN}}(G,D) = \mathbb{E}_{\vx,\vy}\left[\log{D(\vx, \vy)}\right] + \mathbb{E}_{\vx}\left[\log{\left(1 - D(\vx, G(\vx)\right)}\right],
\end{equation}
where $\vx$ is the input, $\vy$ is the target, $G$ tries to minimize this objective against an adversarial $D$ that tries to maximize it, i.e., $G^{\ast}= \arg\min_{G}\max_{D}\mathcal{L}_{\text{cGAN}}(G,D)$. In addition, a weighted $\ell_1$ loss function is applied to train the generator's output close to the target with less blurring compared to $\ell_2$ loss,
\begin{equation}
\mathcal{L}_{\ell_1}=\mathbb{E}_{\vx,\vy}\left[||\vw \cdot (\vy - G(\vx))||_1 \right],
\label{eqn:gradientweightloss}
\end{equation}
where $\vw$ is a weight map calculated by the Sobel edge detector to emphasize edges \citep{stimpel2019projection}.
\subsubsection{Rebinning}
In the CBCT system, we denote the source-to-isocenter distance by $d_0$ and the source-to-detector distance by $d_1$. Due to perspective projection, the anatomical structures at the midsagittal plane, which passes through the isocenter, have a magnification factor of $d_1/d_0$. To remove this magnification factor, the acquired CBCT projections are rebinned into a virtual detector (VD) located at the midsagittal plane. Such rebinning removes the magnification for structures in the midsagittal plane. However, structures in other sagittal planes still have different magnification factors, although these factors are reduced by rebinning. Therefore, the perspective deformation remains.

\subsubsection{Patch selection}

Cone-beam projections and cephalograms typically have a large image size. To avoid high computation burden, patch-wise learning is applied. In this work, the input of the generator $G$ is a patch from a cone-beam projection while the target output is the corresponding patch from the paired cephalogram. 

\begin{figure}[t]
\centering
\begin{minipage}{0.44\linewidth}
\subfigure[General case]{
\includegraphics[width = 1\linewidth]{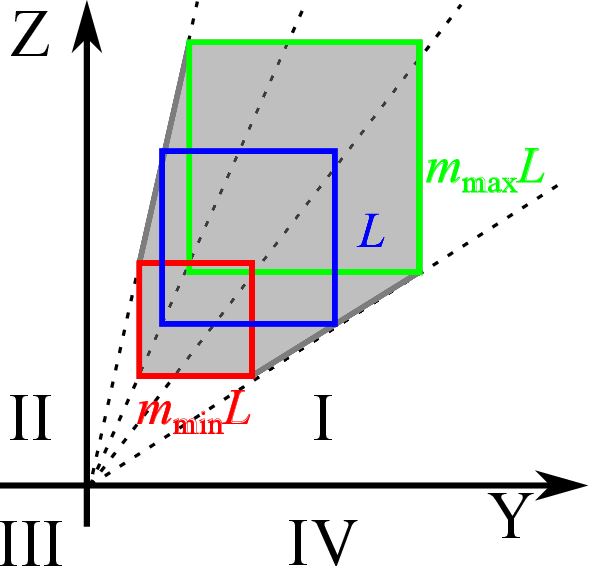}
}
\end{minipage}
\hspace{5pt}
\begin{minipage}{0.44\linewidth}
\subfigure[Special case]{
\includegraphics[width = 1\linewidth]{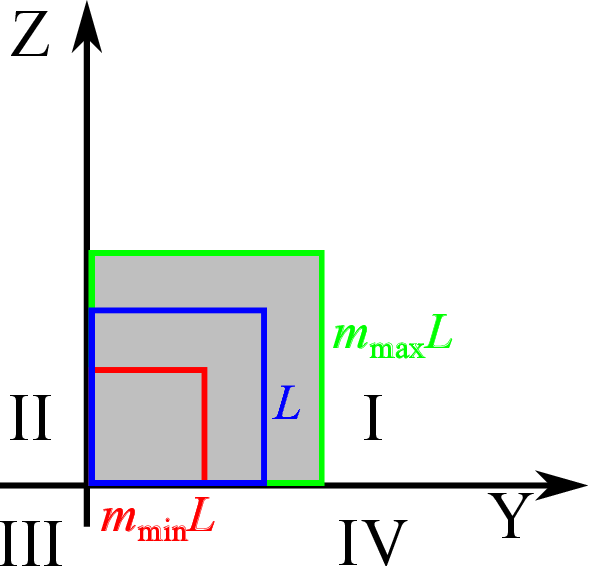}
}
\end{minipage}
\caption{The cone-beam projections of volume-domain patches on the VD. The blue square corresponds to the projected patch without any magnification, while the red square corresponds to the projected patch with the minimum magnification and the green square corresponds to the projected patch with the maximum magnification. The grey area corresponds to the union set of all the projected patches with different magnification \modified{factors} between $m_{\min}$ and $m_{\max}$. (a) is a general case where the left bottom corners of the projected patches are inside the first quadrant, while (b) is a special case where the patch corners are located at origin. }
\label{Fig:perspectivePatch}
\end{figure}

Note that due to perspective deformation the patch pairs need to be carefully selected. In the 3D patient volume, a 2D square patch can be determined by its vertex location, edge length and orientation (direction of its normal vector). Here we consider patches all oriented along the $X$ axis. Hence we can denote a 2D square patch with a left bottom vertex location $(x, y, z)$ and an edge length $L$ by $p_\text{volume}(x,y,z,L)$.
Now we consider a set of parallel patches which share the same $Y$ and $Z$ coordinates, $y_0$ ($y_0\geq 0$) and $z_0$ ($z_0 \geq 0$) respectively, for the left bottom vertexes and the same edge length $L_0$, while the $X$ coordinate $x$ can shift between $x_{\min}$ and $x_{\max}$. Such a patch is denoted by $p_\text{volume}(x,y_0,z_0,L_0)$, where $x_{\min} \leq x \leq x_{\max}$. In cone-beam projection, the anatomical structures in such a patch have a magnification factor of $m = d_0/(d_0 - x)$ in the VD. Hence, the projection of this patch has a corresponding left bottom corner vertex $(m\cdot y_0, m\cdot z_0)$ and an edge length $m\cdot L_0$. Since the magnification factor $m$ varies between $m_{\max}=d_0/(d_0 - x_{\min})$ and $m_{\min} = d_0/(d_0 - x_{\max})$, the cone-beam projections of all the patches $p_\text{volume}(x, y_0,z_0,L_0)$, where $x_{\min} \leq x \leq x_{\max}$, are located in a hexagon, as displayed in the grey area in Fig.\,\ref{Fig:perspectivePatch}(a). However, the orthogonal projections of all the patches $p_\text{volume}(x, y_0,z_0,L_0)$, where $x_{\min} \leq x \leq x_{\max}$, are located in a square patch with the corresponding left bottom vertex $(y_0, z_0)$ and the edge length $L_0$, as displayed in the blue square in Fig.\,\ref{Fig:perspectivePatch}(a). Therefore, it is difficult to find exact matching patch pairs in the general case for such a hexagon-to-square mapping.

However, in the special case of $y_0=0$ and $z_0=0$, this hexagon area becomes a square, as displayed in Fig.\,\ref{Fig:perspectivePatch}(b). But the grey square area and the blue square area in Fig.\,\ref{Fig:perspectivePatch}(b) have different edge lengths. This issue can be relieved by choosing a large patch size $L$, e.g., each patch being one quadrant, so that the area between the blue square and the grey square has zero values since human heads are compact. 
%In general case, if $y_0=-L/2$ and $z_0=-L/2$, the cone-beam projection of this central patch remains a square area. But all other patches fail to map to square areas. 
Accordingly, in this work, we divide each CBCT projection into four patches according to the four quadrants. \modified{With such patch selection, paired patch-to-patch translation is feasible.}

\subsubsection{One model for multi-quadrant patches}
The perspective deformation is inhomogeneous. For the patches in the first quadrant, the anatomical structures near the left bottom corner have the minimum deformation while those near the right top corner have the most deformation. However, for the patches in the second quadrant, the anatomical structures near the right bottom corner have the minimum deformation while those near the left top corner have the most deformation. Therefore, an individual model needs to be trained for each quadrant due to different perspective deformation patterns. However, it is likely that the four models will learn (or rather ``memorize'') quadrant-specific features, which may not be related to perspective deformation. To mitigate this problem and to reduce the computation burden of training four models as well, the symmetry property is utilized. If the patches from the second quadrant are flipped horizontally, then the patches have the same perspective deformation as those from the first quadrant. Similarly, we can apply horizontal + vertical flipping and vertical flipping respectively for the patches from the third and the fourth quadrants to get the same deformation pattern. With such flipping operations, all the patches from different quadrants can be used together to train one model. Such a model is expected to learn the common features in these four-quadrant patches, i.e. perspective deformation, instead of quadrant-specific features.

\subsubsection{Dual projections to one cephalogram synthesis}
\begin{figure}[t]
\centering
\includegraphics[width = 0.6\linewidth]{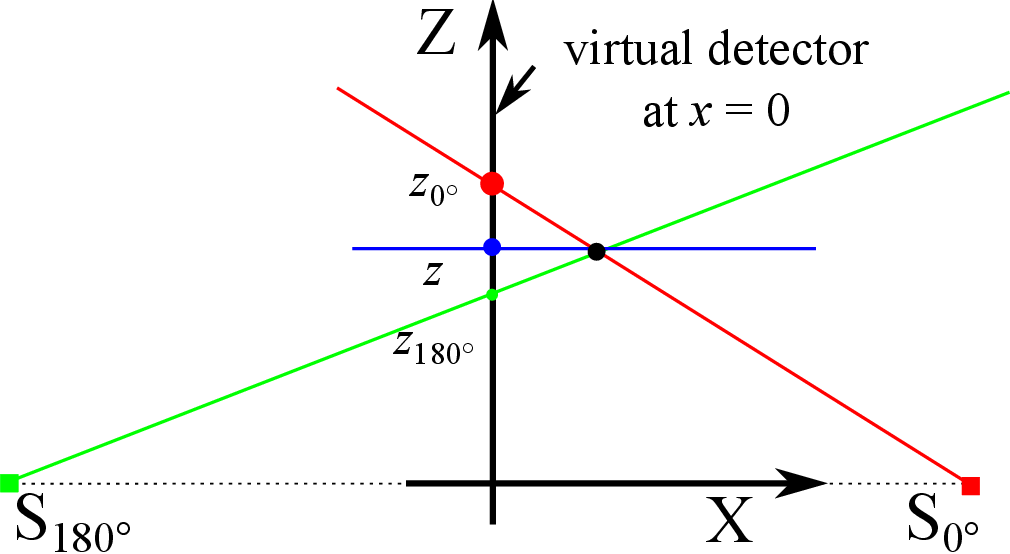}
\caption{The benefit of dual projections in localizing anatomical structures in cone-beam projections visualized in the $X$-$Z$ plane. The VD is located in the plane of $x=0$. The cone-beam projections (the red and green points) of the black point $(x, y, z)$ from the $0^\circ$ and $180^\circ$ X-ray sources to the VD have the heights of $z_{0^\circ}$ and $z_{180^\circ}$ respectively, while its orthogonal projection (the blue point) has the height of $z$ with the relation $z_{0^\circ} < z < z_{180^\circ}$.}
\label{Fig:dualProjection}
\end{figure}

In orthogonal projection, if the parallel-beam rays are rotated by $180^\circ$, the acquired projection is the same as the original projection after a horizontal flip. Therefore, in parallel-to-cone projection conversion, using an additional $180^\circ$ projection is entirely redundant. However, in cone-to-parallel projection conversion in this work, $180^\circ$ projections can provide additional information together with $0^\circ$ projections due to the following two factors: a) The isocenter (or rotation axis) of a dental CBCT system is not perfectly locate at the midsagittal plane of a head; b) Human heads are not perfectly symmetric with respect to (w.\,r.\,t.) the midsagittal plane. Therefore, using dual projections is beneficial in localizing anatomical structures with perspective deformation. To illustrate this benefit, a sketch of the dual cone-beam projections of a point visualized in the $X$-$Z$ plane is displayed in Fig.\,\ref{Fig:dualProjection}. The cone-beam projections (the red and green points) of the black point $(x, y, z)$ from the $0^\circ$ and $180^\circ$ X-ray sources to the VD (located at the $Y$-$Z$ plane with $x=0$) have the heights of $z_{0^\circ}$ and $z_{180^\circ}$, respectively, while the orthogonal projection (the blue point) of the black point has the height of $z$. It is clear that the value of $z$ is between $z_{0^\circ}$ and $z_{180^\circ}$.
%, if the point is not located in the plane of $x=0$. 
This relation indicates that the orthogonal projection of an anatomical structure must be between the locations of its $0^\circ$ and $180^\circ$ cone-beam projections. 

It is worth noting that using projections other than the $0^\circ$ and $180^\circ$ projections, e.g. $1^\circ$ or $90^\circ$, will introduce additional deformation caused by angular rotations. Therefore, only $0^\circ$ and $180^\circ$ these two angles are chosen.

\begin{figure}[t]
\centering
\begin{minipage}{0.38\linewidth}
\subfigure[$0^\circ$ CBCT projection]{
\includegraphics[width = 1\linewidth]{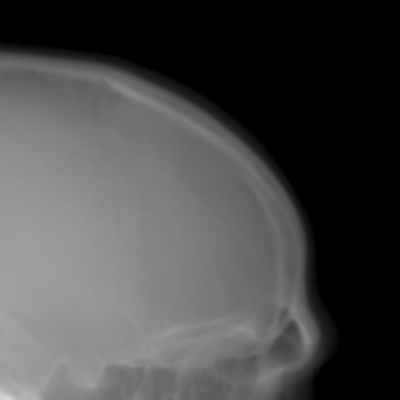}
}
\end{minipage}
\begin{minipage}{0.38\linewidth}
\subfigure[$180^\circ$ CBCT projection]{
\includegraphics[width = 1\linewidth]{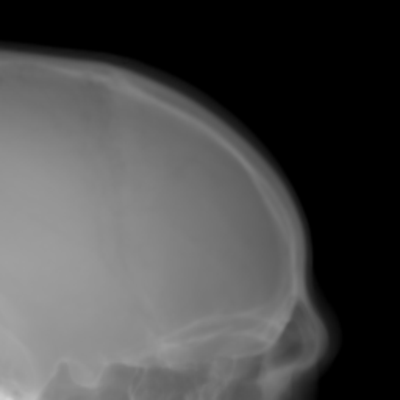}
}
\end{minipage}

\begin{minipage}{0.38\linewidth}
\subfigure[RGB input patch]{
\includegraphics[width = \linewidth]{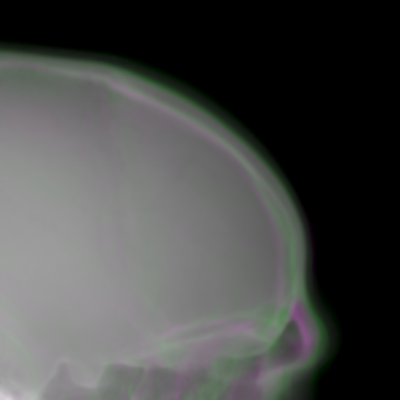}
}
\end{minipage}
\begin{minipage}{0.38\linewidth}
\subfigure[Target patch]{
\includegraphics[width = \linewidth]{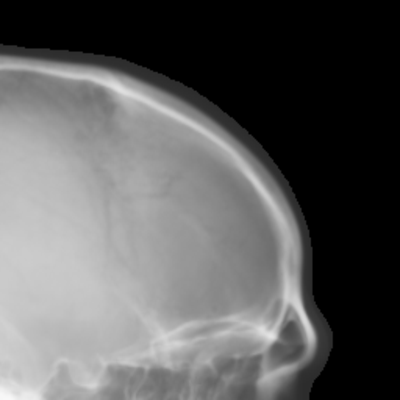}
}
\end{minipage}
\caption{One patch example for dual projections to one cephalogram synthesis: (a) the first quadrant patch from the $0^\circ$ rebinned cone-beam projection; (b) the first quadrant patch (horizontally flipped) from the $180^\circ$ rebinned cone-beam projection; (c) the RGB patch using (a) for the red and blue \modified{channels} and (b) for the green \modified{channel}, where the colourful \modified{areas} highlight the difference between (a) and (b); (d) the target patch synthesized by our proposed volume-to-cephalogram method.}
\label{Fig:RGBPatch}
\end{figure}

To combine such dual projection information, we convert the patches from $0^\circ$ and $180^\circ$ cone-beam projections to 3-channel patches forming RGB color patches. The $0^\circ$ patch is used for the red and blue channels, while the $180^\circ$ patch is used for the green channel. The $0^\circ$ patch instead of the $180^\circ$ patch takes two channels, since the target cephalograms are also acquired in the $0^\circ$ view in our setting. In the RGB patch, the intensity difference between the $0^\circ$ patch and the $180^\circ$ patch is revealed by the color: grey areas have the same values for the three channels, indicating that the intensity values from the $0^\circ$ and $180^\circ$ patches respectively are close to each other, while green areas indicate that the $180^\circ$ patch has larger values and magenta areas indicate that the $0^\circ$ patch has larger values. An example is displayed in Fig.\,\ref{Fig:RGBPatch}, where Figs.\,\ref{Fig:RGBPatch}(a) and (b) are the $0^\circ$ and $180^\circ$ patches respectively, Fig.\,\ref{Fig:RGBPatch}(c) is the RGB patch as the input of the neural network, and Fig.\,\ref{Fig:RGBPatch}(d) is the corresponding target output of the neural network.

\subsection{Super Resolution}
\begin{figure}[t]
\centering
\includegraphics[width = 0.9\linewidth]{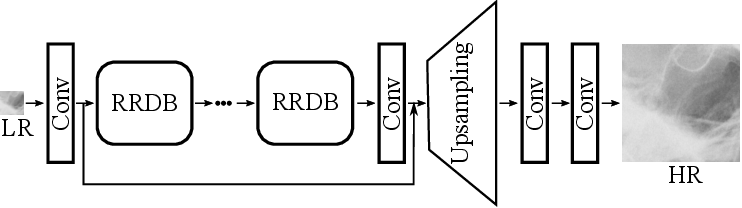}
\caption{The generator architecture in ESRGAN\textsubscript{RRDB} \citep{wang2018esrgan}.}
\label{Fig:SRGAN}
\end{figure}

%In high-end modern dental CBCT systems, the voxel resolution can reach as high as 0.1\,mm/pixel. However, 
In dental CBCT systems, the flat-panel detectors typically have a resolution around 0.3\,mm/pixel. Due to the pursue of fast reconstruction, typically the 3D volume resolution is around 0.5\,mm/pixel \citep{hatvani2018tensor}. In contrast, the image resolution in conventional film-based cephalograms is as high as 0.1\,mm/pixel. Therefore, image resolution in synthetic cephalograms is worse than that in real conventional cephalograms in general. To reduce blur in synthetic cephalograms, deep learning SR techniques are applied.

In this work, we investigate the application of two ESRGANs \citep{zhang2018residual,wang2018esrgan} for SR. For distinction, we refer to \citep{zhang2018residual} as ESRGAN\textsubscript{RDB} and \citep{wang2018esrgan} as ESRGAN\textsubscript{RRDB} respectively, as they utilize RDBs and RRDBs respectively for the basic blocks in the generator. The architecture of the generator in ESRGAN\textsubscript{RRDB} is displayed in Fig.\,\ref{Fig:SRGAN} as an example. Since in this work the scaling factor from LR to HR images is large, 0.5\,mm/pixel to 0.1\,mm/pixel particularly, checkerboard artifacts \citep{odena2016deconvolution} are observed in predicted HR images, although the PixShuffle or deconvolution operation is replaced by upsampling followed by a convolution \citep{odena2016deconvolution}. Therefore, similar to SRCNN \citep{dong2015image}, we choose to use ILR images using bicubic upsampling as the input of the generator along with the removal of the upsampling layer, which effectively reduces checkerboard artifacts. Additional information on network architecture, loss function and training procedure for the ESRGANs are provided in the original publications \citep{zhang2018residual} and \citep{wang2018esrgan}, respectively. In addition, the U-Net has been demonstrated effective for SR in dental imaging \citep{hatvani2018deep}. Therefore, pix2pixGAN using the U-Net generator is also investigated to map ILR images to HR images. Note that the SR task uses an individual neural network because it allows us to train with very small patches. If it is included in the pix2pixGAN for Type II synthesis, very large patches (in our experiments, $1280\times 1280$) are required, which is very computationally expensive.

%One RRDB contains three dense blocks which are connected by residual connections. Each dense block contains five convolutions and four rectified linear units (ReLUs). The ESRGAN uses a low resolution (LR) image as the input and the desired super resolution image as the output.  A $3 \times 3$ convolution operation is applied to the input LR image first. After that, a set of RRDBs is applied followed by the second convolution. The outputs of the first convolution and the second convolution are connected together by a residual connection. After that, a upsampling layer and two convolution layers are applied to get the final output.

\subsection{Automated Cephalometric Landmark Detection}
\begin{figure}[t]
\centering
\begin{minipage}{1\linewidth}
\centering
\includegraphics[width=0.73\linewidth]{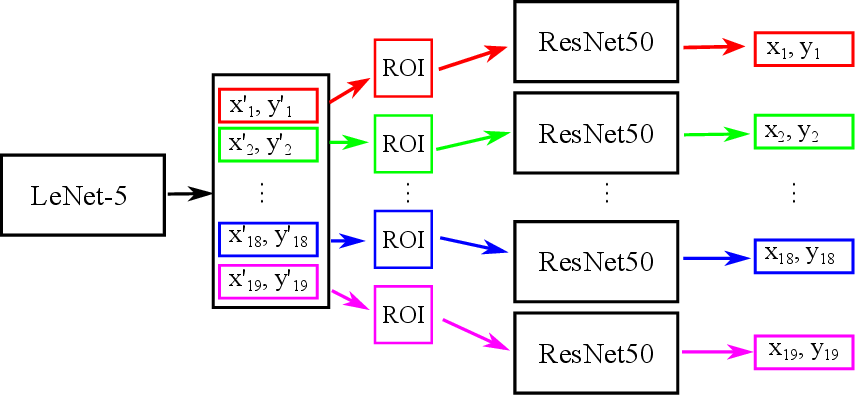}
\end{minipage}
\caption{The combination of LeNet-5 and ResNet50 for landmark detection, where LeNet-5 approximately determines the ROI patches of 19 landmarks and each ResNet50 further determines the final location of each landmark.}
\label{Fig:landmarkDetectionNeuralNetwork}
\end{figure}

For cephalometric landmark detection, we propose a fully automated deep learning method combining LeNet-5 \citep{lecun1998gradient} and ResNet50 \citep{he2016deep}. Our method is an improved version of the latest landmark detection method \citep{song2020automatic} in terms of efficiency. In \citep{song2020automatic}, the ResNet50 is used to detect the location of one landmark in each ROI patch. To obtain the ROI patches, registering a test image to 150 training images to find the closet reference image is necessary. However, this registration step is computationally expensive and can take up to 20 minutes \citep{song2020automatic}. Therefore, we propose to utilize another neural network to determine the region which ResNet50 should pay attention to. In this work, we choose the LeNet-5 \citep{lee2017cephalometric} to obtain such ROI patches. LeNet-5 has a simple architecture, which is efficient and stable for training. Although LeNet-5 is not sufficient to detect the 19 landmarks accurately, it is sufficient to detect \modified{an} ROI patch for a subsequent neural network to work on.

The whole neural network architecture is displayed in Fig.\,\ref{Fig:landmarkDetectionNeuralNetwork}. 
%As original cephalograms have a large image size, they are down-sampled by a factor of 5 as the input of the LeNet-5. Since we only need approximate locations of the landmarks in this step, the accuracy loss caused by the down-sampling operation does not make a big difference, but it rather improves training efficiency. 
The output of the LeNet-5 is a 1-dimensional vector of 38 elements, representing the 38 coordinate candidates of the 19 landmarks. For the $i^\text{th}$ landmark, the predicted position is denoted by $(x'_i, y'_i)$. Each position determines an ROI patch with a size of $512 \times 512$ pixels centred at $(x'_i, y'_i)$ in the HR cephalogram image. The large patch size also relieves the accuracy demand on the LeNet-5. For each ROI patch, a ResNet50 model is trained respectively to predict the accurate position of the corresponding landmark. The final predicted position of each landmark is denoted by $(x_i, y_i)$ for the $i^\text{th}$ landmark. Here we train 19 different ResNet50 models instead of training one model to predict 19 landmarks simultaneously, as landmark specific features can be extracted by each ResNet50 to achieve higher accuracy.

\begin{table*}[t]
\centering
\caption{\modified{The methods/architectures and datasets used in our cephalometric analysis pipeline.}}
\label{Tab:pipelineInformation}
\begin{footnotesize}
\begin{tabular}{|l|c|c|c|c|}
\hline
Content & Method/ & \multicolumn{2}{c|}{Training data} & Test data \\
\cline{3-5}
\ & Architecture & Input & Target & Input \\
\hline
Type I synthesis& Regression  & CQ500  & ISBI cephalograms & CQ500\\
& & CBCT volumes & &CBCT volumes\\
\hline
Type II synthesis& Pix2pixGAN & CQ500  & Type I synthesis & CQ500  \\
 & & CBCT projections & from CQ500 volumes &CBCT projections \\
 \hline
SR & Pix2pixGAN  &ILR ISBI  &HR ISBI  &CQ500 Type-I or \\
& (ESRGAN) & cephalograms & cephalograms & Type-II cephalograms \\
\hline
Landmark  & LeNet-5 +  & ISBI &ISBI & CQ500 Type-I or\\
detection & ResNet50 &cephalograms &cephalometric coordinates & Type-II cephalograms\\
\hline
\end{tabular}
\end{footnotesize}
\end{table*}

\begin{figure*}[t]
\centering
\begin{minipage}[t]{0.24\linewidth}
\subfigure[Orthogonal RayCast, \modified{8.32}]{
\includegraphics[width=\linewidth]{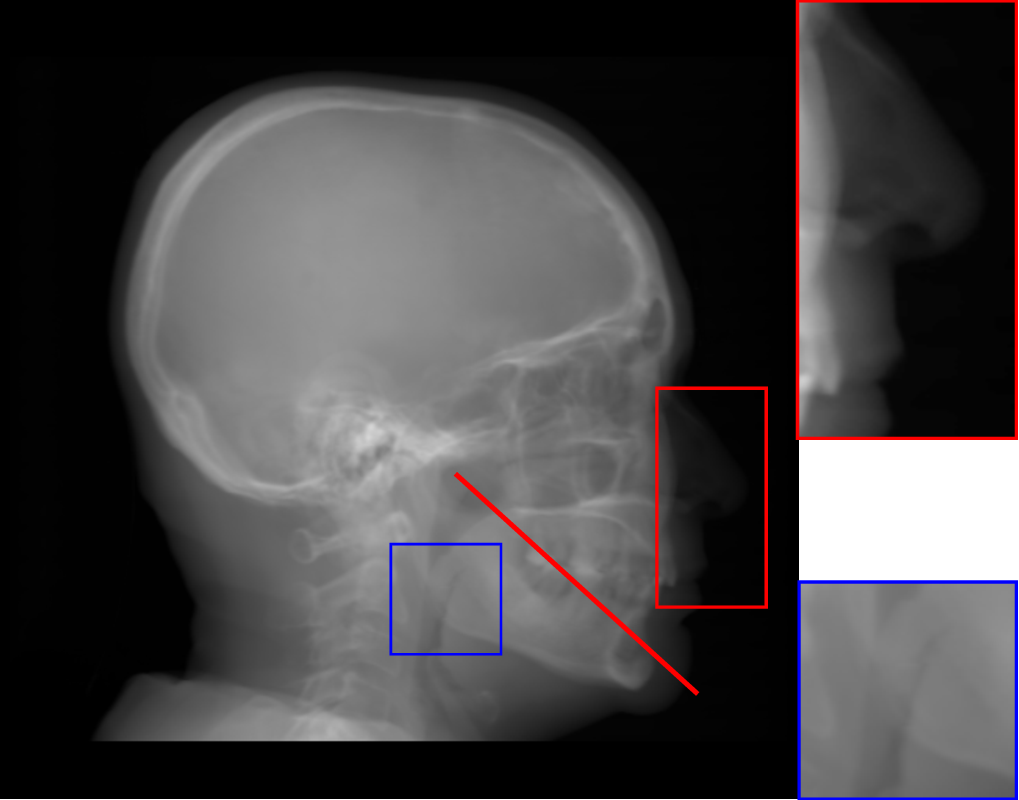}
}
\end{minipage}
\begin{minipage}[t]{0.24\linewidth}
\subfigure[Perspective RayCast, \modified{8.23}]{
\includegraphics[width=\linewidth]{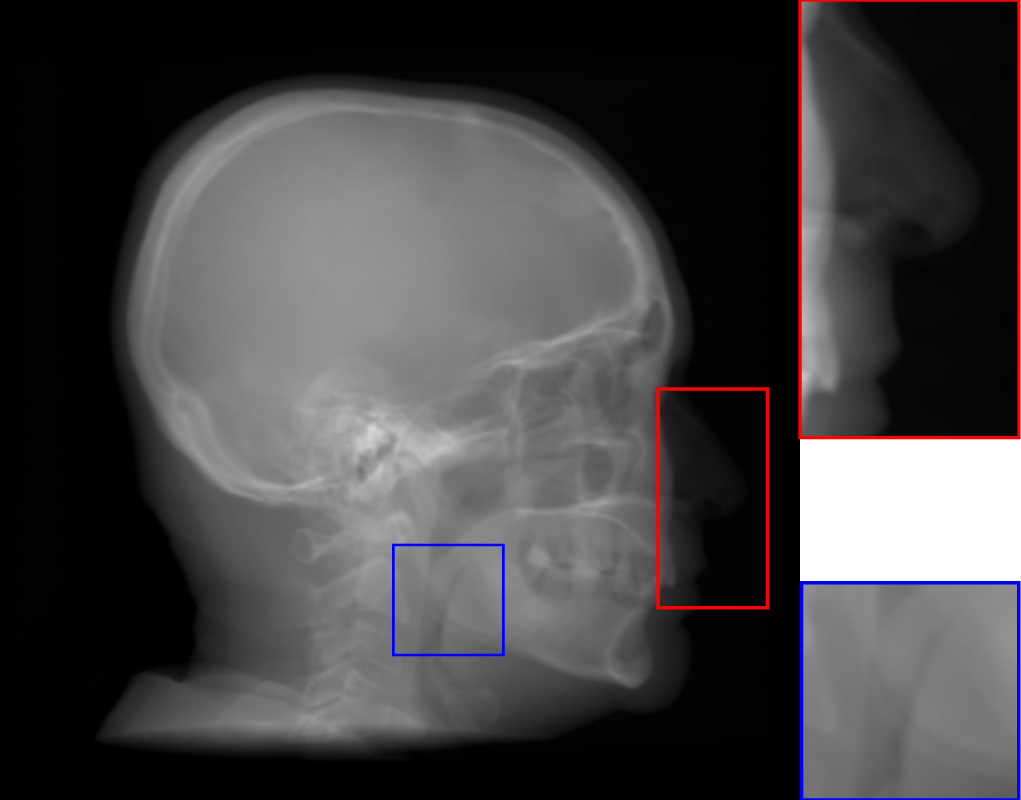}
}
\end{minipage}
\begin{minipage}[t]{0.24\linewidth}
\subfigure[MIP100, \modified{7.59}]{
\includegraphics[width=\linewidth]{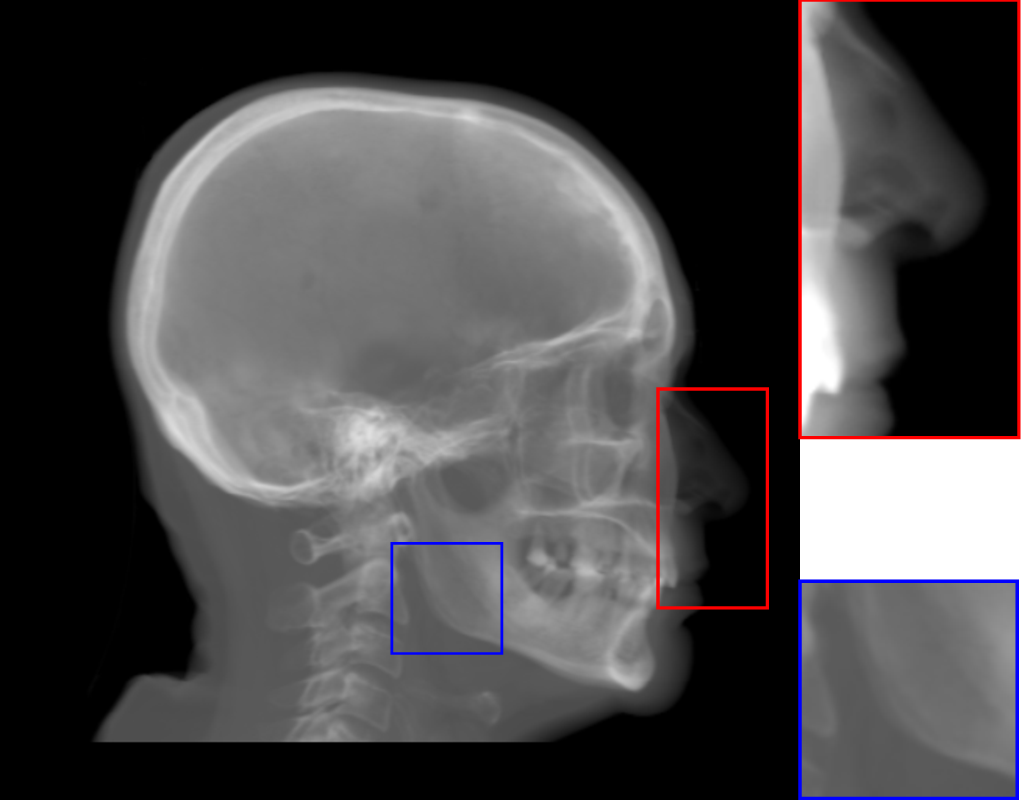}
}
\end{minipage}
\begin{minipage}[t]{0.24\linewidth}
\subfigure[CycleGAN, \modified{5.53}]{
\includegraphics[width=\linewidth]{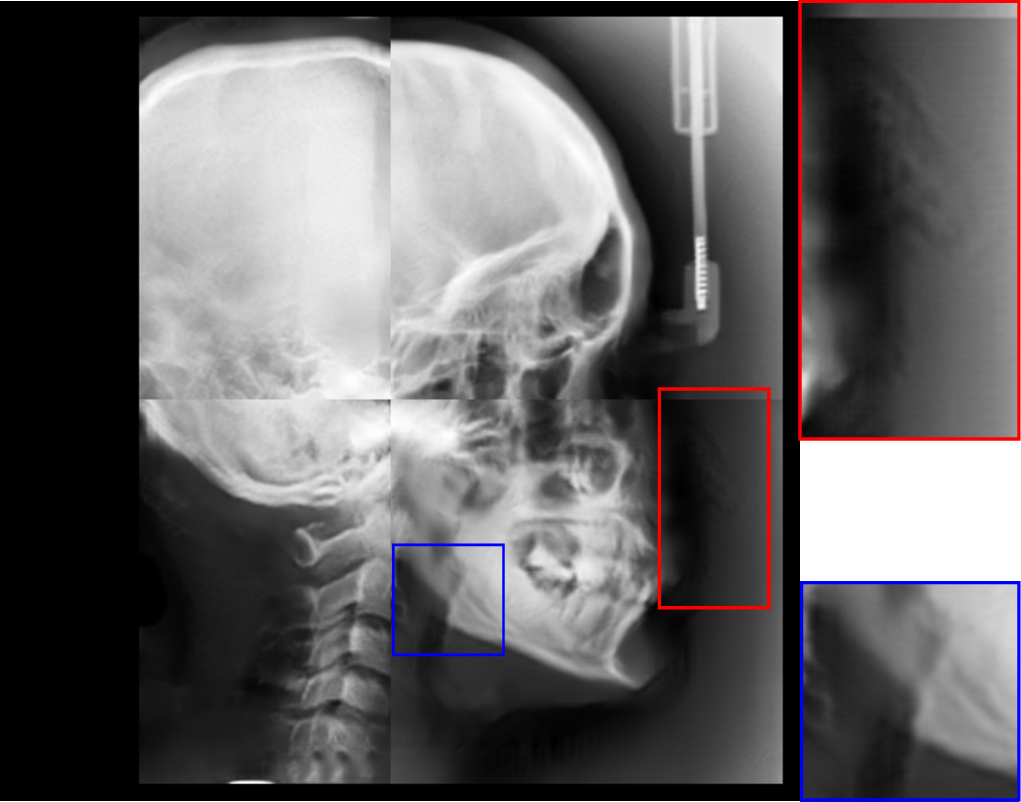}
}
\end{minipage}

\begin{minipage}[t]{0.24\linewidth}
\subfigure[Orthogonal RayCast, enhanced, \modified{7.78}]{
\includegraphics[width=\linewidth]{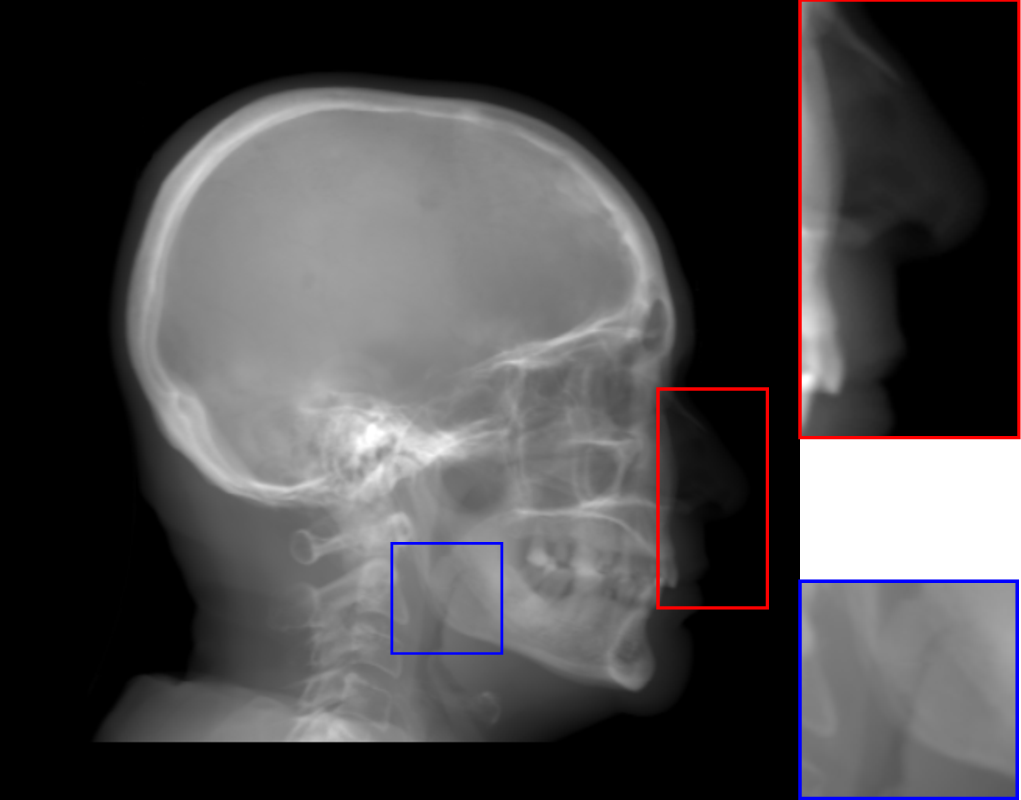}
}
\end{minipage}
\begin{minipage}[t]{0.24\linewidth}
\subfigure[Original sigmoid transform, \modified{6.93}]{
\includegraphics[width=\linewidth]{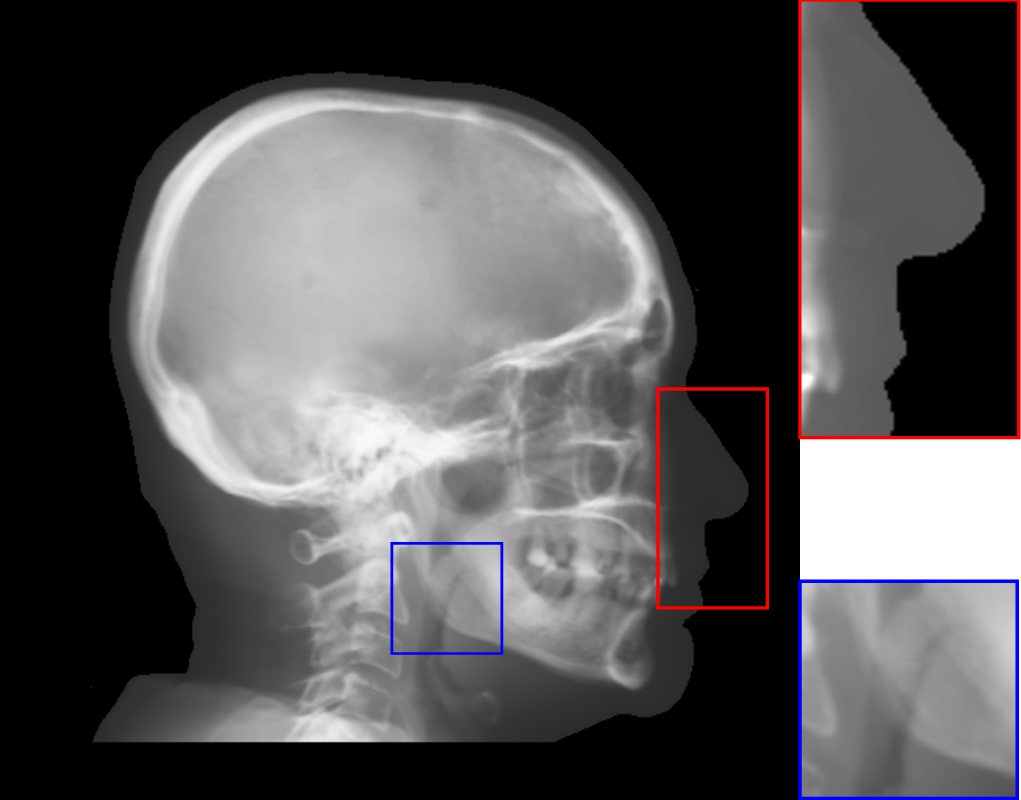}
}
\end{minipage}
\begin{minipage}[t]{0.24\linewidth}
\subfigure[Proposed, \modified{6.83}]{
\includegraphics[width=\linewidth]{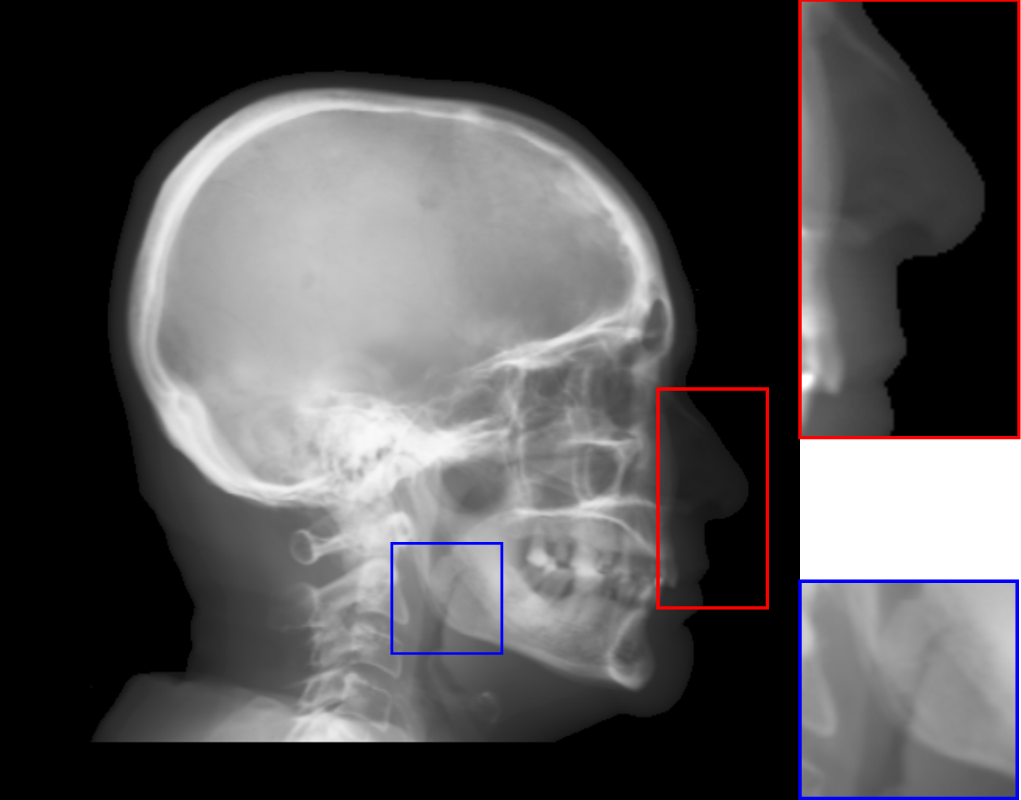}
}
\end{minipage}
\begin{minipage}[t]{0.24\linewidth}
\subfigure[Proposed with perspective projection, \modified{6.77}]{
\includegraphics[width=\linewidth]{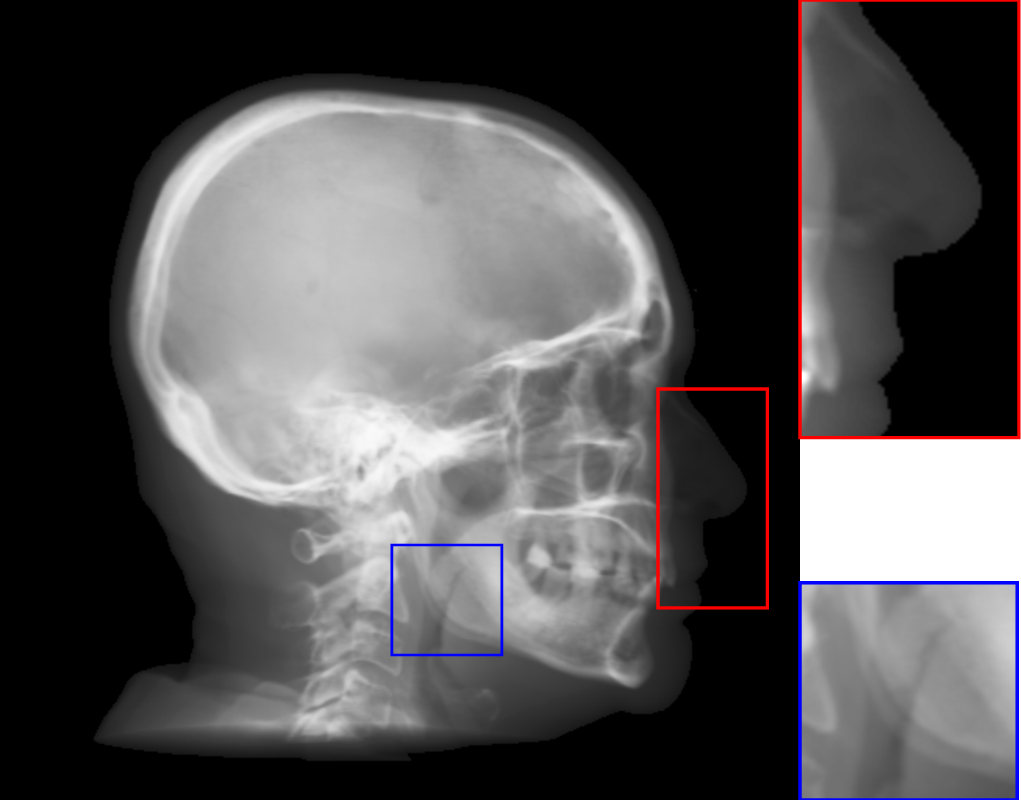}
}
\end{minipage}
\caption{Synthetic cephalogram examples from 3D CBCT volumes with different methods. (a)-(d) are comparison results while (e)-(h) are the (intermediate) results of our proposed method: (a) RayCast using orthogonal projection; (b) RayCast using perspective projection; (c) MIP using the largest \modified{100} pixels along each orthogonal ray; (d) \modified{CycleGAN using patches from (a) and ISBI real conventional cephalograms;} (e) RayCast using orthogonal projection from the skeleton enhanced volume; (f) original sigmoid transform of (e) using Eqn.\,(\ref{eqn:sigmoidTransform}) with air background recovery; (g) modified sigmoid transform of (e) using Eqn.\,(\ref{eqn:twoStepSigmoidTransform}), the final Type I synthetic cephalogram using orthogonal projection; (h) the final Type I synthetic cephalogram using perspective projection. The nose ROIs are redisplayed in the intensity window [0, 125] for better visualization. The mandible angle ROIs are for the visualization of the difference between orthogonal projection and perspective projection. \modified{The perceptual index of each synthetic cephalogram is displayed in the corresponding subcaption.}}
\label{Fig:cephalogramComparison}
\end{figure*}

\subsection{Experimental Setup}
The proposed methods are demonstrated on the CQ500 head CT dataset \citep{chilamkurthy2018development}. It consists of 491 scans, whereby 5 complete head scans are used for testing purposes.

\subsubsection{Parameters for Type I cephalogram synthesis}
For skeleton enhancement, the weight $a$ in Eqn.\,(\ref{eqn:skeletonEnhancement}) is 1.3 as aforementioned. For sigmoid transform in Eqn.\,(\ref{eqn:sigmoidTransform}), the base intensity value $c_1$ is 40 and the saturation parameter $c_2$ is 5, the intensity shift parameter $t$ is 2.6, and the scaling factor $s$ is 1.5. For the modified sigmoid transform, $c_3$ and $c_4$ are 18 and 23, respectively. The thresholding parameter $\tau_1$ and $\tau_2$ are 0.1 and 1.2, respectively. The synthetic cephalograms $\vg^{\ast}$ have an image size of $512 \times 512$ with a pixel resolution of 0.5\,mm\,$\times$\,0.5\,mm.

\subsubsection{Parameters for Type II cephalogram synthesis}
The CBCT projections are simulated using a ray driven method with a sampling rate of 3/mm. The source-to-isocenter distance and the source-to-detector distance of the CBCT system are 950\,mm and 650\,mm, respectively. Dental CBCT systems use flat panel detectors typically with a pixel size up to 0.1\,mm. To save computation time, in this work the detector has $512\times 512$ rebinned pixels with a pixel resolution of 0.73\,mm $\times$ 0.73\,mm. The $180^\circ$ projections are horizontally flipped to have the same orientation as the $0^\circ$ projections. Afterwards, both the $0^\circ$ and $180^\circ$ projections are rebinned to the VD located at the isocenter to reduce magnification. As a result, the rebinned projections have an image size of $512 \times 512$ with a pixel resolution of 0.5\,mm\,$\times$\,0.5\,mm. Note that finer resolution is typically available in practice. In this work, we choose 0.5\,mm\,$\times$\,0.5\,mm so that we can reuse the same SR models from Type I synthesis. The rebinned projections are further divided to four patches with a patch size of $256 \times 256$ according to the four quadrants. 
%To reduce stitching artifacts, the patch boundaries have an overlap width of 10 pixels. 
The patches from the other quadrants are flipped horizontally or vertically to have the same perspective deformation as those from the first quadrant. Afterwards, the patches are converted to RGB patches as the input of the neural network, where the intensity range [0, 6] is linearly mapped to [0, 255]. In total, 1840 patches are generated. Among them, 1600 patches are used for training, 40 patches for validation, and 200 patches for test. For training, 300 epochs with the Adam optimizer are used. The initial learning rate is 0.0002 with a decay rate of 0.999. The weight for the $\ell_1$ loss is 100. 
%\citep{isola2017image}

\subsubsection{Parameters for super resolution}
The SR models are trained on the ISBI Challenge training dataset \citep{wang2015evaluation,wang2016benchmark}. The original cephalograms have an image size of $1935\times 2400$ with a pixel resolution of 0.1\,mm\,$\times$\,0.1\,mm. The original cephalograms are down-sampled with a factor of 5 using averaging down-sampling to have an image size of $387\times 480$ with a pixel resolution of 0.5\,mm\,$\times$\,0.5\,mm. In addition, the original cephalograms are also down-sampled with a factor of 10 and further up-sampled with a factor of 2. The resulted images also have a pixel resolution of 0.5\,mm$\times$0.5\,mm, but with more blurry structures. This operation is carried out to have different levels of blur in the training images, since our test images from the CQ500 dataset are acquired from different CT scanners, leading to different resolutions. For SR models using ILR images as the input, the LR images are up-sampled with a factor of 5 using bicubic up-sampling to have a pixel resolution of 0.1\,mm $\times$ 0.1\,mm. The LR patches have an image size of $64\times 64$, while the ILR and HR patches have an image size of $320\times 320$. For each cephalogram among the ISBI datasets, we generate 42 patches. In total, we have 6300 patches for training, 420 patches for validation, and 2100 patches for test. For each method, 100 epochs are used for training with the Adam optimizer. For pix2pixGAN in the SR task, no weight is applied for the $\ell_1$ loss, i.e., $\vw = \boldsymbol{1}$ in Eqn.\,(\ref{eqn:gradientweightloss}).

%\todo{For the two ESRGANs, other training parameters are the same as those in \citep{zhang2018residual,wang2018esrgan} respectively. For pix2pixGAN, other parameters are the same as those for Type II cephalogram synthesis but with $\vw = \boldsymbol{1}$ in Eqn.\,(\ref{eqn:gradientweightloss}).}

\subsubsection{Parameters for landmark detection}
We train the proposed network in two steps, one for the LeNet-5 and the other for the ResNet50. For the LeNet-5 part, the 150 down-sampled images from the ISBI training dataset and the corresponding given landmark locations are used for training. The loss function is mean absolute error (MAE). The Adam optimizer is used. The initial learning rate is 0.005 with a decay rate of 0.999. In total, 200 epochs are used for training. For the ResNet50 part, a $512 \times 512$ patch is generated for each given landmark position. The detected landmark is located randomly at the corresponding patch. \modified{For data augmentation, this random patch selection process is repeated 400 times.} Overall, 60000 patches are used for training each model. 19 models are trained for the 19 landmarks respectively. MAE is used as the loss function and the Adam optimizer is used. The initial learning rate is 0.01 with a decay rate of 0.999. In total, 100 epochs are used for training. Due to the large amount of training data, in every 10 epochs, 6000 patches are randomly chosen for training.

%\subsection{CycleGAN comparison}
%As a comparison, CycleGAN \citep{zhu2017unpaired} is applied to synthesize cephalograms from 3D CBCT volumes and 2D projections as well. For synthesis from 3D CBCT volumes, the input training images of CycleGAN are orthogonal RayCast images generated from the CQ500 dataset, and the target training images are the real cephalograms from the ISBI dataset. The images have a pixel resolution of 0.5\,mm\,\times\,0.5\,mm.

\subsubsection{Image quality metrics}
For Type II synthesis results (Fig.\,\ref{Fig:dualProjectionResults}) and SR results on the ISBI Test1 data (Fig.\,\ref{Fig:SRResultsISBI}), conventional image quality metrics of root-mean-square error (RMSE), PSNR and structure similarity (SSIM) index are used, since ground truth images are available. For Type I synthesis results (Fig.\,\ref{Fig:cephalogramComparison}) and SR results on the test CQ500 data (Fig.\ref{Fig:SRSyhthesis}), such conventional image quality metrics are not applicable due to the lack of ground truth images. Therefore, we choose a non-reference image quality metric called perceptual index \citep{blau20182018}. It is calculated from the non-reference metrics of Ma's score \citep{ma2017learning} and natural image quality evaluator (NIQE) \citep{mittal2012making}, i.e., perceptual index = $\frac{1}{2}\left((10- \text{Ma}) + \text{NIQE}\right)$. A lower perceptual index represents a better perceptual quality. For landmark detection, the SDRs in 2\,mm, 2.5\,mm, 3\,mm and 4\,mm precision ranges are reported.

\subsubsection{Pipeline summary}
To give a high-level summary of the whole data processing pipeline, the methods/architectures and datasets used in the different parts of the pipeline are listed in Tab.\,\ref{Tab:pipelineInformation}.

\section{Results}
\subsection{Results of Type I Cephalogram Synthesis}

The synthetic cephalograms generated by different methods are displayed in Fig.\,\ref{Fig:cephalogramComparison}. Figs.\,\ref{Fig:cephalogramComparison}(a) and (b) show the cephalograms synthesized by the orthogonal \citep{moshiri2007accuracy} and perspective \citep{kumar2007comparison} RayCast methods, respectively, which are the most widely used methods for cephalogram synthesis from CBCT volumes. 
%In Fig.\,\ref{Fig:cephalogramComparison}(a), anatomical structures on both sides of the midsaggital plane overlap well in the orthogonal projection. In contrast, 
Comparing Fig.\,\ref{Fig:cephalogramComparison}(b) to Fig.\,\ref{Fig:cephalogramComparison}(a), due to different magnification factors of structures at different positions in perspective projection, anatomical structures on both sides of the midsaggital plane cannot overlap well, for example, the projections of the left and right mandible (gonial) angles in the zoom-in ROI in Fig.\,\ref{Fig:cephalogramComparison}(b). In Figs.~\ref{Fig:cephalogramComparison}(a) and (b), the skeleton structures, soft tissues and airways are well observed. However, the image contrast in these two synthetic cephalograms are different from conventional cephalograms (Fig.\,\ref{Fig:sigmoidCurve}(a)). 
\modified{The cephalogram synthesized by MIP using the largest 100 pixels along each orthogonal ray is displayed in Fig.~\ref{Fig:cephalogramComparison}(c). In this subfigure, skeleton structures are well observed since they have high intensity. Nevertheless, low intensity structures, e.g. the throat airway, might disappear.}
\modified{The CycleGAN \citep{zhu2017unpaired} synthetic cephalogram is shown in Fig.~\ref{Fig:cephalogramComparison}(d). Compared with the conventional cephalogram example in Fig.\,\ref{Fig:sigmoidCurve}(a), it has the closest image contrast. Hence, it achieves the best perceptual index of 5.53. However, some detailed anatomical structures are incorrect. For example, the nose has extremely low intensity in the zoom-in ROI. It also suffers from geometric distortion. As a consequence, the stitching artifacts are visible.}
%The cephalograms synthesized by MIP using the largest 50 pixels and 100 pixels along each orthogonal ray are displayed in Figs.~\ref{Fig:cephalogramComparison}(c) and (d), respectively. In both images, skeleton structures are well observed since they have high intensities. Comparing Fig.\,\ref{Fig:cephalogramComparison}(d) to (c), more anatomical details are added. Nevertheless, in both images, low intensity structures, e.g. the throat airway, might disappear.

Figure~\ref{Fig:cephalogramComparison}(e)-(g) are the Type I synthetic cephalograms of different steps using orthogonal projection. Fig.\,\ref{Fig:cephalogramComparison}(e) is the orthogonal RayCast celphalogram synthesized from the enhanced CBCT volume using Eqn.\,(\ref{eqn:skeletonEnhancement}). Compared with Fig.\,\ref{Fig:cephalogramComparison}(a), skeleton structures in Fig.\,\ref{Fig:cephalogramComparison}(e) have higher contrast. Fig.\,\ref{Fig:cephalogramComparison}(f) is obtained by applying the original sigmoid transform in Eqn.\,(\ref{eqn:sigmoidTransform}) to Fig.\,\ref{Fig:cephalogramComparison}(e), where the skeleton structures are further enhanced. Moreover, the appearance of Fig.\,\ref{Fig:cephalogramComparison}(f) is very close to conventional cephalograms. However, the soft tissues like the nose and lips have an almost constant intensity value, as displayed in the zoom-in ROI in Fig\,\ref{Fig:cephalogramComparison}(f).
The final Type I synthetic cephalogram with orthogonal projection is displayed in Fig.\,\ref{Fig:cephalogramComparison}(g). With the proposed modified sigmoid transform in Eqn.\,(\ref{eqn:twoStepSigmoidTransform}), the contrast in the soft tissues is brought back, as displayed in the zoom-in ROI.
For comparison, the final Type I synthetic cephalogram with perspective projection is displayed in Fig.\,\ref{Fig:cephalogramComparison}(h). \modified{Compared with Fig.\,\ref{Fig:cephalogramComparison}(g), the mandible angles in the zoom-in ROI of Fig.\,\ref{Fig:cephalogramComparison}(h) are not overlapped well due to perspective magnification.}

\begin{figure}[t]
\centering
\includegraphics[width=\linewidth]{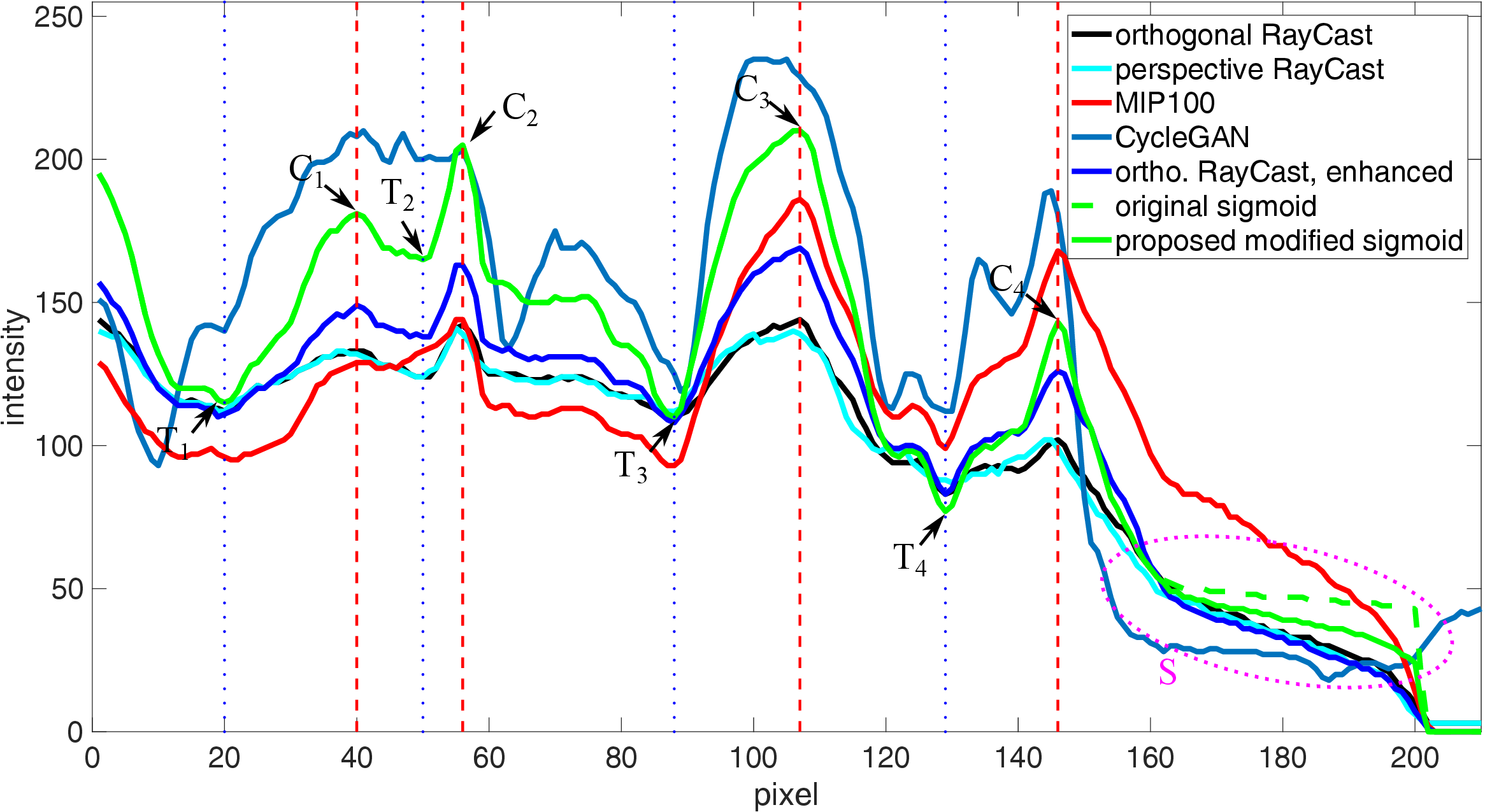}
\caption{\modified{Intensity profiles of a chosen line from Figs.\,\ref{Fig:cephalogramComparison}(a)-(g). The position of the line is marked in Fig.\,\ref{Fig:cephalogramComparison}(a). The major crests and troughs in the curve of our proposed method are marked by $\text{C}_i$ and $\text{T}_i$ where $i = 1,2,3,4$, respectively. Their positions are indicated by the red-dashed and blue-dotted vertical lines respectively. The soft tissue region is marked by S. }}
\label{Fig:TypeILineProfiles}
\end{figure}
\begin{figure*}[!t]
\centering

\begin{small}
\begin{minipage}{0.16\linewidth}
\centering
$0^\circ$ projection
\end{minipage}
\begin{minipage}{0.16\linewidth}
\centering
$180^\circ$ projection
\end{minipage}
\begin{minipage}{0.16\linewidth}
\centering
RGB projection
\end{minipage}
\begin{minipage}{0.16\linewidth}
\centering
Target
\end{minipage}
\begin{minipage}{0.16\linewidth}
\centering
1-projection output
\end{minipage}
\begin{minipage}{0.16\linewidth}
\centering
2-projection output
\end{minipage}
\end{small}

\begin{minipage}{0.16\linewidth}
\subfigure[]{
\includegraphics[width = \linewidth]{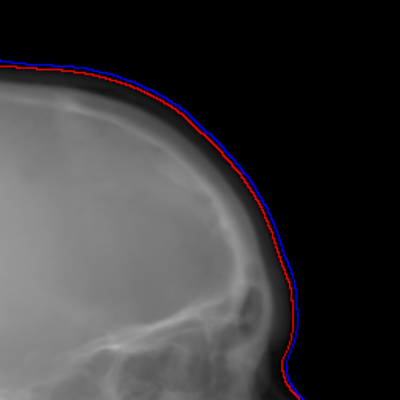}
}
\end{minipage}
\begin{minipage}{0.16\linewidth}
\subfigure[]{
\includegraphics[width = \linewidth]{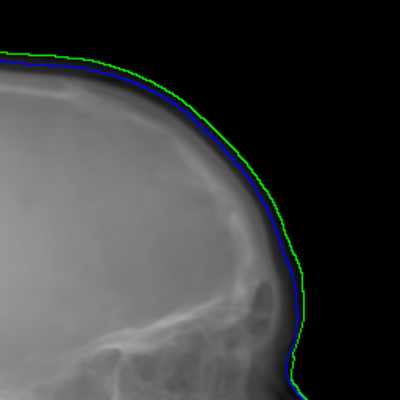}
}
\end{minipage}
\begin{minipage}{0.16\linewidth}
\subfigure[]{
\includegraphics[width = \linewidth]{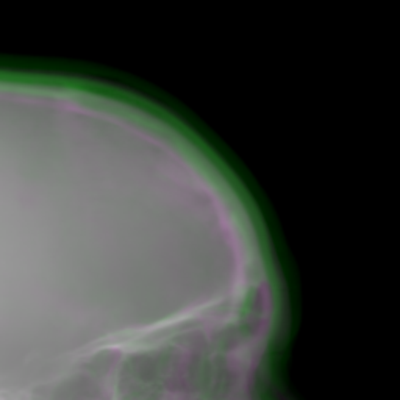}
}
\end{minipage}
\begin{minipage}{0.16\linewidth}
\subfigure[]{
\includegraphics[width = \linewidth]{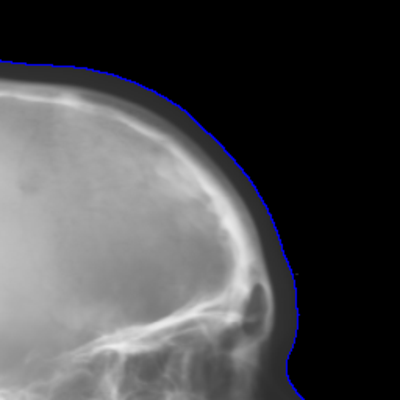}
}
\end{minipage}
\begin{minipage}{0.16\linewidth}
\subfigure[13.56, 24.66, \modified{0.983}]{
\includegraphics[width = \linewidth]{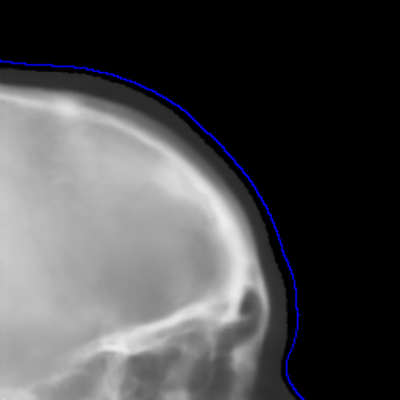}
}
\end{minipage}
\begin{minipage}{0.16\linewidth}
\subfigure[6.98, 30.26, \modified{0.993}]{
\includegraphics[width = \linewidth]{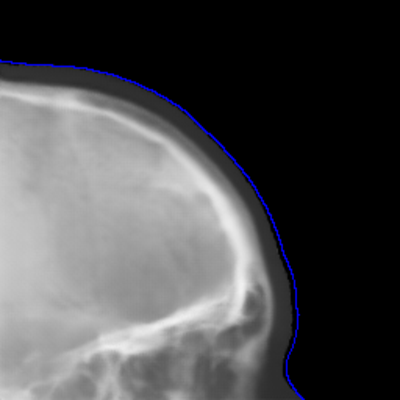}
}
\end{minipage}

\begin{minipage}{0.16\linewidth}
\subfigure[]{
\includegraphics[width = \linewidth]{patch360_R}
}
\end{minipage}
\begin{minipage}{0.16\linewidth}
\subfigure[]{
\includegraphics[width = \linewidth]{patch360_G}
}
\end{minipage}
\begin{minipage}{0.16\linewidth}
\subfigure[]{
\includegraphics[width = \linewidth]{patch360_RGB}
}
\end{minipage}
\begin{minipage}{0.16\linewidth}
\subfigure[]{
\includegraphics[width = \linewidth]{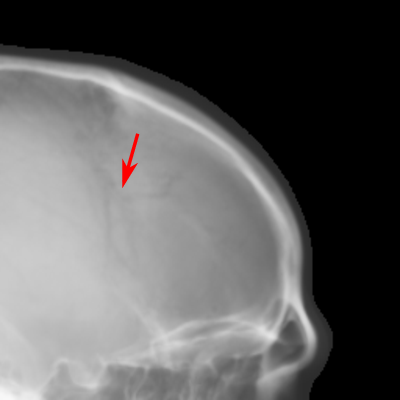}
}
\end{minipage}
\begin{minipage}{0.16\linewidth}
\subfigure[5.35, 32.43, \modified{0.996}]{
\includegraphics[width = \linewidth]{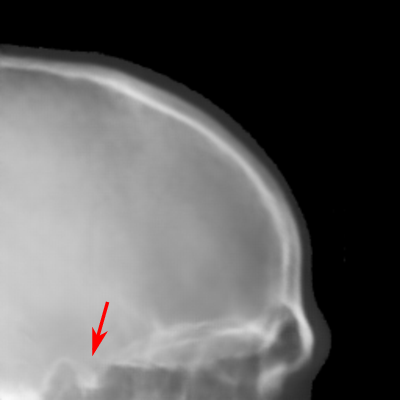}
}
\end{minipage}
\begin{minipage}{0.16\linewidth}
\subfigure[4.12, 34.70, \modified{0.998}]{
\includegraphics[width = \linewidth]{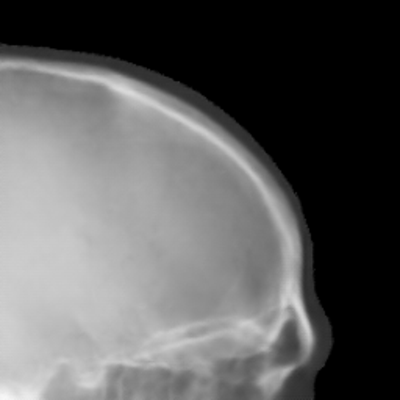}
}
\end{minipage}

\begin{minipage}{0.16\linewidth}
\subfigure[]{
\includegraphics[width = \linewidth]{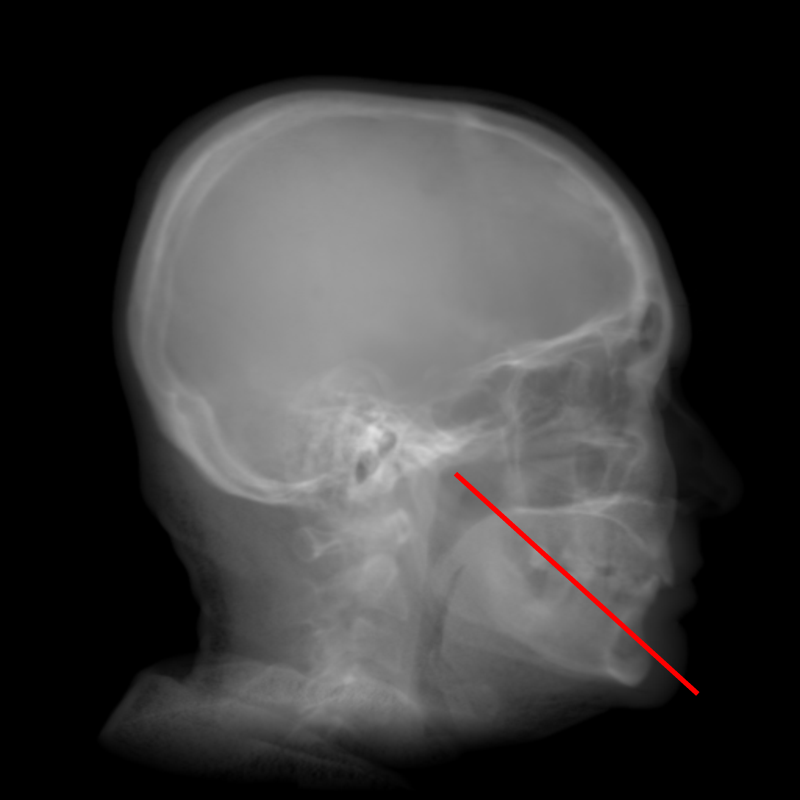}
}
\end{minipage}
\begin{minipage}{0.16\linewidth}
\subfigure[]{
\includegraphics[width = \linewidth]{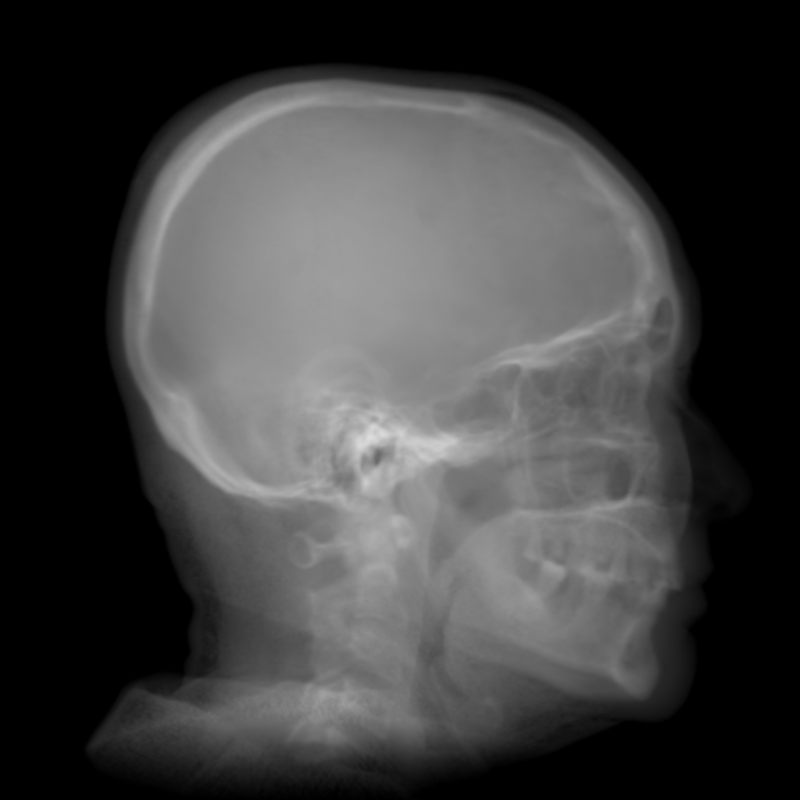}
}
\end{minipage}
\begin{minipage}{0.16\linewidth}
\subfigure[]{
\includegraphics[width = \linewidth]{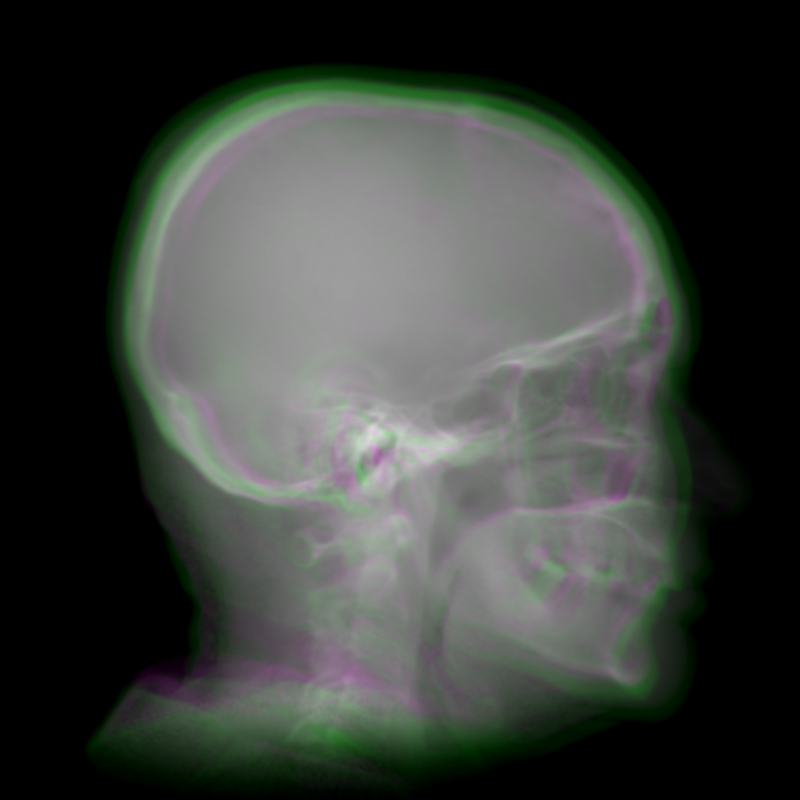}
}
\end{minipage}
\begin{minipage}{0.16\linewidth}
\subfigure[]{
\includegraphics[width = \linewidth]{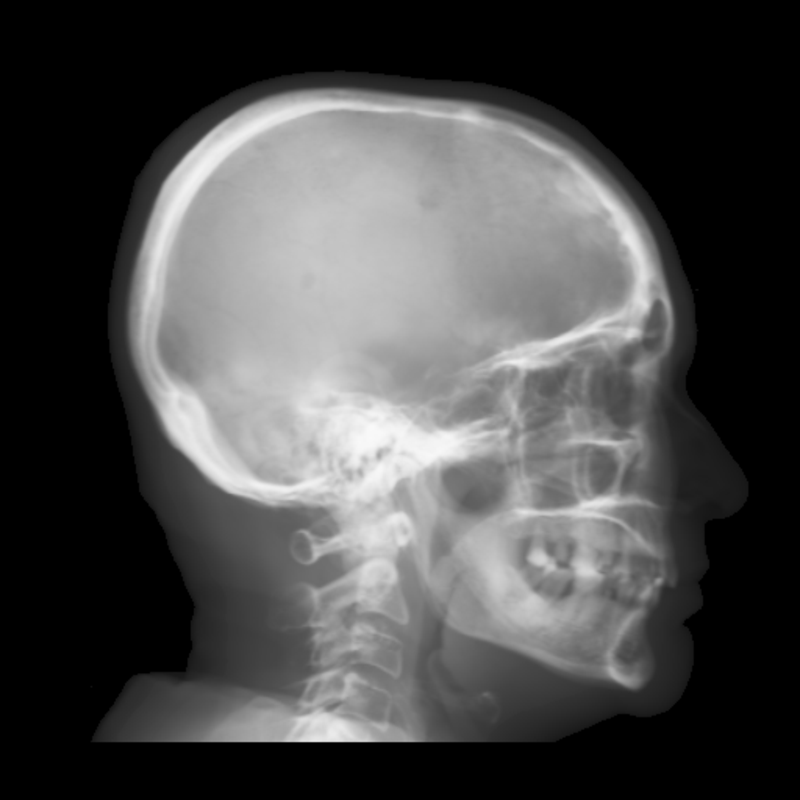}
}
\end{minipage}
\begin{minipage}{0.16\linewidth}
\subfigure[13.88, 24.73, \modified{0.990}]{
\includegraphics[width = \linewidth]{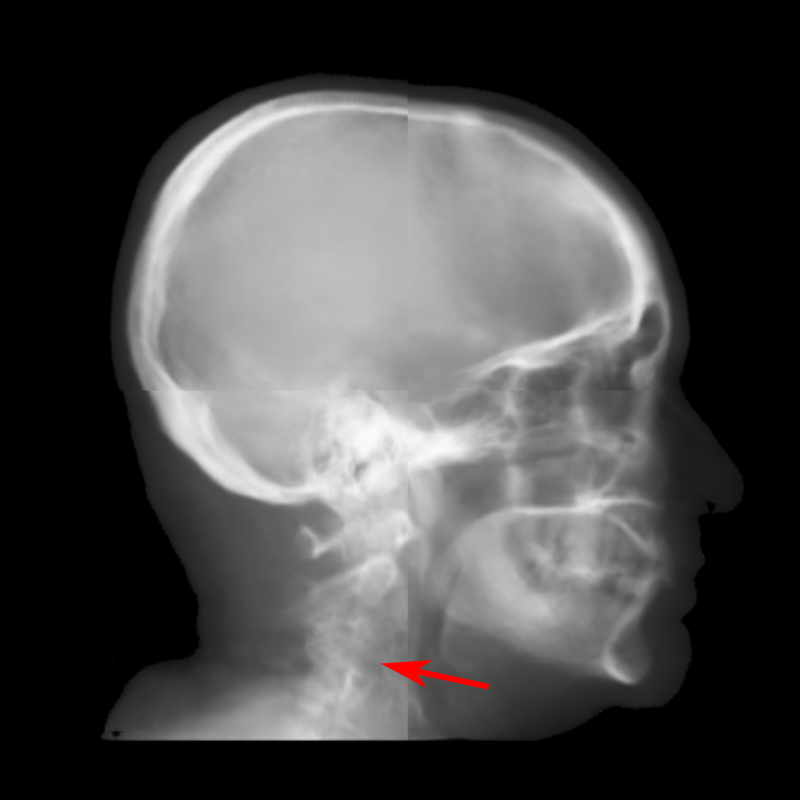}
}
\end{minipage}
\begin{minipage}{0.16\linewidth}
\subfigure[8.41, 28.45, \modified{0.994}]{
\includegraphics[width = \linewidth]{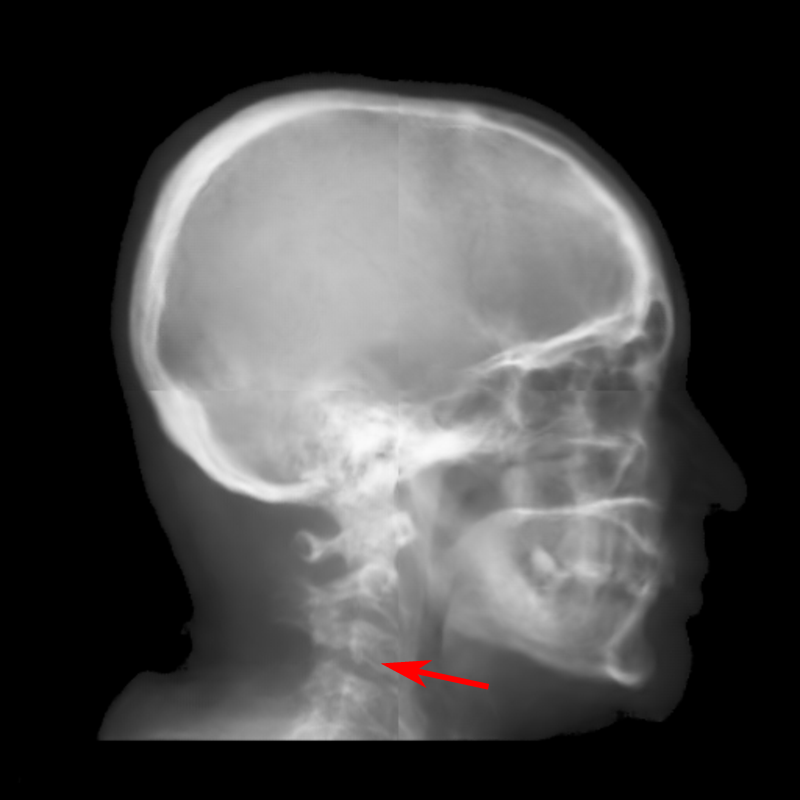}
}
\end{minipage}
\caption{Synthetic cephalogram examples from 2D CBCT projections. The top two rows are two patches respectively, while the bottom row is a complete stitched cephalogram. In the top row, the blue curve is the outline of the target patch, while the red and green curves are the outlines for the $0^\circ$ and $180^\circ$ projections, respectively. In the middle row, the cranial sutures indicated by the arrow in (j) are not visible in (k) nor in (l); the sella turcica indicated by the arrow in (k) is distorted. 
{In the bottom row, the vertebrae indicated by the arrow in (r) are more accurate than those in (q). The line in (m) \modified{marks} the position for line profiles in Fig.\,\ref{Fig:lineProfile}.}
%In the bottom row, large displacement is observed in the ROI marked by the red box in (q) due to patch stitching, while such displacement is smaller in (r). 
For the 1-projection output and the 2-projection output, \modified{the RMSE (left), PSNR (mid) and SSIM (right)} values w.\,r.\,t. the target are displayed in the corresponding subcaptions.}
\label{Fig:dualProjectionResults}
\end{figure*}

For quantification, the intensity profiles of a chosen line from Figs.\,\ref{Fig:cephalogramComparison}(a)-(g), the position of which is marked in Fig.\,\ref{Fig:cephalogramComparison}(a), are plotted in Fig.\,\ref{Fig:TypeILineProfiles}. The major crests and troughs in the curve of our proposed method are marked by $\text{C}_i$ and $\text{T}_i$ where $i = 1,2,3,4$, respectively. At the position $\text{C}_4$, it is clear that the line profiles of orthogonal RayCast and perspective RayCast have low contrast. The profile of MIP100 has high contrast for the crests and troughs. However, the soft tissue part marked by S has too large intensity. The profile of CycleGAN has high contrast as well. Nevertheless, it also introduces undesired crests and troughs, for example, those between $\text{T}_4$ and $\text{C}_4$. In addition, the position of $\text{C}_4$ is also shifted. Compared with the original orthogonal RayCast, the profile of orthogonal RayCast from the enhanced volume has better contrast at all the positions of $\text{C}_1 - \text{C}_4$. With the original sigmoid transform, the contrast is further enhanced. However, at the soft tissue part marked by S, it almost has a constant value. The profile of our proposed modified sigmoid transform overlaps with that of the original sigmoid transform except for the soft tissue part, where the contrast of the soft tissue is brought back.

\subsection{Results of Type II Cephalogram Synthesis}

\begin{figure}[!h]
\centering
\begin{minipage}{0.025\linewidth}
\centering
\scriptsize{\rotatebox[origin=c]{90}{1-projection output}}
\end{minipage}
\begin{minipage}{0.31\linewidth}
\subfigure[\modified{19.67, 21.43, 0.956}]{
\includegraphics[width = \linewidth]{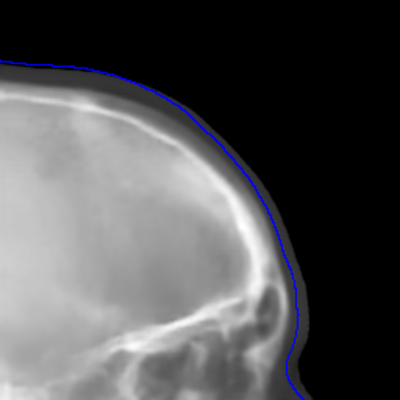}
}
\end{minipage}
\begin{minipage}{0.31\linewidth}
\subfigure[\modified{11.07, 26.12, 0.984}]{
\includegraphics[width = \linewidth]{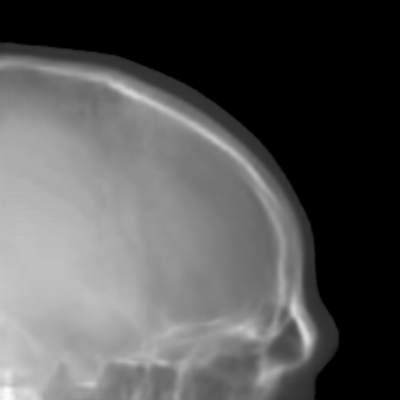}
}
\end{minipage}
\begin{minipage}{0.31\linewidth}
\subfigure[\modified{11.25, 26.90, 0.984}]{
\includegraphics[width = \linewidth]{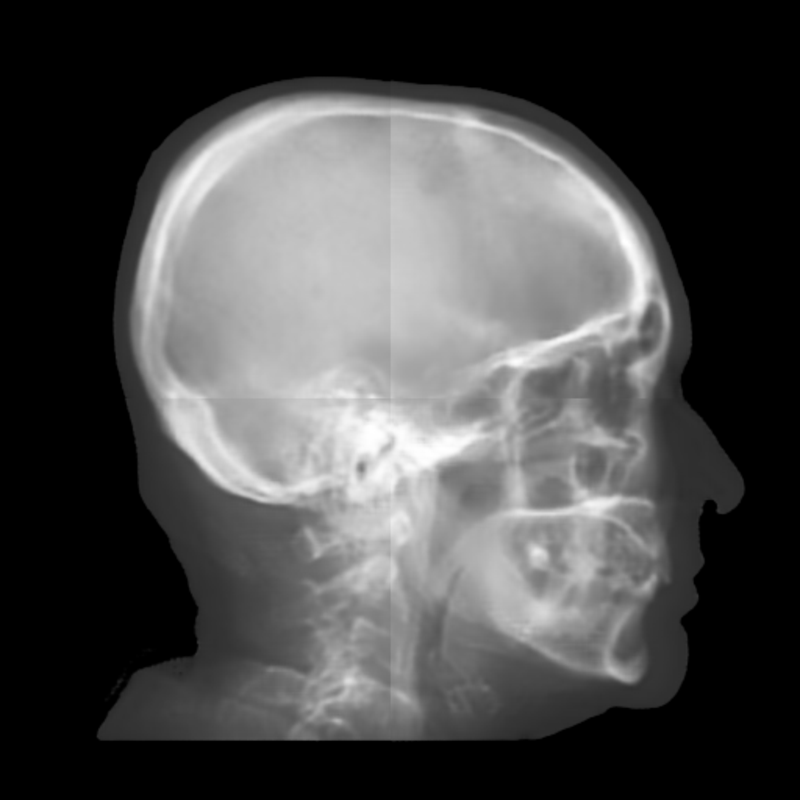}
}
\end{minipage}

\begin{minipage}{0.025\linewidth}
\centering
\scriptsize{\rotatebox[origin=c]{90}{2-projection output}}
\end{minipage}
\begin{minipage}{0.31\linewidth}
\subfigure[\modified{20.72, 20.98, 0.953}]{
\includegraphics[width = \linewidth]{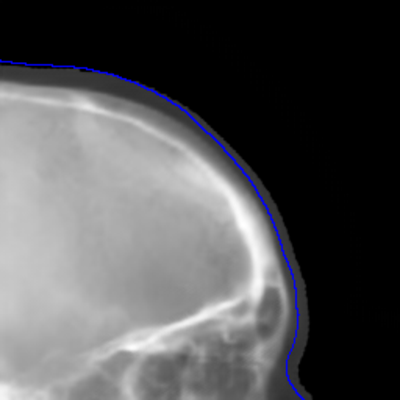}
}
\end{minipage}
\begin{minipage}{0.31\linewidth}
\subfigure[\modified{9.46, 27.49, 0.988}]{
\includegraphics[width = \linewidth]{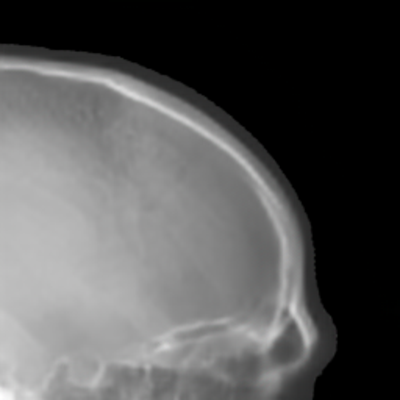}
}
\end{minipage}
\begin{minipage}{0.31\linewidth}
\subfigure[\modified{10.42, 27.57, 0.989}]{
\includegraphics[width = \linewidth]{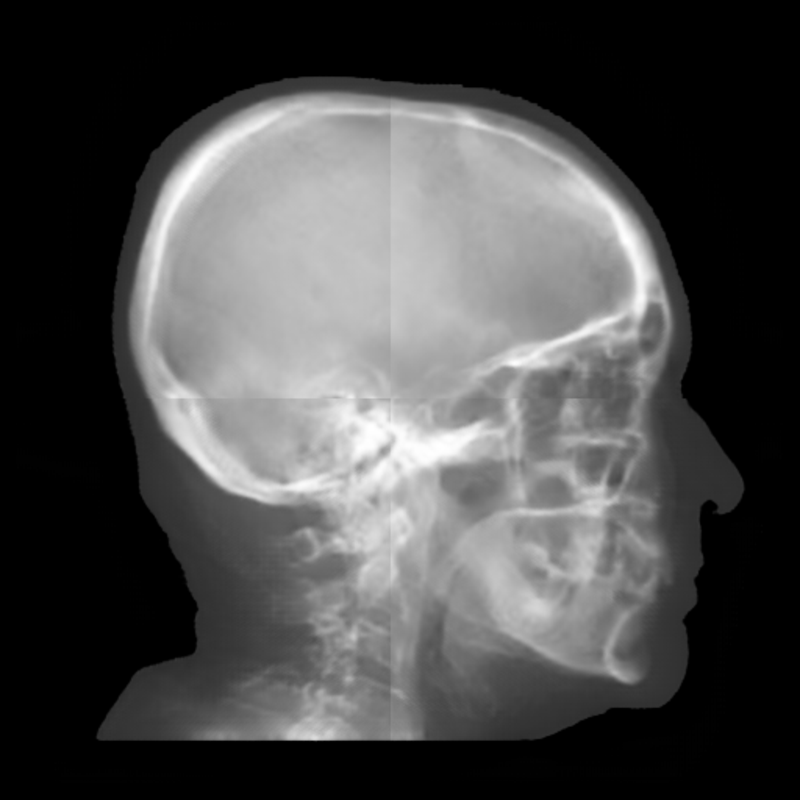}
}
\end{minipage}
\caption{\modified{Synthetic cephalogram examples from 2D CBCT projections by CycleGAN. The 1st to 3rd columns correspond to the 1st to 3rd rows in Fig.\,\ref{Fig:dualProjectionResults}, respectively. The RMSE (left), PSNR (mid) and SSIM (right) values w.\,r.\,t. the target are displayed in the corresponding subcaptions.}}
\label{Fig:cycleGANResults}
\end{figure}

\begin{figure}[!h]
\centering
\includegraphics[width=\linewidth]{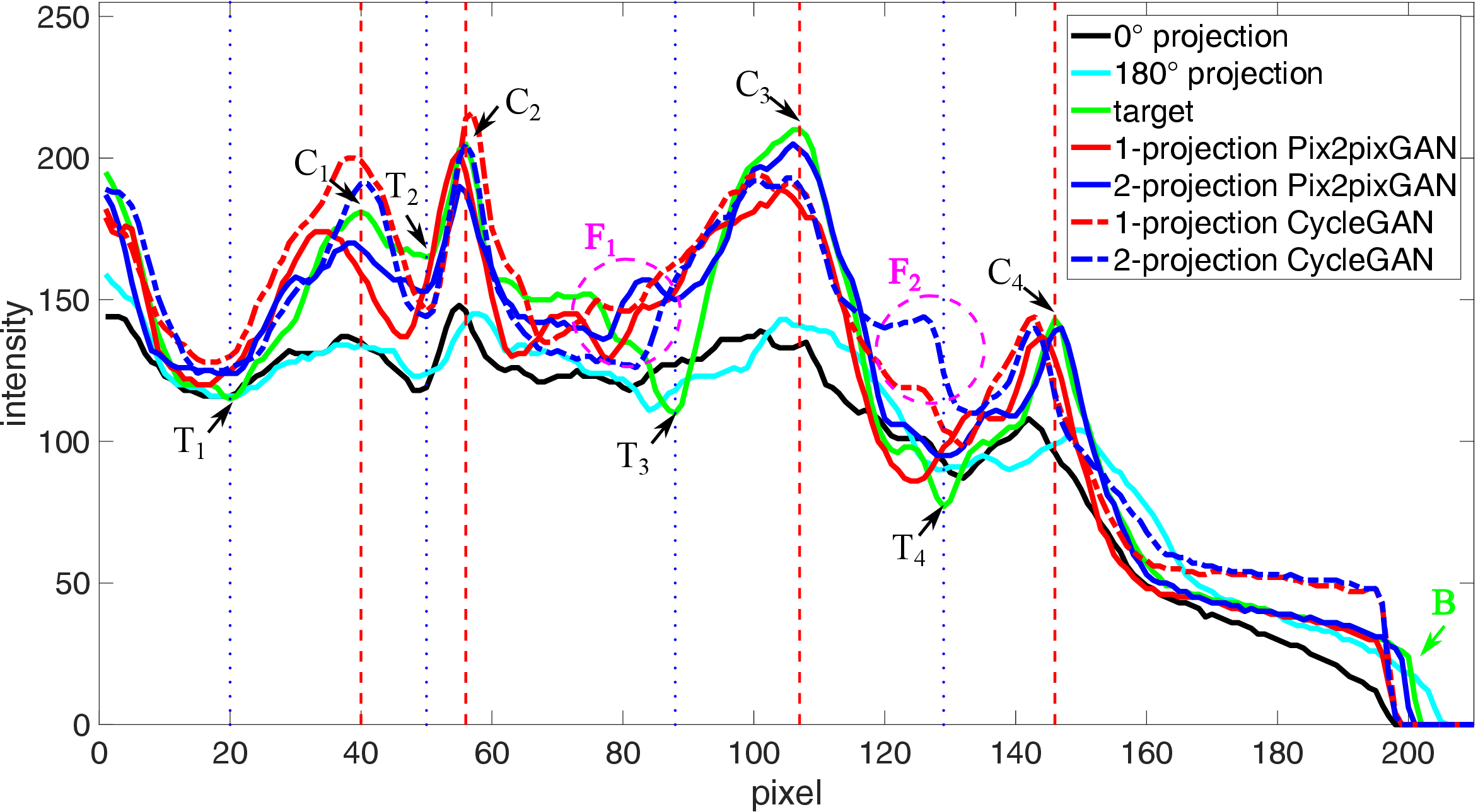}
\caption{Intensity profiles of a chosen line from Figs.\,\ref{Fig:dualProjectionResults}(m)-(r), \modified{Fig.\,\ref{Fig:cycleGANResults}(c) and Fig.\,\ref{Fig:cycleGANResults}(f)}. The position of the line is marked in Fig.\,\ref{Fig:dualProjectionResults}(m). The boundary area is marked by $\text{B}$. The major crests and troughs in the target curve are marked by $\text{C}_i$ and $\text{T}_i$ where $i = 1,2,3,4$, respectively, while a circular region near $\text{T}_3$ contains small crests and troughs is marked by \modified{$\text{F}_\text{1}$. The profiles of CycleGAN in the circular region $\text{F}_\text{2}$ have large error.} }
\label{Fig:lineProfile}
\end{figure}

The cephalogram synthesis results of two patches and one complete stitched cephalogram are displayed in Fig.\,\ref{Fig:dualProjectionResults}. In the top row, the blue curve is the outline of the target patch, while the red and green curves are the outlines for the $0^\circ$ and $180^\circ$ projections, respectively. Consistent with the relationship in Fig.\,\ref{Fig:dualProjection}, the blue curve is between the red and green curves. Since the $180^\circ$ projection (green channel) has larger area than the $0^\circ$ projection (red and blue channels), the region near the boundary appears green in the RGB input patch in Fig.\,\ref{Fig:dualProjectionResults}(c). Fig.\,\ref{Fig:dualProjectionResults}(d) is the target output. Fig.\,\ref{Fig:dualProjectionResults}(e) is the output using the $0^\circ$ projection only, where the outline has large deviation from the target blue curve. In contrast, in Fig.\,\ref{Fig:dualProjectionResults}(f) where the dual projections are used, the outline is closer to the target blue curve. In the middle row, the sella turcica in the 1-projection output has large distortion, as indicated by the arrow in Fig.\,\ref{Fig:dualProjectionResults}(k), compared with that in the target patch. On the contrary, the sella turcica in the 2-projection output (Fig.\,\ref{Fig:dualProjectionResults}(l)) preserves its shape. These observations highly demonstrate the benefit of using dual projections. 

\begin{figure*}[h]
\centering
\begin{small}
\begin{minipage}{0.135\linewidth}
\centering
Reference
\end{minipage}
\begin{minipage}{0.135\linewidth}
\centering
Bicubic
\end{minipage}
\begin{minipage}{0.135\linewidth}
\centering
%ESRGAN\textsubscript{RDB}, LR
RDB, LR
\end{minipage}
\begin{minipage}{0.135\linewidth}
\centering
RDB, ILR
\end{minipage}
\begin{minipage}{0.135\linewidth}
\centering
%ESRGAN\textsubscript{RRDB}, LR
RRDB, LR
\end{minipage}
\begin{minipage}{0.135\linewidth}
\centering
RRDB, ILR
\end{minipage}
\begin{minipage}{0.135\linewidth}
\centering
pix2pixGAN
\end{minipage}
\end{small}

\begin{minipage}{0.135\linewidth}
\subfigure[]{
\includegraphics[width=\linewidth]{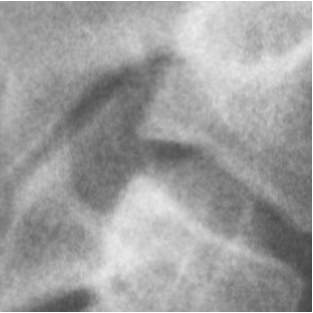}
}
\end{minipage}
\begin{minipage}{0.135\linewidth}
\subfigure[6.41, 31.43,\,\modified{0.972}]{
\includegraphics[width=\linewidth]{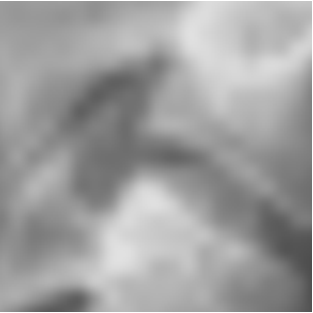}
}
\end{minipage}
\begin{minipage}{0.135\linewidth}
\subfigure[20.90, 21.16,\,\modified{0.955}]{
\includegraphics[width=\linewidth]{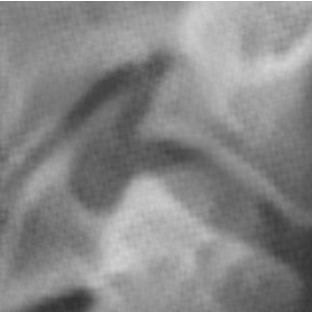}
}
\end{minipage}
\begin{minipage}{0.135\linewidth}
\subfigure[10.61, 27.05,\,\modified{0.970}]{
\includegraphics[width=\linewidth]{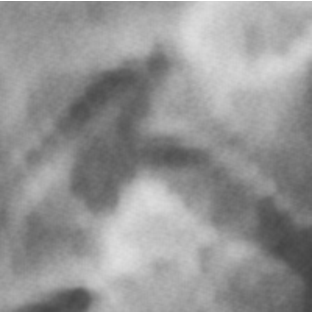}
}
\end{minipage}
\begin{minipage}{0.135\linewidth}
\subfigure[7.38, 30.21,\,\modified{0.968}]{
\includegraphics[width=\linewidth]{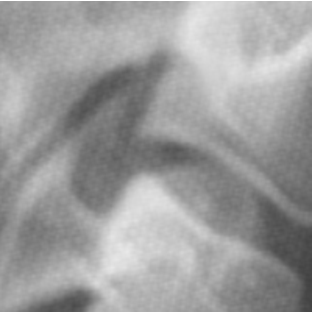}
}
\end{minipage}
\begin{minipage}{0.135\linewidth}
\subfigure[9.07, 28.41,\,\modified{0.970}]{
\includegraphics[width=\linewidth]{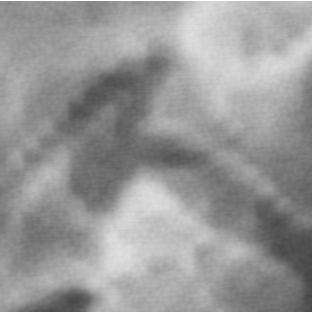}
}
\end{minipage}
\begin{minipage}{0.135\linewidth}
\subfigure[6.72, 31.03,\,\modified{0.972}]{
\includegraphics[width=\linewidth]{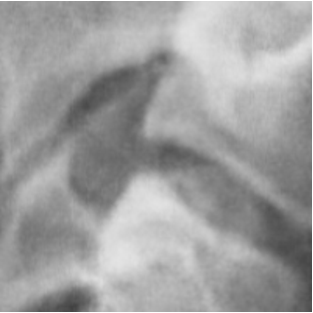}
}
\end{minipage}

\begin{minipage}{0.135\linewidth}
\subfigure[]{
\includegraphics[width=\linewidth]{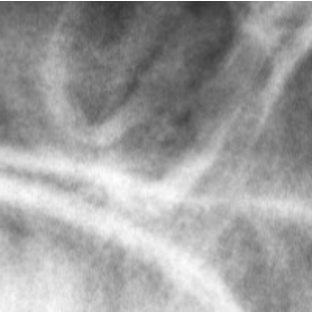}
}
\end{minipage}
\begin{minipage}{0.135\linewidth}
\subfigure[7.23, 30.92,\,\modified{0.982}]{
\includegraphics[width=\linewidth]{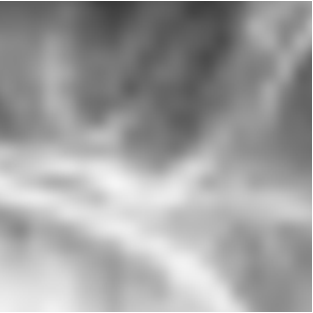}
}
\end{minipage}
\begin{minipage}{0.135\linewidth}
\subfigure[20.76, 21.75,\,\modified{0.968}]{
\includegraphics[width=\linewidth]{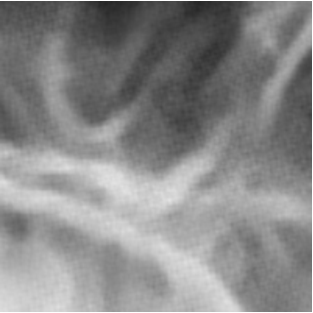}
}
\end{minipage}
\begin{minipage}{0.135\linewidth}
\subfigure[10.76, 27.46,\,\modified{0.981}]{
\includegraphics[width=\linewidth]{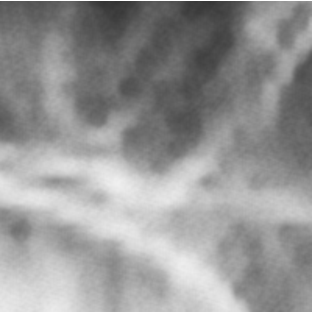}
}
\end{minipage}
\begin{minipage}{0.135\linewidth}
\subfigure[8.04, 29.99,\,\modified{0.980}]{
\includegraphics[width=\linewidth]{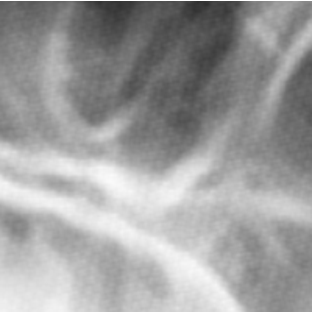}
}
\end{minipage}
\begin{minipage}{0.135\linewidth}
\subfigure[9.10, 28.91,\,\modified{0.981}]{
\includegraphics[width=\linewidth]{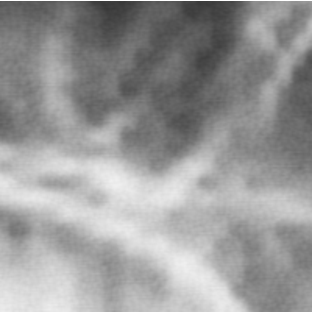}
}
\end{minipage}
\begin{minipage}{0.135\linewidth}
\subfigure[7.30, 30.83,\,\modified{0.982}]{
\includegraphics[width=\linewidth]{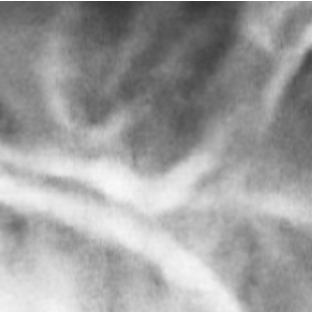}
}
\end{minipage}

%\begin{minipage}{0.135\linewidth}
%\subfigure[]{
%\includegraphics[width=\linewidth]{146targets200}
%}
%\end{minipage}
%\begin{minipage}{0.135\linewidth}
%\subfigure[]{
%\includegraphics[width=\linewidth]{146bicubic200}
%}
%\end{minipage}
%\begin{minipage}{0.135\linewidth}
%\subfigure[]{
%\includegraphics[width=\linewidth]{146RDN200}
%}
%\end{minipage}
%\begin{minipage}{0.135\linewidth}
%\subfigure[]{
%\includegraphics[width=\linewidth]{146RDNss200}
%}
%\end{minipage}
%\begin{minipage}{0.135\linewidth}
%\subfigure[]{
%\includegraphics[width=\linewidth]{146RRDN200}
%}
%\end{minipage}
%\begin{minipage}{0.135\linewidth}
%\subfigure[]{
%\includegraphics[width=\linewidth]{146RRDNss200}
%}
%\end{minipage}
%\begin{minipage}{0.135\linewidth}
%\subfigure[]{
%\includegraphics[width=\linewidth]{146pix2pixGAN200}
%}
%\end{minipage}
\caption{SR test examples on the ISBI Test1 data. The \modified{RMSE (left), PSNR (mid), and SSIM (right)} values are displayed in the corresponding subcaptions.}
\label{Fig:SRResultsISBI}
\end{figure*}

\begin{figure*}
\centering
\begin{small}
\begin{minipage}{0.16\linewidth}
\centering
Bicubic
\end{minipage}
\begin{minipage}{0.16\linewidth}
\centering
%ESRGAN\textsubscript{RDN}, LR
RDB, LR
\end{minipage}
\begin{minipage}{0.16\linewidth}
\centering
RDB, ILR
\end{minipage}
\begin{minipage}{0.16\linewidth}
\centering
%ESRGAN\textsubscript{RRDB}, LR
RRDB, LR
\end{minipage}
\begin{minipage}{0.16\linewidth}
\centering
RRDB, ILR
\end{minipage}
\begin{minipage}{0.16\linewidth}
\centering
pix2pixGAN
\end{minipage}
\end{small}

\begin{minipage}{0.16\linewidth}
\subfigure[\modified{10.02}]{
\includegraphics[width=\linewidth]{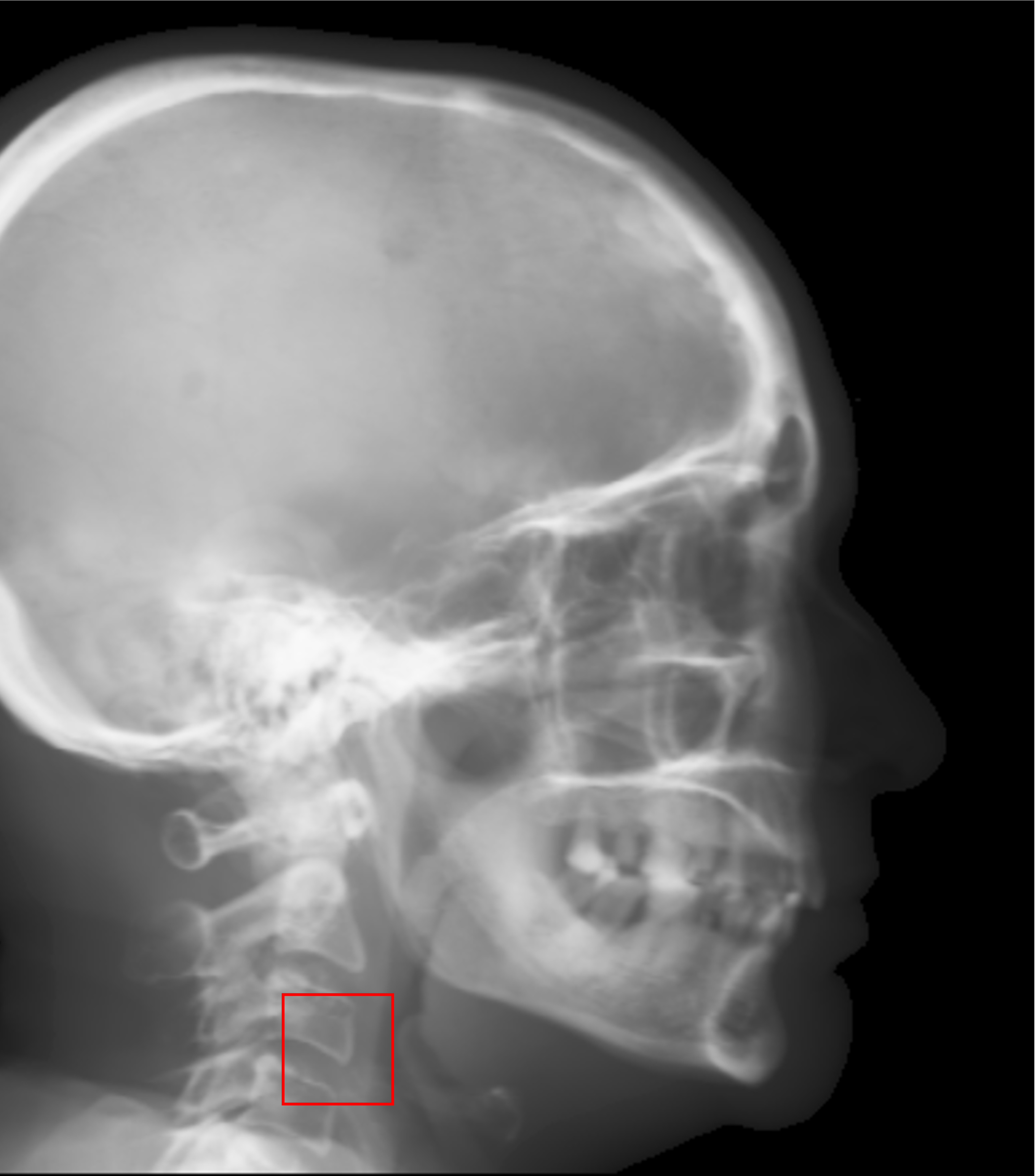}
}
\end{minipage}
\begin{minipage}{0.16\linewidth}
\subfigure[\modified{5.43}]{
\includegraphics[width=\linewidth]{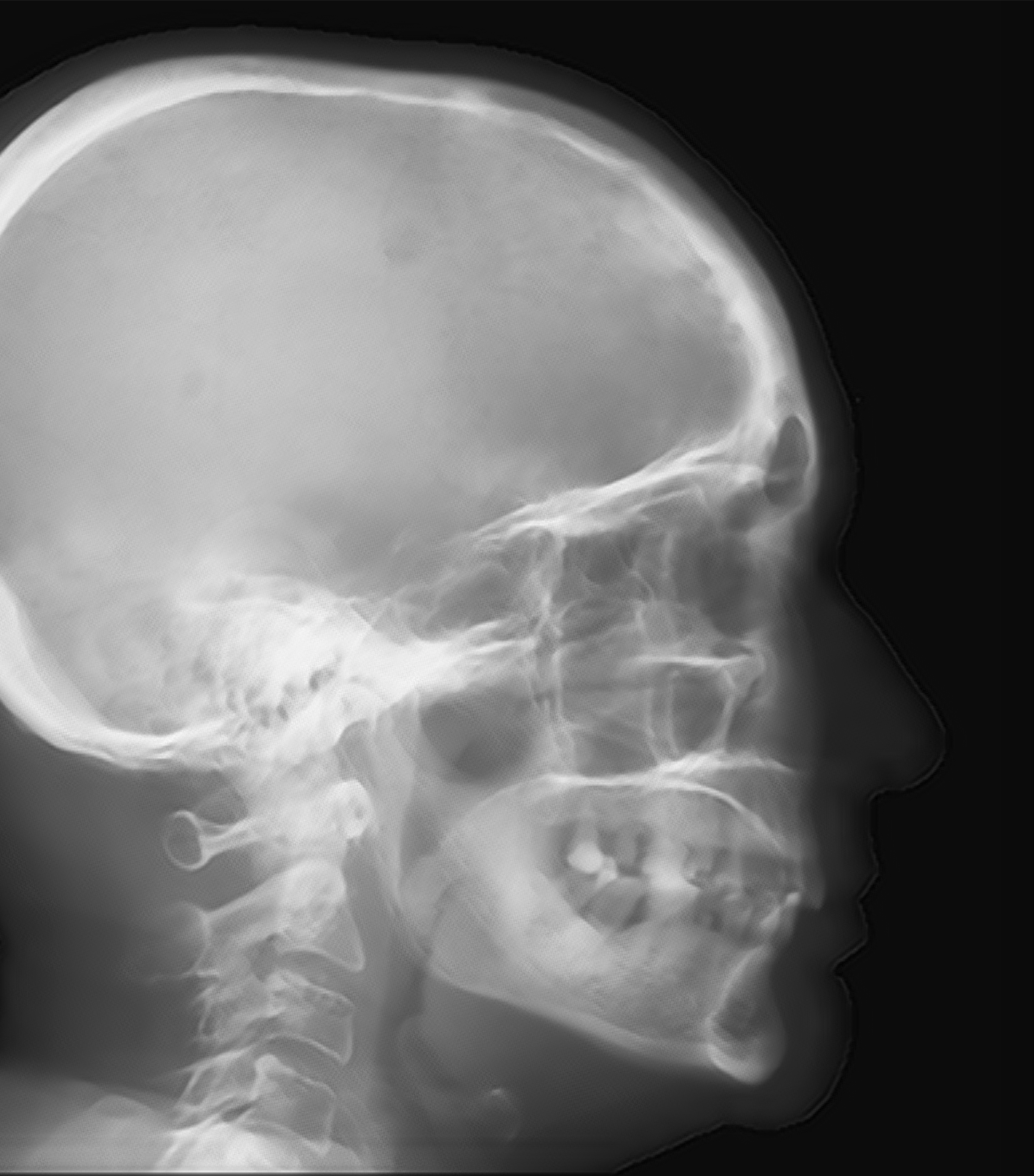}
}
\end{minipage}
\begin{minipage}{0.16\linewidth}
\subfigure[\modified{4.67}]{
\includegraphics[width=\linewidth]{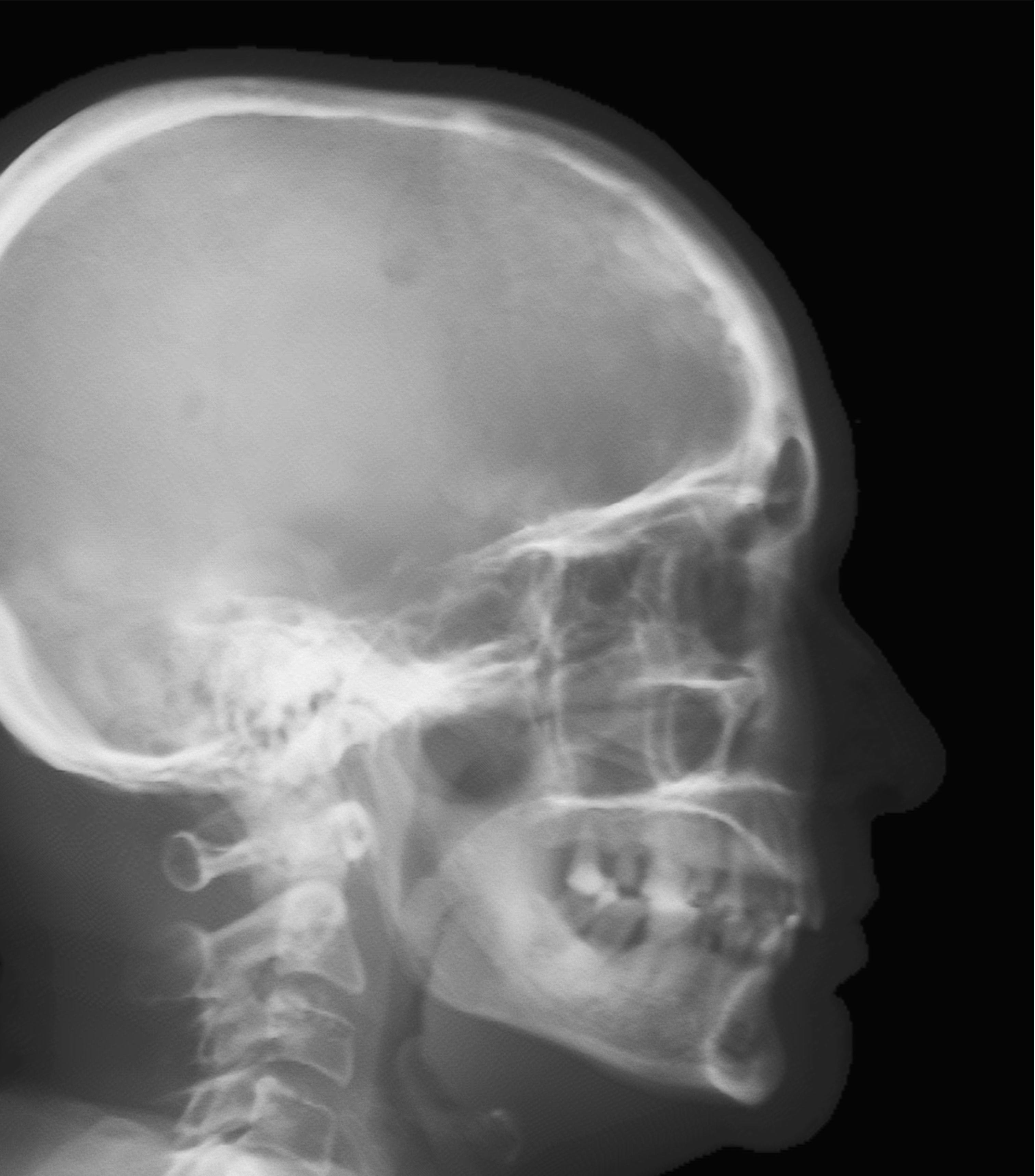}
}
\end{minipage}
\begin{minipage}{0.16\linewidth}
\subfigure[\modified{5.43}]{
\includegraphics[width=\linewidth]{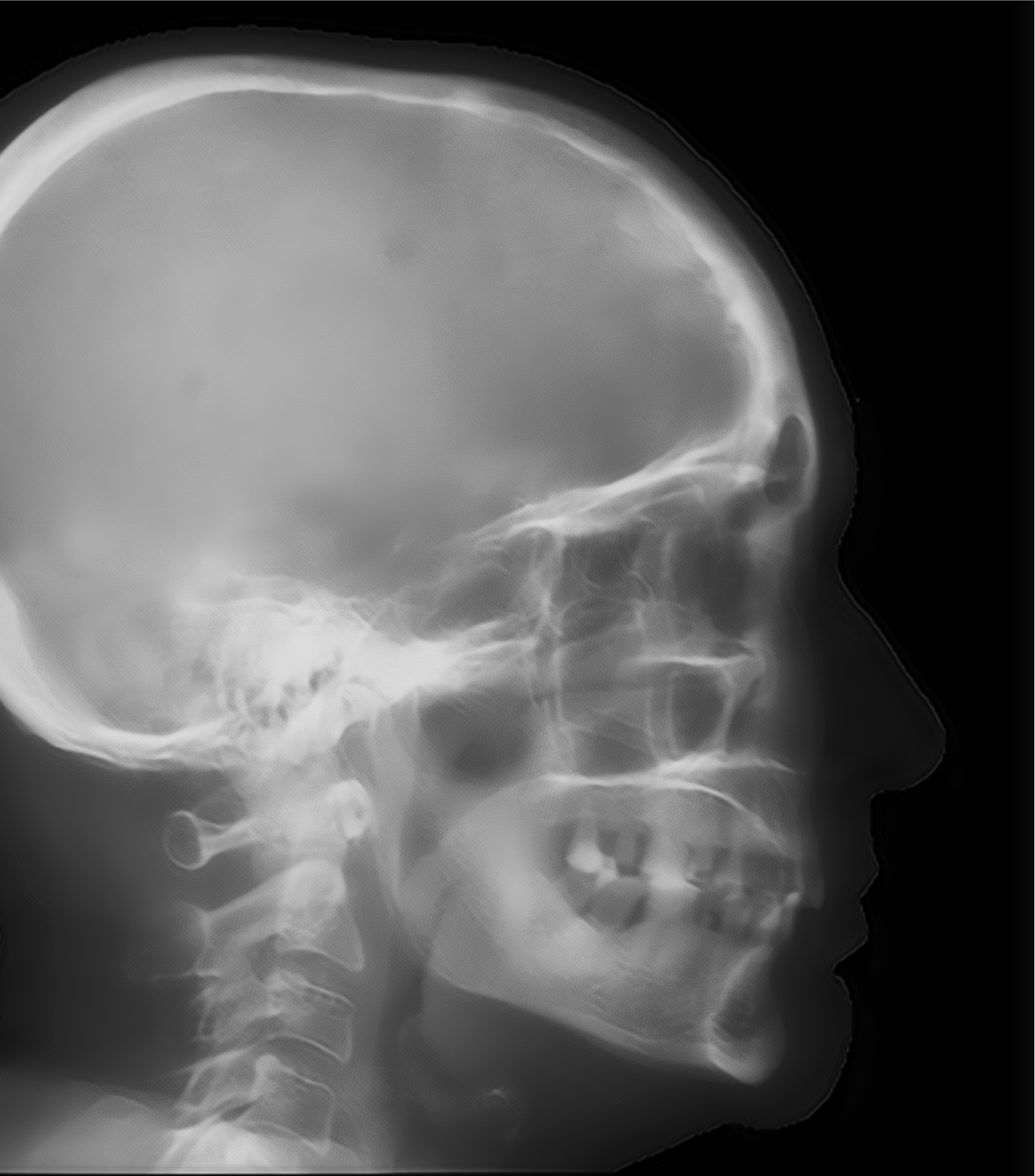}
}
\end{minipage}
\begin{minipage}{0.16\linewidth}
\subfigure[\modified{4.92}]{
\includegraphics[width=\linewidth]{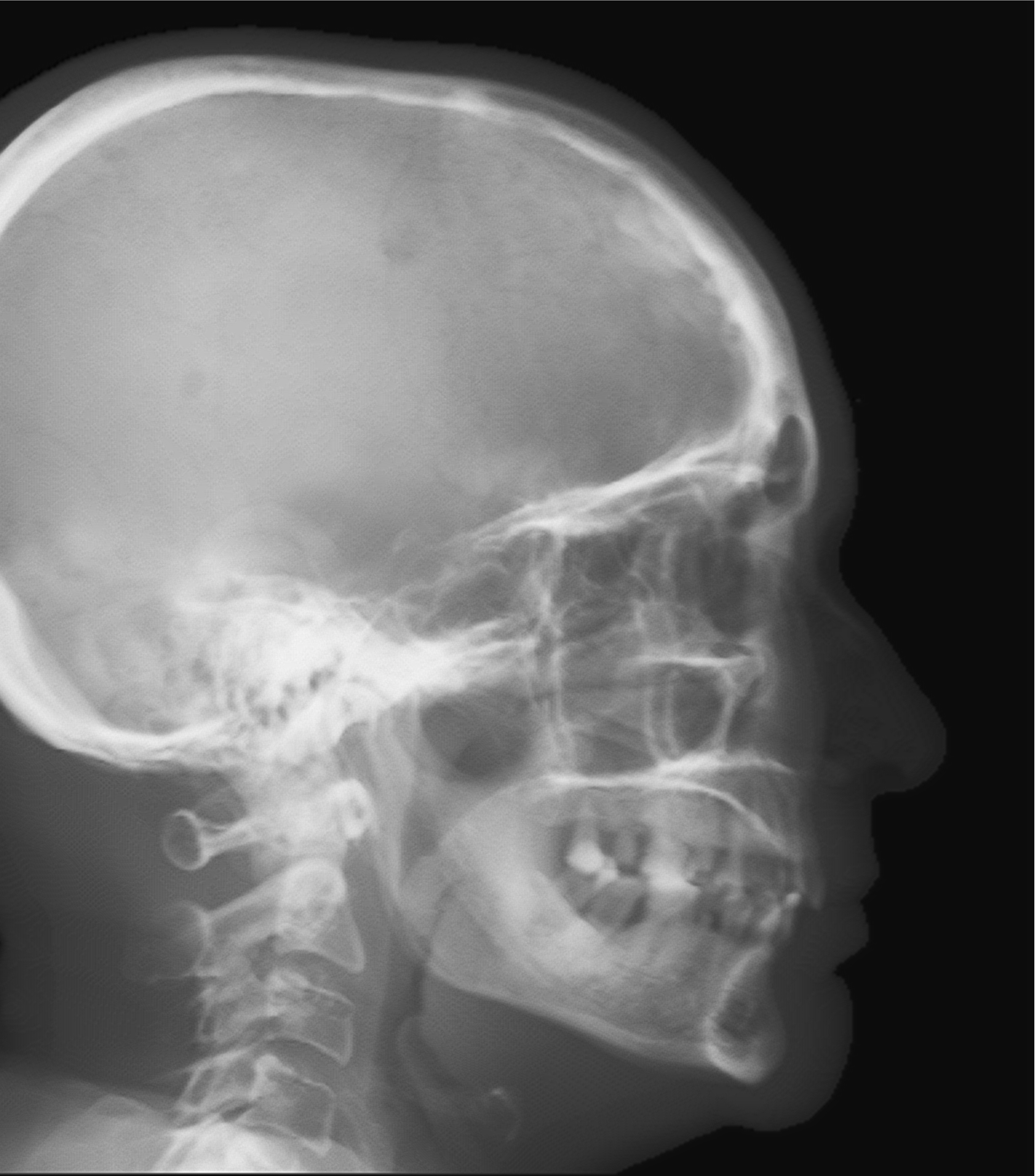}
}
\end{minipage}
\begin{minipage}{0.16\linewidth}
\subfigure[\modified{4.07}]{
\includegraphics[width=\linewidth]{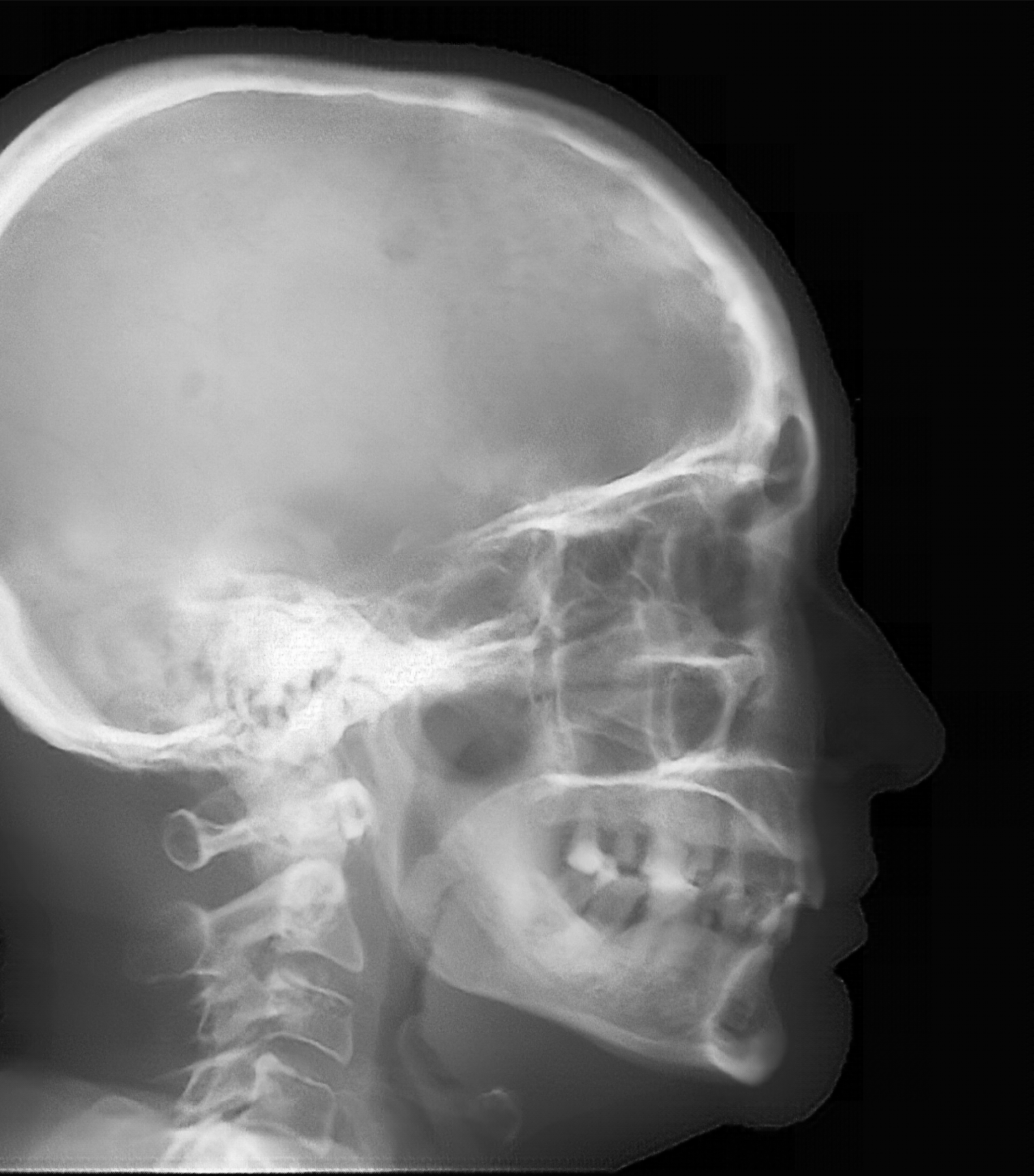}
}
\end{minipage}

\begin{minipage}{0.16\linewidth}
\subfigure[]{
\includegraphics[width=\linewidth]{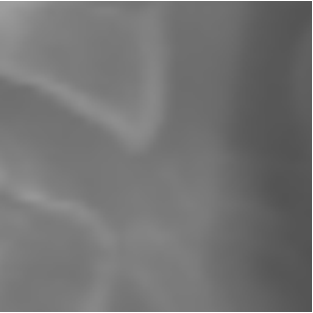}
}
\end{minipage}
\begin{minipage}{0.16\linewidth}
\subfigure[]{
\includegraphics[width=\linewidth]{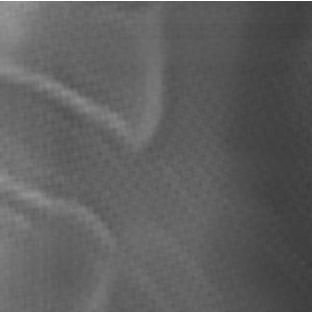}
}
\end{minipage}
\begin{minipage}{0.16\linewidth}
\subfigure[]{
\includegraphics[width=\linewidth]{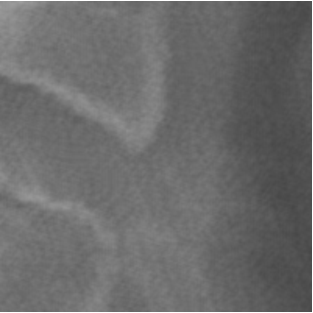}
}
\end{minipage}
\begin{minipage}{0.16\linewidth}
\subfigure[]{
\includegraphics[width=\linewidth]{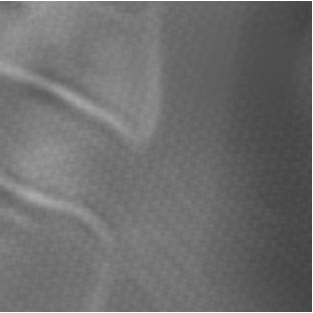}
}
\end{minipage}
\begin{minipage}{0.16\linewidth}
\subfigure[]{
\includegraphics[width=\linewidth]{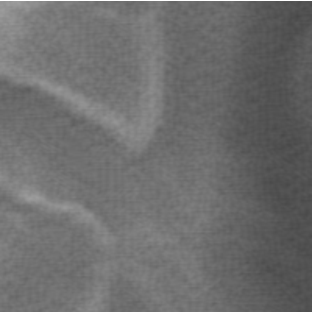}
}
\end{minipage}
\begin{minipage}{0.16\linewidth}
\subfigure[]{
\includegraphics[width=\linewidth]{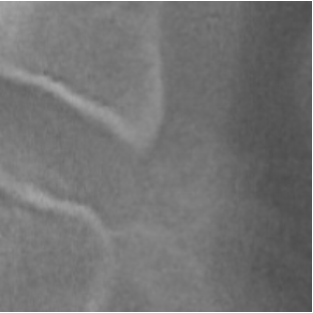}
}
\end{minipage}
\caption{\modified{SR} results on synthesized cephalogram from 3D volumes. The position of the example ROI patches in the bottom row is marked in (a). \modified{The perceptual indices of the top images are displayed in the corresponding subcaptions.}}
\label{Fig:SRSyhthesis}
\end{figure*}

In the bottom row, the results of one complete cephalogram are displayed. Compared with Fig.\,\ref{Fig:dualProjectionResults}(q), some structures like the vertebrae indicated by the arrow in Fig.\,\ref{Fig:dualProjectionResults}(r) are more accurate. For quantification, the intensity profiles of a chosen line from Figs.\,\ref{Fig:dualProjectionResults}(m)-(r), the position of which is marked in Fig.\,\ref{Fig:dualProjectionResults}(m), are plotted in Fig.\,\ref{Fig:lineProfile}. For the boundary area marked by $\text{B}$, the profile of the 2-projection output is the closest one to the target profile compared with others, which is consistent with the results in Fig.\,\ref{Fig:dualProjectionResults}(a)-(f). The major crests and troughs in the target curve are marked by $\text{C}_i$ and $\text{T}_i$ where $i = 1,2,3,4$, respectively, while a circular region near $\text{T}_3$ contains small crests and troughs is marked by $\text{F}_\text{1}$. The intensity differences between crests and troughs in the target profile and the 2-projection output profile, as well as the 1-projection output profile, are larger than those in the $0^\circ$ and $180^\circ$ projections, indicating image contrast improvement in our synthetic cephalograms. In the $\text{F}_\text{1}$ region, the 2-projection output profile, as well as the 1-projection output file, has deviation from the target profile. Nevertheless, all other major crests and troughs in the 2-projection output profile are concurrent with those of the target profile. Compared with those of the 1-projection output, the crest and trough positions of the 2-projection output are more accurate, especially for $\text{C}_1$, $\text{C}_4$, $\text{T}_2$, and $\text{T}_4$. This highlights the benefit of using dual projections in learning perspective deformation. 

\modified{For comparison, the synthetic cephalogram examples from 2D CBCT projections by CycleGAN \citep{zhu2017unpaired} are displayed in Fig.\,\ref{Fig:cycleGANResults} and the corresponding intensity profiles are also plotted in Fig.\,\ref{Fig:lineProfile}. The RMSE, PSNR and SSIM values tell us that using dual projections has no significant difference from using one projection only for CycleGAN. Comparing Fig.\,\ref{Fig:cycleGANResults}(d) with Fig.\,\ref{Fig:dualProjectionResults}(f), the outline in the CycleGAN output has larger deviation to the target blue curve than that in the pix2pixGAN output. The RMSE, PSNR and SSIM values of Fig.\,\ref{Fig:cycleGANResults}(f) are also worse than those of Fig.\,\ref{Fig:dualProjectionResults}(r). In Fig.\,\ref{Fig:lineProfile}, the intensity profile of the 2-projection CycleGAN output has apparent deviation from the target profile in the $\text{F}_2$ region. In addition, the $\text{C}_4$ crest position of the 2-projection CycleGAN output is about 2.5\,mm away from the target position, while that of the 2-projection pix2pixGAN is in the right place. These observations demonstrate the superiority of pix2pixGAN to CycleGAN in learning perspective deformation.}

\begin{table}[h]
\caption{Quantitative evaluation of different methods for Cephalogram synthesis from 2D CBCT projections.}
\label{Tab:SynthesisFromProjections}
\begin{footnotesize}
\begin{tabular}{cccc}
Method & one projection, & dual projections, & dual projections, \\
\ & one model  & one model  & four models\\
\hline
RMSE & 10.04 & 5.47 & 5.01 \\
PSNR & 28.03 & 33.10 & 33.83 \\
\modified{SSIM} & \modified{0.992} & \modified{0.997} & \modified{0.998}\\
\end{tabular}
\end{footnotesize}
\end{table}

The average \modified{RMSE, PSNR and SSIM} values of all the test patches are displayed in Tab.\,\ref{Tab:SynthesisFromProjections}. \modified{When one model is used} for four quadrants with one projection as the input of the pix2pixGAN, the average RMSE and PSNR values are 10.04 and 28.03 respectively. \modified{When one model is used} for four quadrants with dual projections, the image quality of synthetic cephalograms is significantly improved with RMSE = 5.47 and PSNR = 33.10. \modified{The average SSIM index is also improved from 0.992 to 0.997.} Therefore, using dual-projection RGB patches for training is superior to using one-projection only. While four models \modified{are used} for four respective quadrants with dual projections, the average \modified{RMSE, PSNR and SSIM} values are slightly improved without significance. Hence, using one model for four quadrant patches is applicable according to the symmetry property of perspective deformation.

%\begin{figure}
%\centering
%\begin{small}
%\begin{minipage}{0.32\linewidth}
%\centering
%Target
%\end{minipage}
%\begin{minipage}{0.32\linewidth}
%\centering
%One model for one quadrant
%\end{minipage}
%\begin{minipage}{0.32\linewidth}
%\centering
%One model for four quadrants
%\end{minipage}
%\end{small}

%\begin{minipage}{0.32\linewidth}
%\subfigure[]{
%\includegraphics[width = \linewidth]{patch3Test1Target}
%}
%\end{minipage}
%\begin{minipage}{0.32\linewidth}
%\subfigure[8.68, 29.15]{
%\includegraphics[width = \linewidth]{patch3Test1FourModels}
%}
%\end{minipage}
%%\begin{minipage}{0.32\linewidth}
%%\subfigure[Target]{
%%\includegraphics[width = \linewidth]{patch3Test1TwoModels}
%%}
%%\end{minipage}
%\begin{minipage}{0.32\linewidth}
%\subfigure[9.07, 28.77]{
%\includegraphics[width = \linewidth]{patch3Test1OneModel}
%}
%\end{minipage}
%\caption{The results of one example patch using one model for four quadrant or one model for one quadrant patches. The RMSE (left) and PSNR (right) values w.\,r.\,t. the target are displayed in the corresponding subcaptions.}
%\label{Fig:oneModelVSFourModels}
%\end{figure}

\subsection{Results of Super Resolution}

The SR results on the test patches of the ISBI Test1 data are displayed in Fig.\,\ref{Fig:SRResultsISBI}. Compared with the reference patches, the bicubic interpolation patches have blurry structures. Especially, the skeleton edges in Fig.\,\ref{Fig:SRResultsISBI}(b) suffer from jagging artifacts due to the large sampling scale. The structures in the outputs of ESRGAN\textsubscript{RDB} and ESRGAN\textsubscript{RRDB} using LR patches as the input have sharp edges. However, a certain level of checkerboard artifacts are observed. In the results of ESRGAN\textsubscript{RDB} and ESRGAN\textsubscript{RRDB} using ILR patches (i.e., the bicubic interpolation patches displayed in Figs.~\ref{Fig:SRResultsISBI}(b) and (i)) as the input, jagging artifacts remain at the edges. In the results of pix2pixGAN, high resolution structures are recovered without the introduction of jagging nor checkerboard artifacts. The quantitative evaluation results on all the test patches in Tab.\,\ref{Tab:ResultsOfSR} also indicate that pix2pixGAN achieves the best image quality, with the best RMSE of 4.8, \modified{PSNR of 32.5 and SSIM of 0.966}.

\begin{table}[h]
\centering
\caption{The quantitative evaluation of different SR methods on the ISBI Test1 data.}
\label{Tab:ResultsOfSR}
\begin{footnotesize}
\begin{tabular}{lccccc}
\hline
\multirow{2}{*}{Methods }&RDB &RDB &RRDB &RRDB & \multirow{2}{*}{pix2pixGAN}\\
 & LR & ILR & LR & ILR & \\
\hline
RMSE & 16.8 &9.7 &10.0 &8.5 &4.8 \\
PSNR   &22.4 &27.2 &26.5 &28.0 &32.5 \\
\modified{SSIM} & \modified{0.909} & \modified{0.928} & \modified{0.924}& \modified{0.948}& \modified{0.966}\\
\hline
\end{tabular}
\end{footnotesize}
\end{table}

The SR techniques are also applied to synthetic cephalograms. The results of one Type I synthetic cephalogram example are displayed in Fig.\,\ref{Fig:SRSyhthesis}. To visualize details better, an ROI patch is chosen for each method, whose position is marked in Fig.\,\ref{Fig:SRSyhthesis}(a). Consistent with the results on the ISBI test patches, the patches of ESRGAN\textsubscript{RDB} and ESRGAN\textsubscript{RRDB} using LR patches as the input also suffer from checkerboard artifacts in Figs.~\ref{Fig:SRSyhthesis}(h) and (j). In addition, some undesired bright/dark artifacts \modified{occur} in Fig.\,\ref{Fig:SRSyhthesis}(j). In Figs.~\ref{Fig:SRSyhthesis}(i) and (k), apparent jagging artifacts are no longer observed for ESRGAN\textsubscript{RDB} and ESRGAN\textsubscript{RRDB} using ILR patches as the input, since the edges in the bicubic interpolation patches are smooth without jaggies. As expected, the patch predicted by pix2pixGAN has realistic appearance without the introduction of jagging nor checkerboard artifacts. Due to its superior performance to ESRGAN\textsubscript{RDB} and ESRGAN\textsubscript{RRDB} using ILR on the ISBI test patches, in this work we choose pix2pixGAN as the SR method.
%& bicubic & 3.9 & 35.1

\subsection{Results of Landmark Detection}
\begin{figure}[h]
\centering
\begin{minipage}{\linewidth}
\centering
%\subfigure[Testset1]{
\includegraphics[width = 0.81\linewidth]{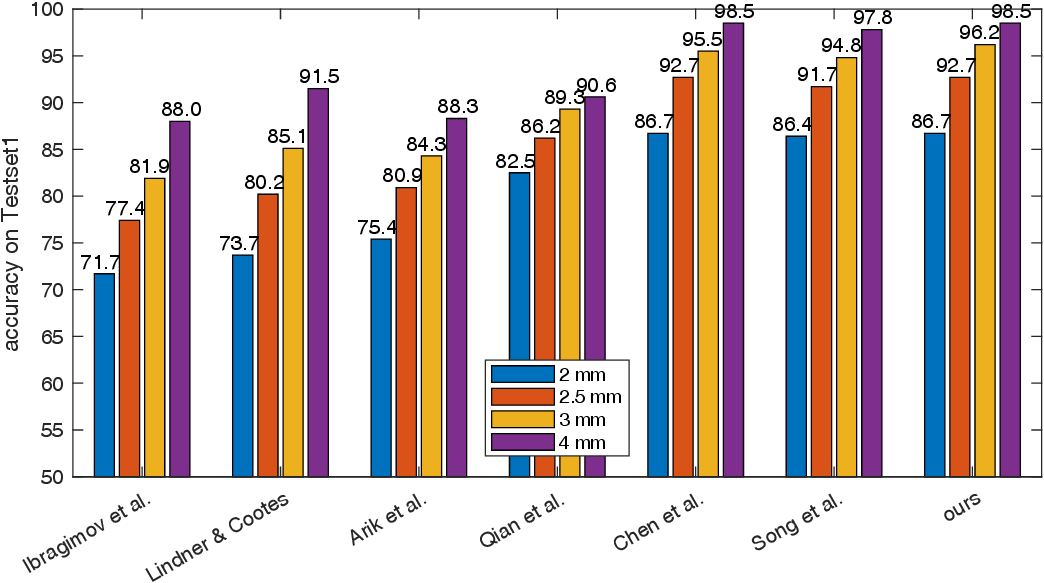}
%}
\end{minipage}

\vspace{3pt}

\begin{minipage}{\linewidth}
\centering
%\subfigure[Testset2]{
\includegraphics[width = 0.81\linewidth]{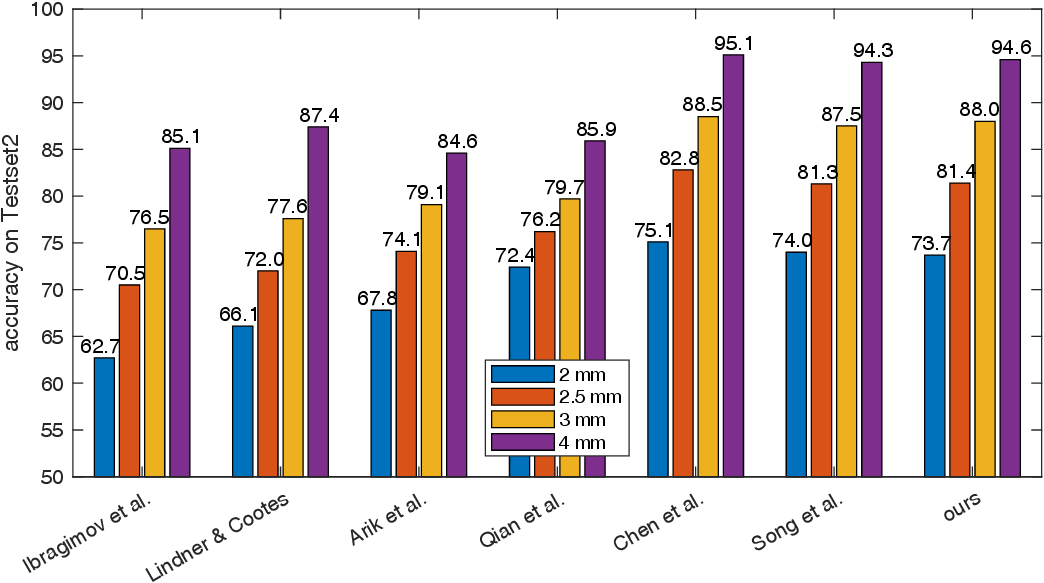}
%}
\end{minipage}
\caption{Accuracy comparison of different cephalometric landmark detection algorithms \citep{ibragimov2014automatic,lindner2015fully,arik2017fully,qian2019cephanet,chen2019cephalometric,song2020automatic} on ISBI Testset1 and Testset2.}
\label{Fig:LandmarkDetectionComparison}
\end{figure}
To validate the efficacy of our proposed automatic landmark detection algorithm, it is compared with other state-of-the-art algorithms on the benchmark ISBI data (Testset1 and Testset2). The SDRs of different algorithms \citep{ibragimov2014automatic,lindner2015fully,arik2017fully,qian2019cephanet,chen2019cephalometric,song2020automatic} in different precision ranges for Testset1 and Testset2 are displayed in Fig.\,\ref{Fig:LandmarkDetectionComparison}(a) and (b), respectively. Our proposed method achieves the 2\,mm-SDRs of 86.7\% and 73.7\% on the ISBI Testset1 and Testset2, respectively, which is comparable to the best accuracy methods \citep{chen2019cephalometric,song2020automatic}. However, our method is more efficient than \citep{chen2019cephalometric} and has a simpler architecture than \citep{song2020automatic}.

\begin{figure}[!h]
\raggedleft
\begin{minipage}{0.025\linewidth}
\scriptsize{\rotatebox[origin=c]{90}{Other syntheses from 3D volumes}}
\end{minipage}
\begin{minipage}{0.31\linewidth}
\subfigure[RayCast]{
\includegraphics[width=\linewidth]{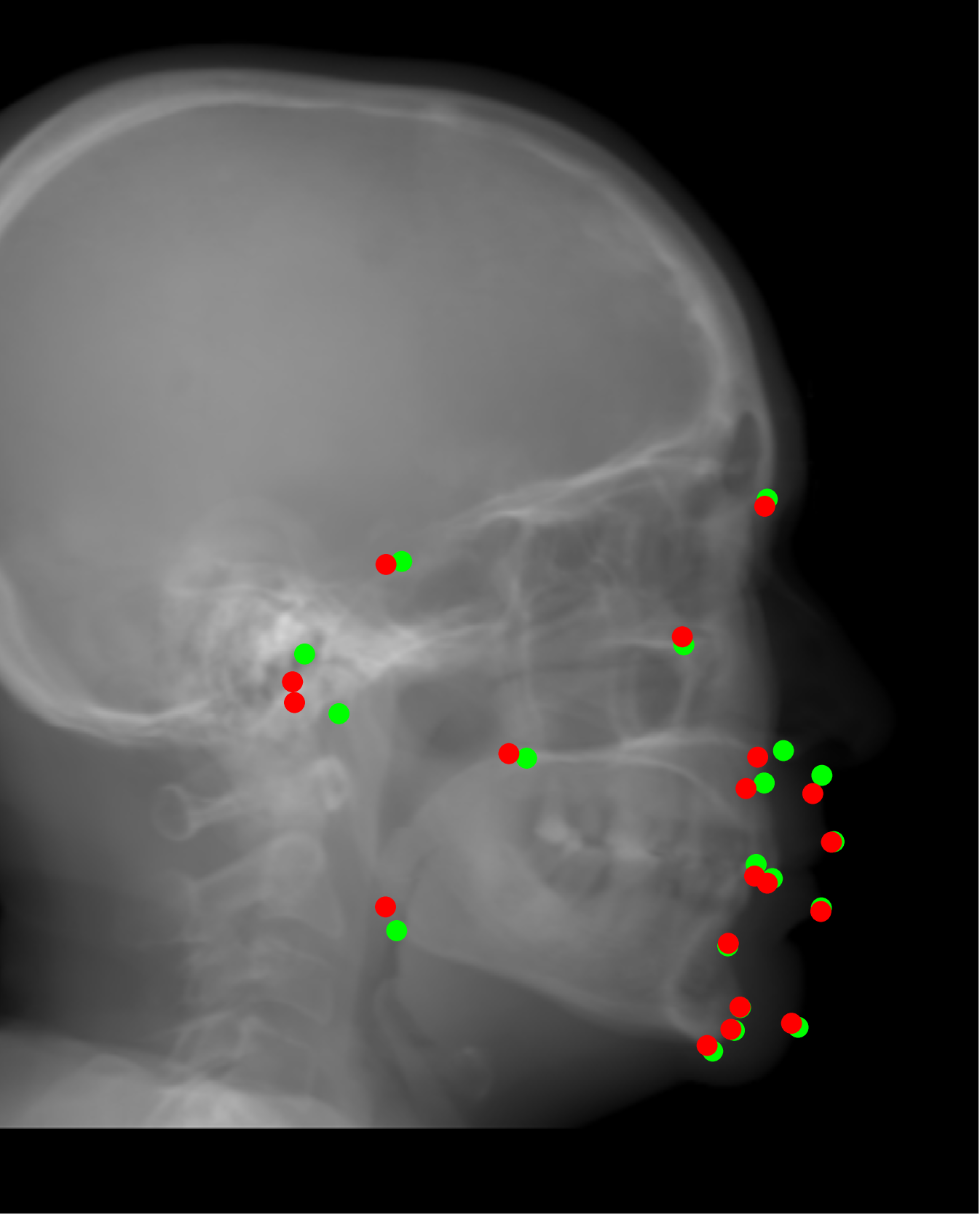}
}
\end{minipage}
\begin{minipage}{0.31\linewidth}
\subfigure[MIP100]{
\includegraphics[width=\linewidth]{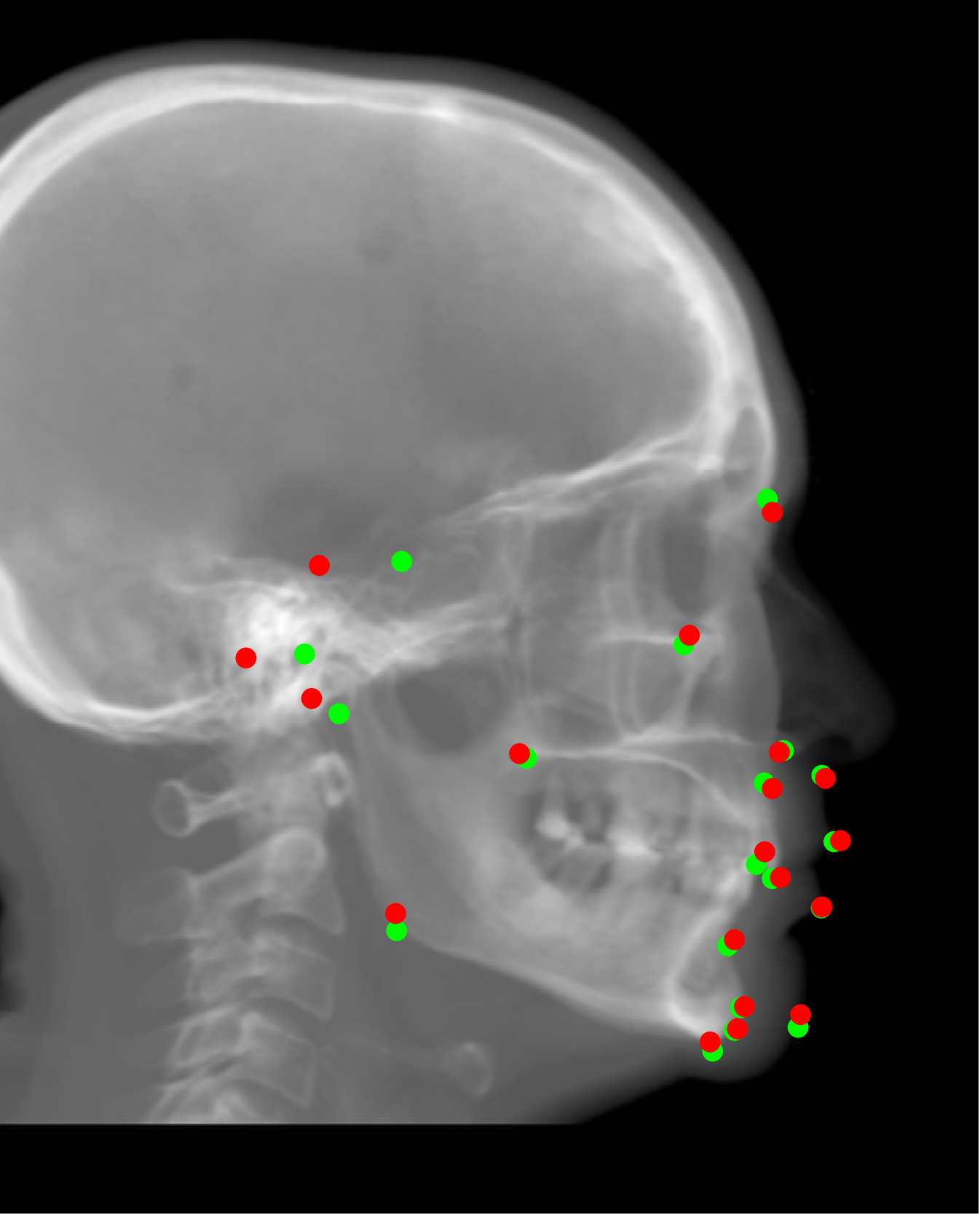}
}
\end{minipage}
\begin{minipage}{0.31\linewidth}
\subfigure[Original sigmoid]{
\includegraphics[width=\linewidth]{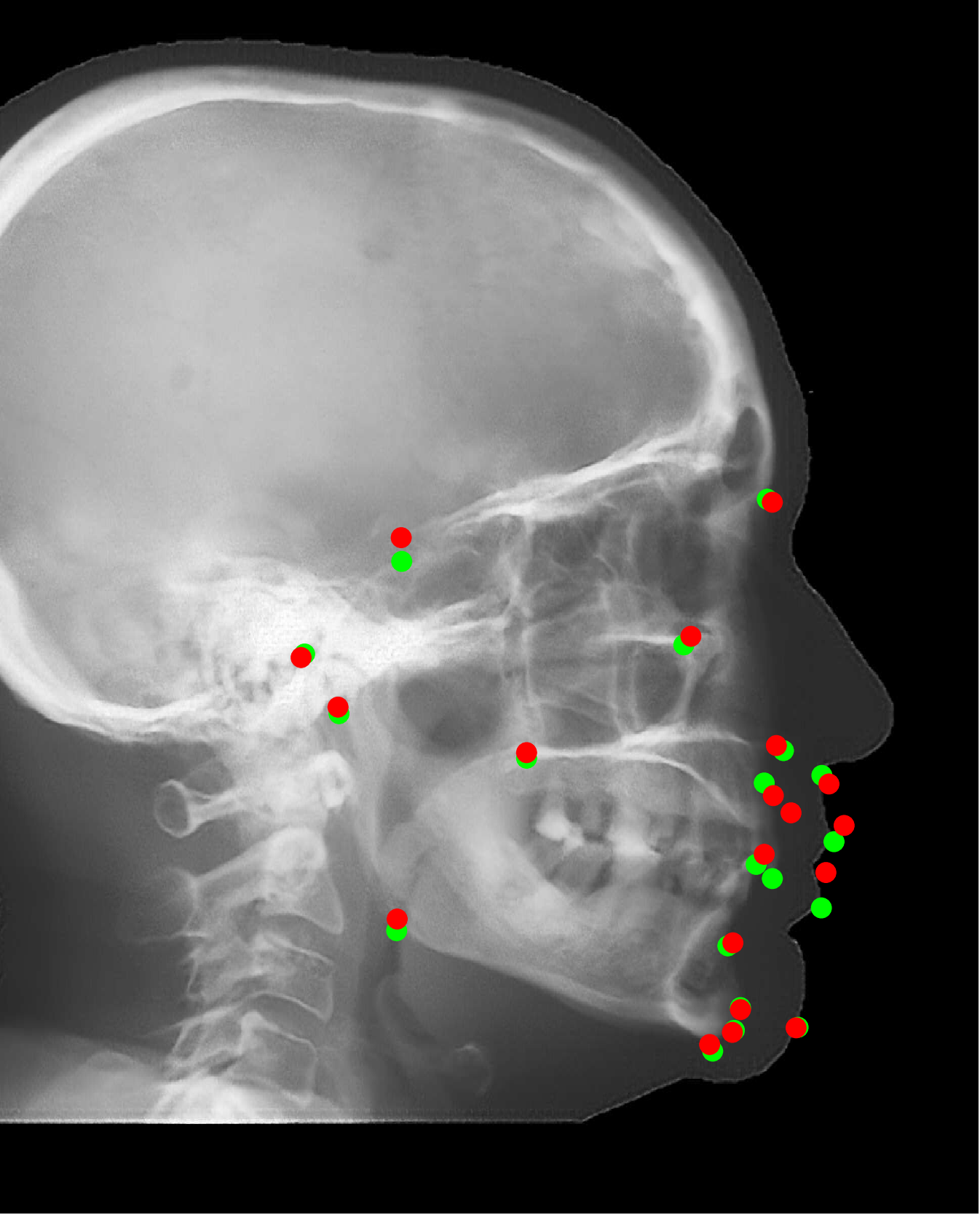}
}
\end{minipage}

\raggedleft
\begin{minipage}{0.025\linewidth}
\scriptsize{\rotatebox[origin=c]{90}{Type I with different SR methods}}
\end{minipage}
\begin{minipage}{0.31\linewidth}
\subfigure[Bicubic]{
\includegraphics[width=\linewidth]{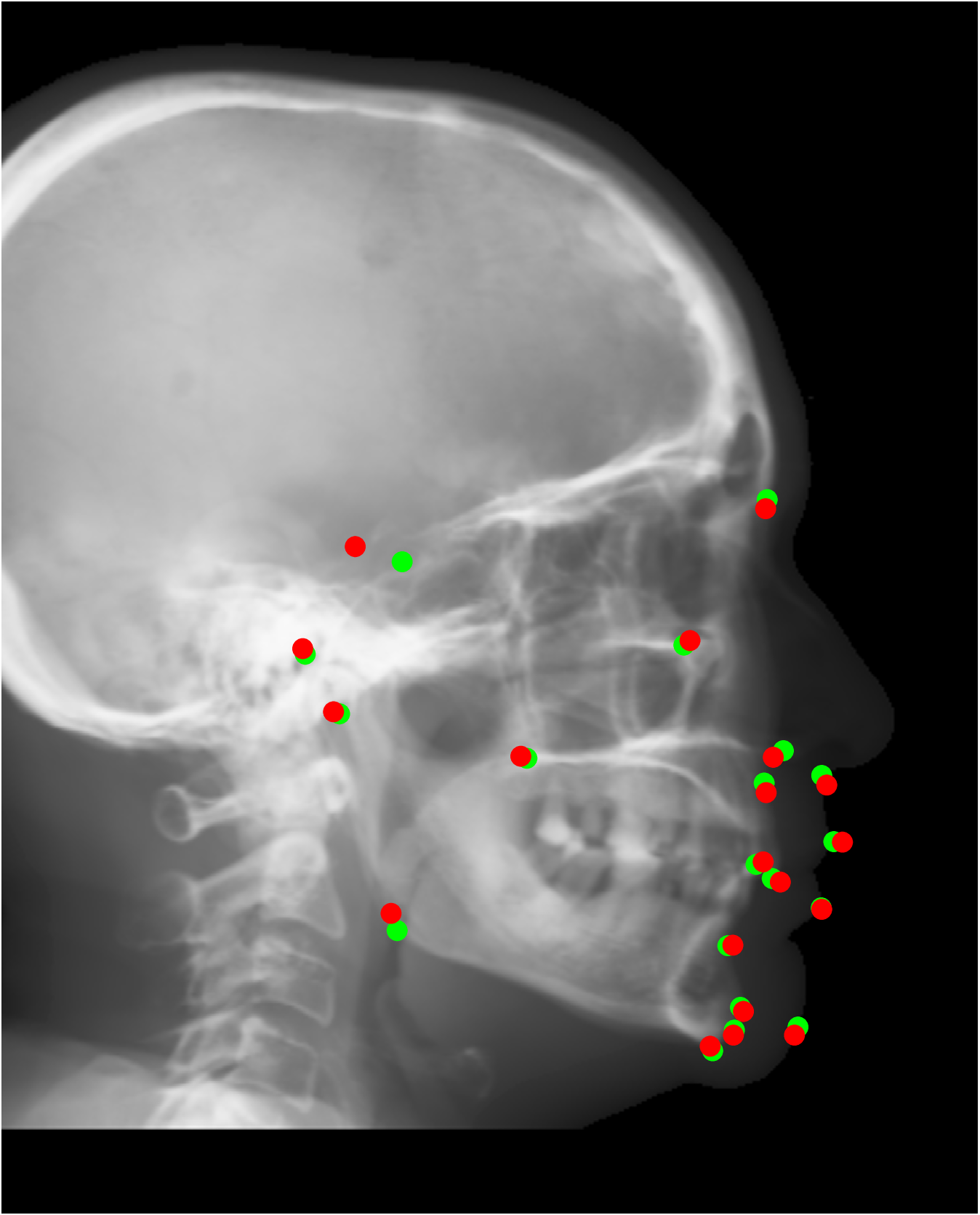}
}
\end{minipage}
\begin{minipage}{0.31\linewidth}
\subfigure[RDN, ILR]{
\includegraphics[width=\linewidth]{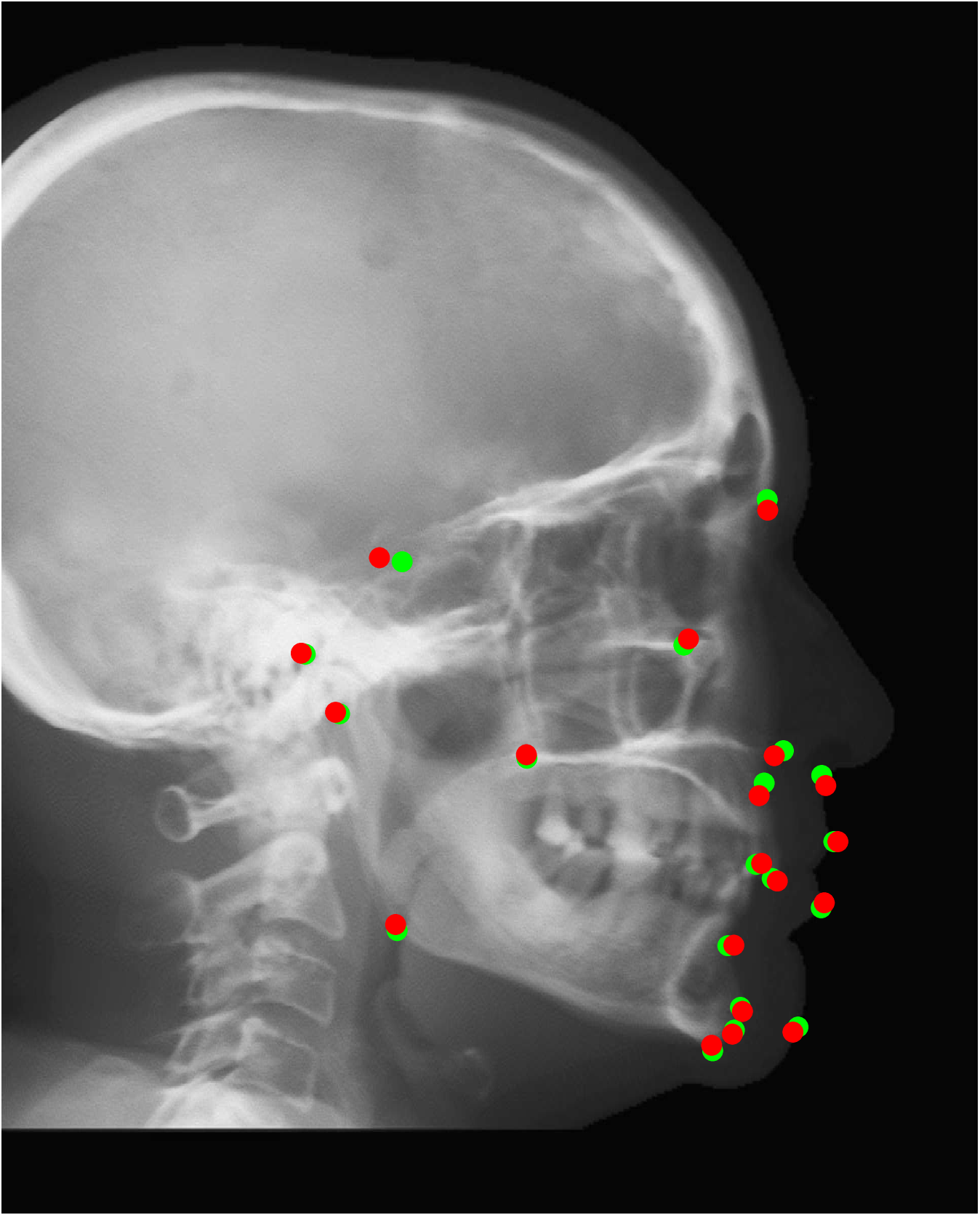}
}
\end{minipage}
\begin{minipage}{0.31\linewidth}
\subfigure[RRDN, ILR]{
\includegraphics[width=\linewidth]{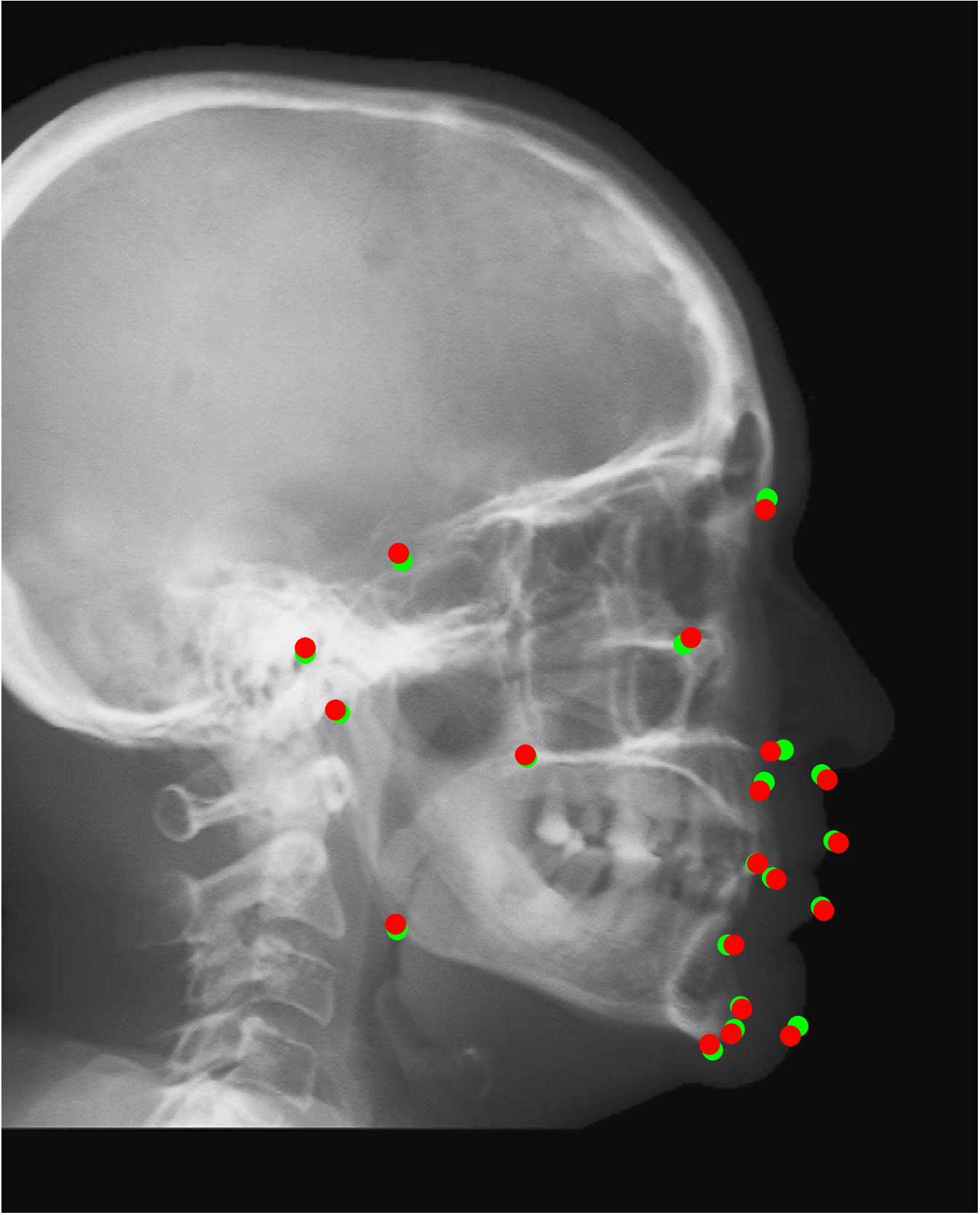}
}
\end{minipage}

\begin{minipage}{0.025\linewidth}
\scriptsize{\rotatebox[origin=c]{90}{Type I synthesis}}
\end{minipage}
\begin{minipage}{0.31\linewidth}
\subfigure[]{
\includegraphics[width=\linewidth]{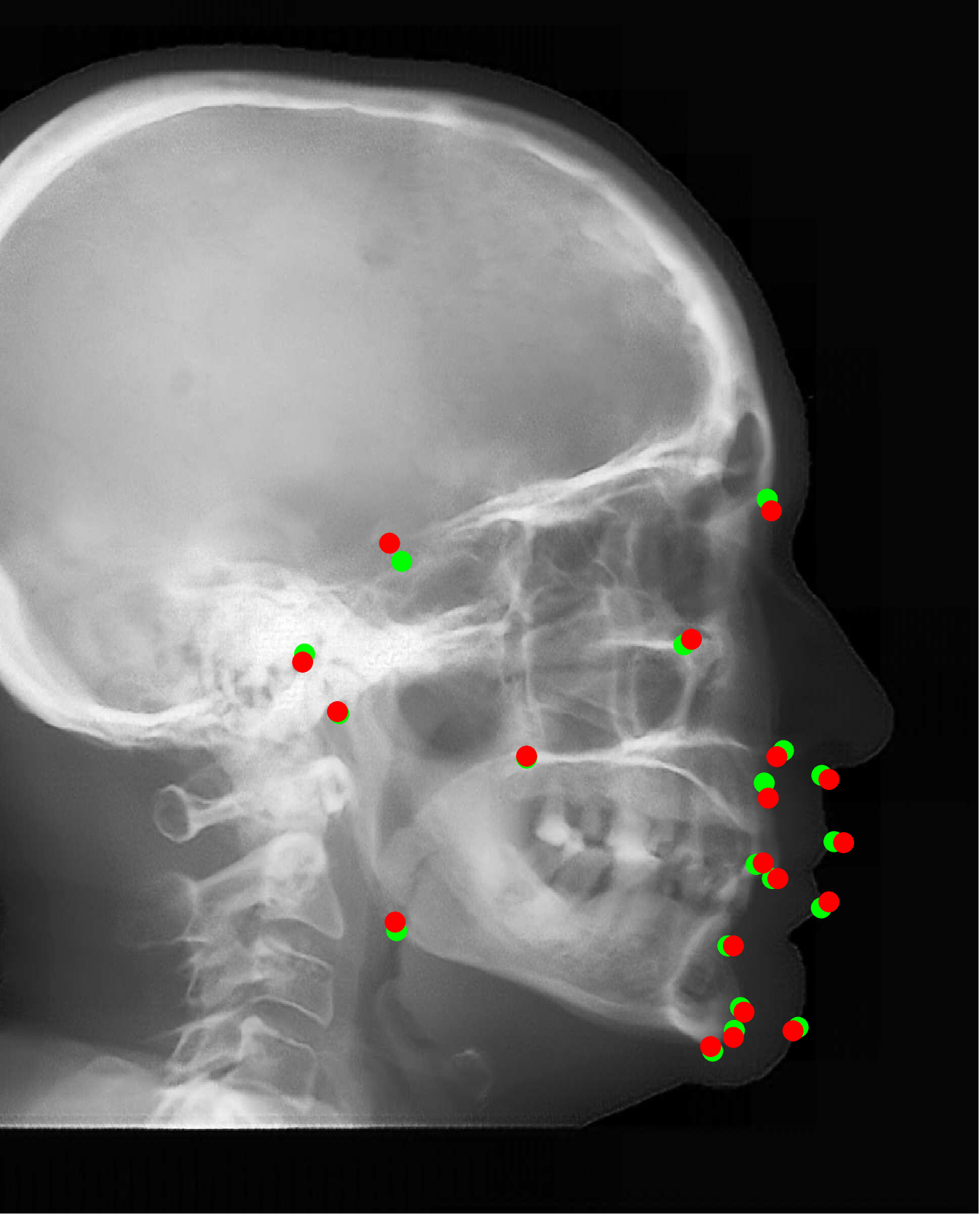}
}
\end{minipage}
\begin{minipage}{0.31\linewidth}
\subfigure[]{
\includegraphics[width=\linewidth]{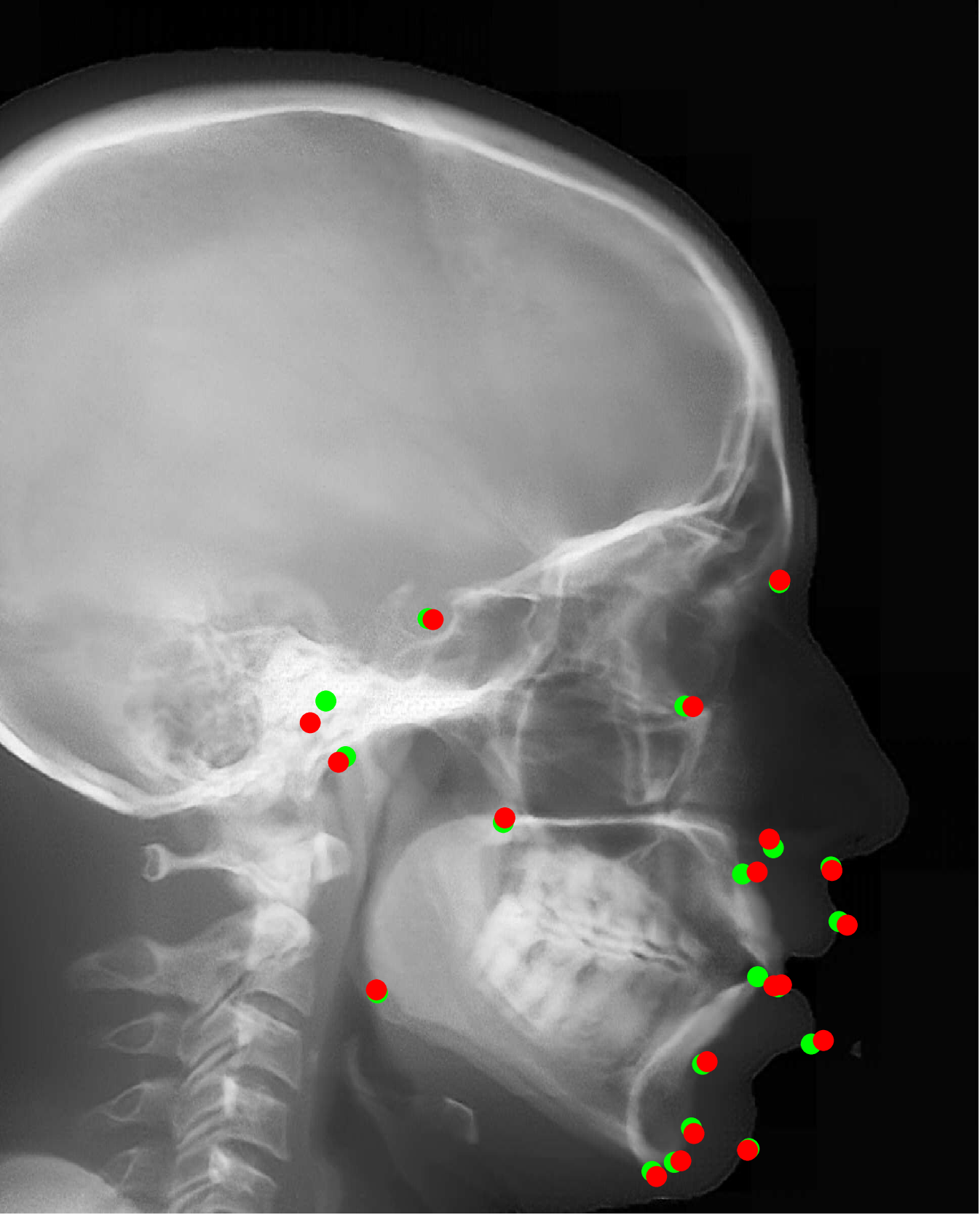}
}
\end{minipage}
\begin{minipage}{0.31\linewidth}
\subfigure[]{
\includegraphics[width=\linewidth]{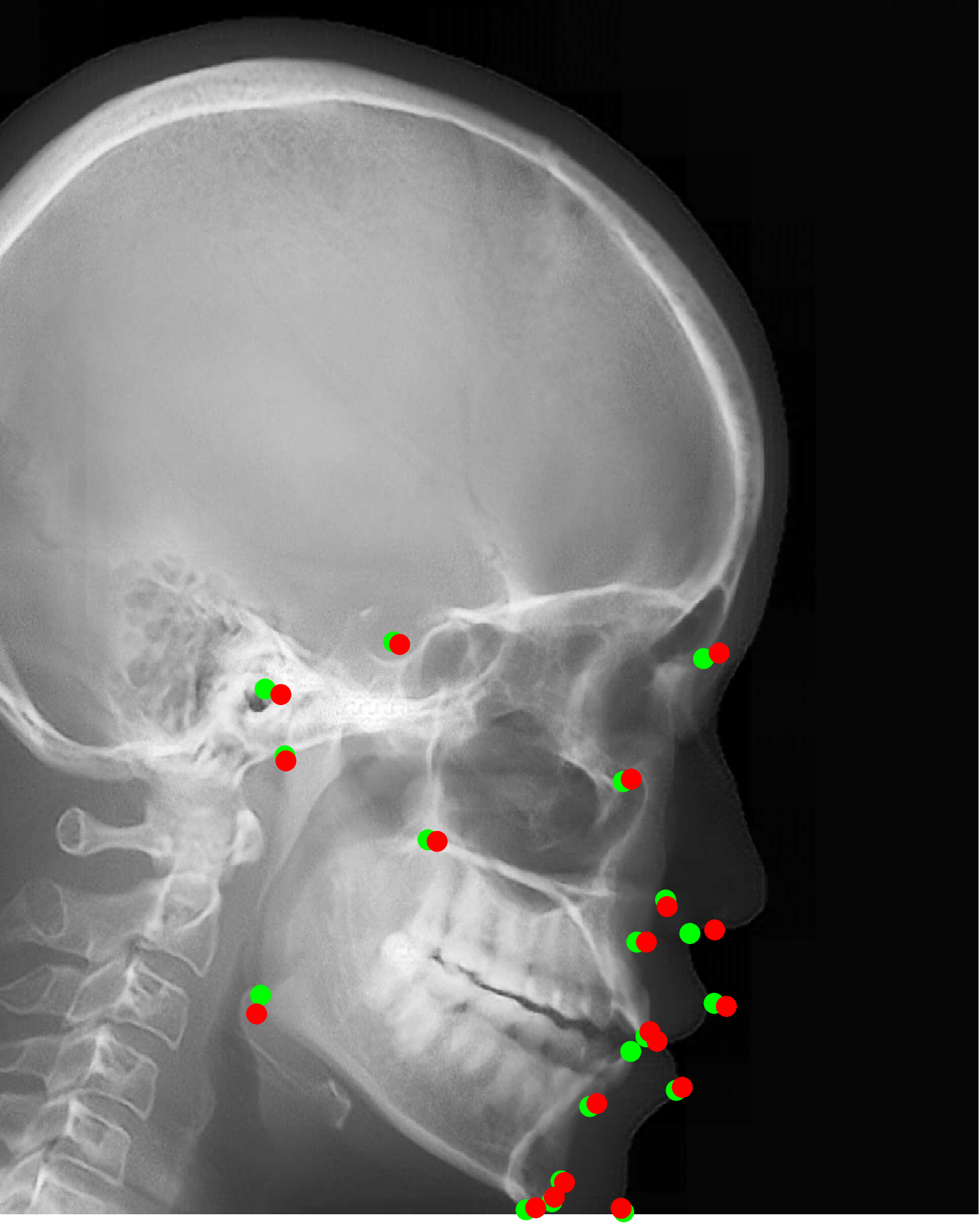}
}
\end{minipage}

\begin{minipage}{0.025\linewidth}
\scriptsize{\rotatebox[origin=c]{90}{Type II synthesis}}
\end{minipage}
\begin{minipage}{0.31\linewidth}
\subfigure[]{
\includegraphics[width=\linewidth]{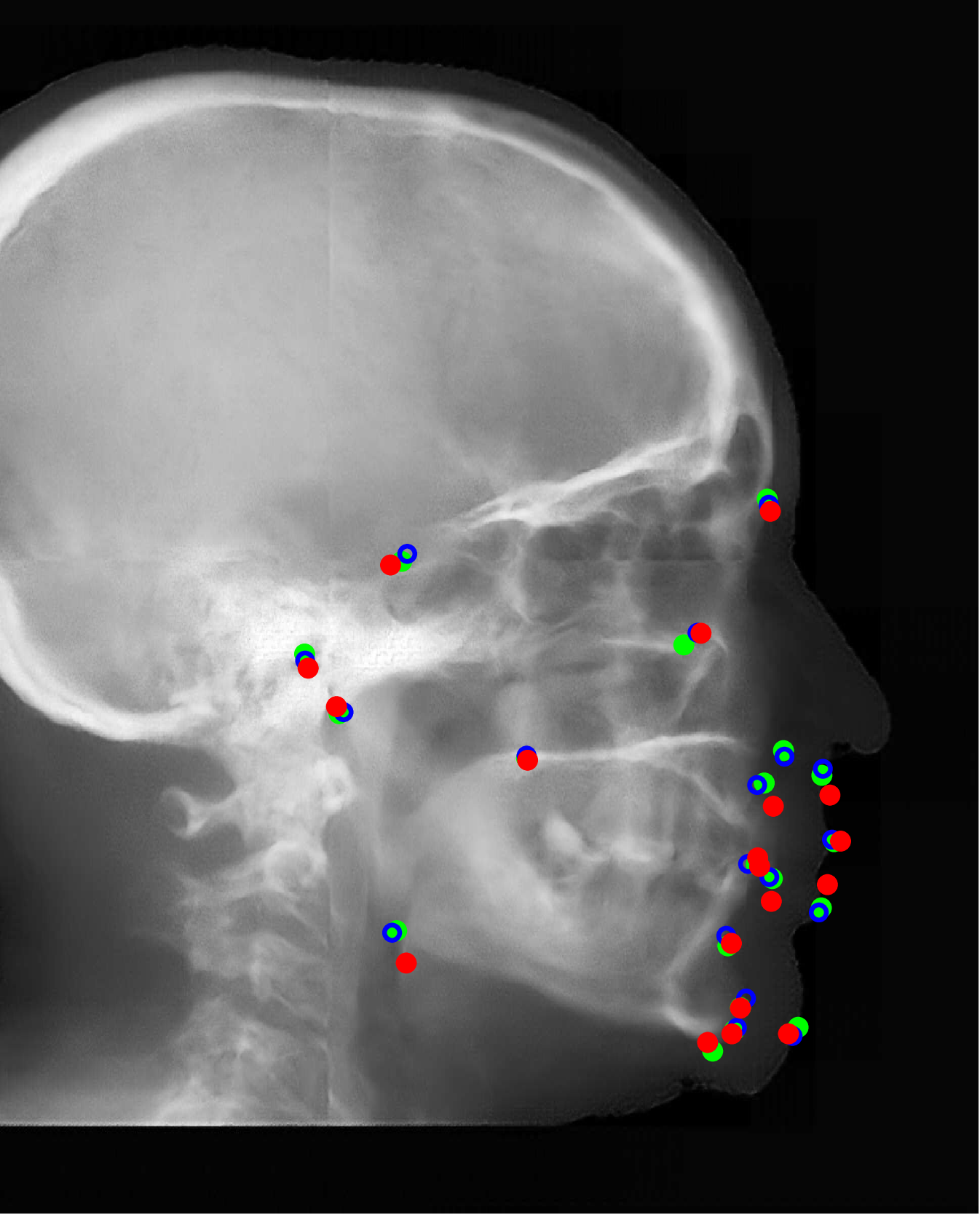}
}
\end{minipage}
\begin{minipage}{0.31\linewidth}
\subfigure[]{
\includegraphics[width=\linewidth]{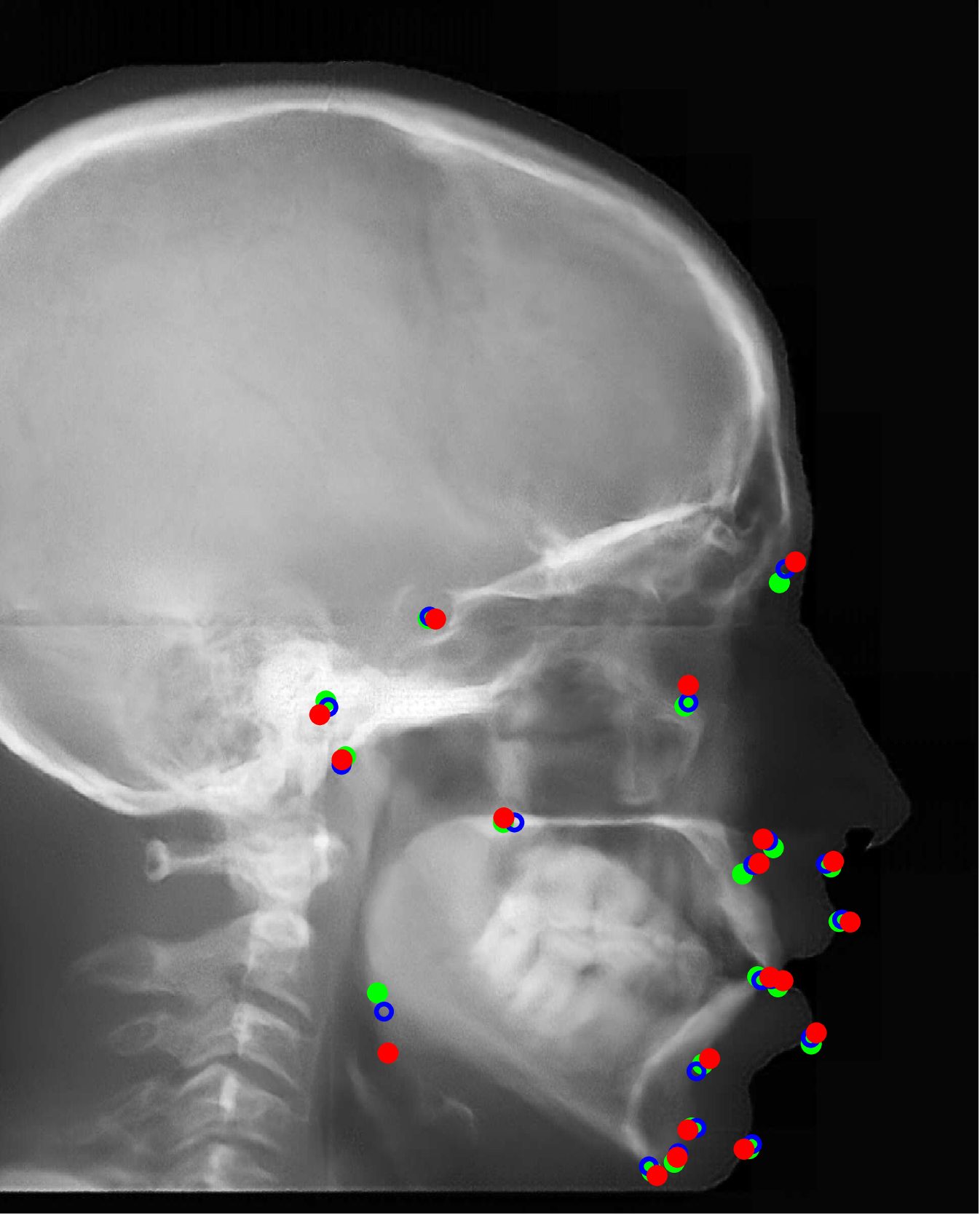}
}
\end{minipage}
\begin{minipage}{0.31\linewidth}
\subfigure[]{
\includegraphics[width=\linewidth]{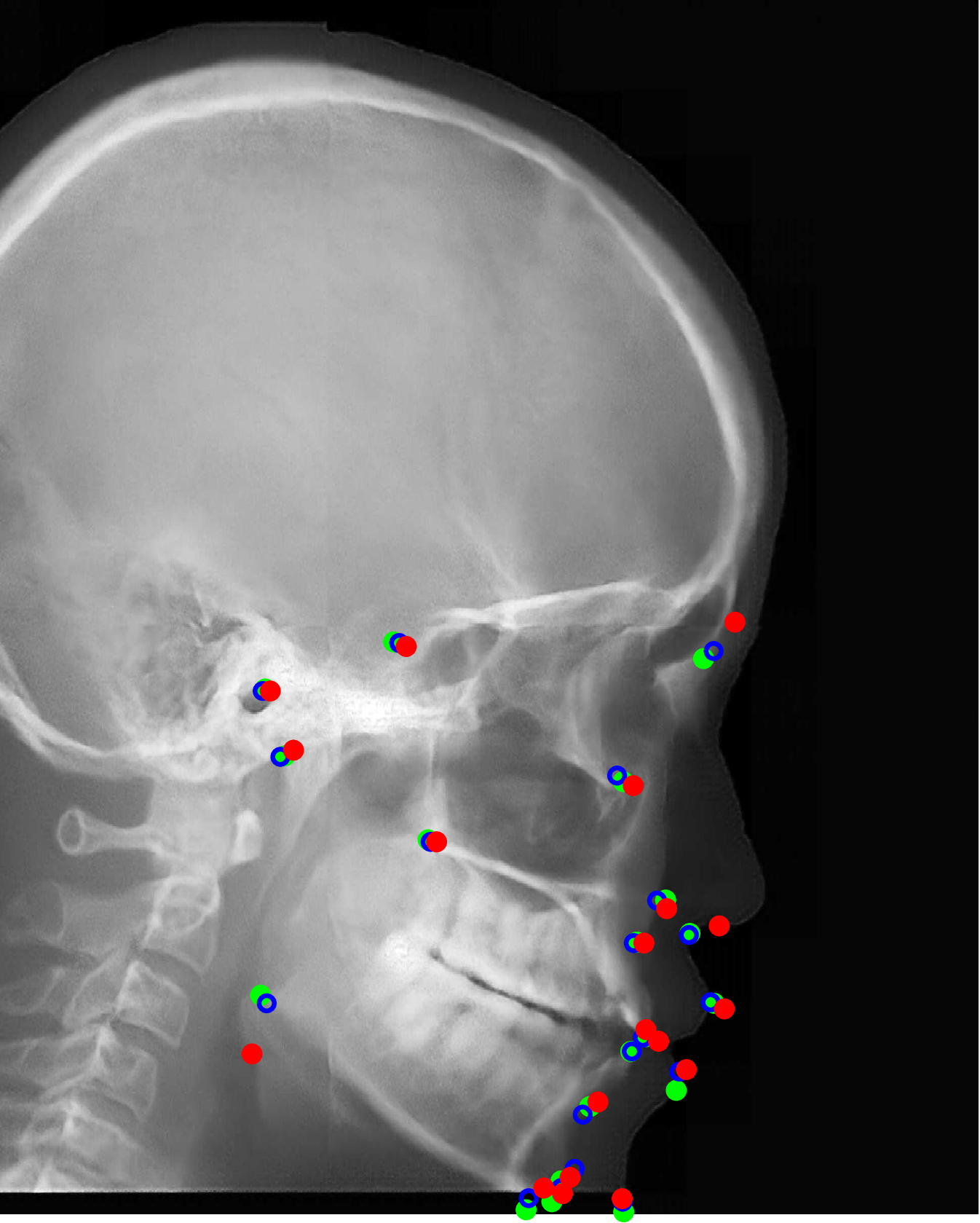}
}
\end{minipage}
\caption{\modified{Landmark detection on synthetic cephalograms. Cephalograms in the top row are obtained by different synthesis methods from 3D volumes for the first patient with pix2pixGAN for SR. The second row shows Type I synthetic cephalograms with different SR methods. The 3rd and bottom rows are Type I and Type II synthetic cephalograms respectively from three example patients with pix2pixGAN for SR.  The green (reference) and blue labels are manual detection landmark positions in Type I and Type II synthetic cephalograms respectively, while the red labels are automated detection landmark positions in each cephalogram. }}
\label{Fig:LandmarkLabelSynthesized}
\end{figure}

Our proposed landmark detection method is applied to detect landmarks in synthetic cephalograms. The results of three example patients are displayed in Fig.\,\ref{Fig:LandmarkLabelSynthesized}. Cephalograms in the top row are obtained by different synthesis methods from 3D volumes for the first patient \modified{with pix2pixGAN for SR. The second row shows Type I synthetic cephalograms with different SR methods. The 3rd and bottom rows are Type I and Type II synthetic cephalograms respectively from three example patients with pix2pixGAN for SR.} The green and blue labels are manual detection landmark positions in Type I and Type II synthetic cephalograms respectively, while the red labels are automated detection landmark positions in each cephalogram. The green labels are used as the reference.
The overall SDRs of the landmarks in different types of synthetic cephalograms on the test patients are displayed in Tab.\,\ref{Tab:landmarkDetection}. The last row SDRs are calculated from manual detection (blue) landmarks in Type II synthetic cephalograms w.\,r.\,t.~reference landmarks, while others are from automatic detection (red) landmarks w.\,r.\,t.~reference landmarks. 

For RayCast in Fig.\,\ref{Fig:LandmarkLabelSynthesized}(a), many automatic detection landmarks exceed the 4\,mm precision range such as the landmarks of anterior nasal spine, gonion, porion and articulare. According to Tab.\,\ref{Tab:landmarkDetection}, overall only 45.6\% landmarks are detected within the 4\,mm precision range by the automatic landmark detection algorithm.
For MIP100 in Fig.\,\ref{Fig:LandmarkLabelSynthesized}(b), the landmarks of sella, porion, articulare have very large deviations from the reference landmarks. Overall, it achieves 47.4\% 2\,mm-SDR and 81.5\% 4\,mm-SDR. 
Figs.\,\ref{Fig:LandmarkLabelSynthesized}(c) and \modified{(g)} are the cephalograms synthesized by our Type I synthesis with the original sigmoid transform and the modified sigmoid transform, respectively. These two synthetic cephalograms have subtle difference in the soft-tissue areas. As a result, the detected incision superius and lower lip positions exceed the 4\,mm range in Fig.\,\ref{Fig:LandmarkLabelSynthesized}(c) while they are well detected within the 2\,mm precision range in Fig.\,\ref{Fig:LandmarkLabelSynthesized}\modified{(g)}. This demonstrates the benefit of the modified sigmoid transform.

\modified{In Figs.\,\ref{Fig:LandmarkLabelSynthesized}(d)-(f), the landmark detection accuracies are very close to each other for the other 18 landmarks except for the sella landmark. The point-to-point errors are 9.77\,mm, 4.57\,mm, and 1.65\,mm respectively in Figs.\,\ref{Fig:LandmarkLabelSynthesized}(d)-(f) for the sella. Tab.\,\ref{Tab:landmarkDetection} indicates that overall 66.7\% landmarks are detected within the 2\,mm precision range for bicubic interpolation. For ESRGAN\textsubscript{RDB} and ESRGAN\textsubscript{RRDB} using ILR patches, the 2\,mm-SDR rises to 70.2\% and 71.9\%, respectively. Nevertheless, pix2pixGAN achieves the best 2\,mm-SDR 75.4\% among different SR methods for Type I synthesis. Therefore, different SR methods have an influence on the landmark detection.}

For Type I synthesis, two more cephalograms are displayed in Figs.\,\ref{Fig:LandmarkLabelSynthesized}\modified{(h) and (i)}. Overall, 93.0\% automatic detection landmarks are overlapped with the reference landmarks within the precision range of 4\,mm, with a few exceptions such as the sella landmark in \modified{(g)}, the porion in \modified{(h)}, and the incision inferius and subnasale landmarks in \modified{(i)} beyond 4\,mm. Particularly, 75.4\% automatic detection landmarks are within the 2\,mm clinical acceptable precision range. This indicates that the automatic landmark detection algorithms learned from real conventional cephalograms can be applied to our Type I synthetic cephalograms. 

For Type II synthetic cephalograms in Figs.\,\ref{Fig:LandmarkLabelSynthesized}\modified{(j)-(l)}, all the manual detection (blue) landmarks are overlapped with the reference landmarks within the distance range of 4\,mm, indicating that the landmark positions in the Type II synthetic cephalograms have no impactful position shift from those in Type I synthetic cephalograms. The majority (80.7\%) of the automatic detection (red) landmarks are also overlapped with the reference landmarks within the 4\,mm precision range. However, more automatic detection landmarks in Type II synthetic cephalograms are outside the 4\,mm range than those in Type I synthetic cephalograms, for example, the anterior nasal spine landmark in \modified{(j)}, the orbitale landmark in \modified{(k)}, the incision inferius landmark in \modified{(l)}, the nasion landmarks in \modified{(k) and (l)}, and the gonion landmarks in \modified{(j)-(l)}. The 2\,mm-SDR also decreases from 75.4\% to 50.9\%. 
%{Nevertheless, the overall SDRs for Type II synthesis are comparable to those for Type I synthesis with the original sigmoid transform.}

\begin{table}[h]
\caption{{SDRs for 2.0\,mm, 2.5\,mm, 3.0\,mm and 4.0\,mm precision ranges.}}
\label{Tab:landmarkDetection}
\centering
\begin{footnotesize}
\begin{tabular}{lcccc}
\modified{Synthesis + SR} \ & 2\,mm & 2.5\,mm & 3\,mm & 4mm \\
\hline
\modified{RayCast + pix2pixGAN} & 31.6 & 38.6 & 38.6 & 45.6\\
\modified{MIP100  + pix2pixGAN} & 47.4 & 59.6 & 70.2 & 81.5 \\
\modified{Original sigmoid  + pix2pixGAN}& 49.1 & 63.1 & 73.7 & 84.2\\
\modified{Type I + Bicubic}  & 66.7 & 78.9 & 86.0 & 91.2\\
\modified{Type I + RDN, ILR} & 70.2 & 78.9 & 86.0 & 94.7\\
\modified{Type I + RRDN, ILR}  & 71.9 & 84.2 & 87.7 & 94.7\\
\modified{Type I  + pix2pixGAN}& 75.4 & 82.5 & 84.2 & 93.0 \\
\modified{CycleGAN (Type II) + pix2pixGAN}& 26.3 & 35.1 & 43.9 & 57.9 \\
\modified{Type II  + pix2pixGAN}& 50.9 & 64.9 & 68.4 & 80.7 \\
\modified{Type II  + pix2pixGAN (manual)} & 73.7 & 87.7 & 91.2 & 100 \\
\end{tabular}
\end{footnotesize}
\end{table}
\section{Discussion}
%For cephalogram synthesis from 3D CBCT volumes, Fig.\,\ref{Fig:cephalogramComparison} demonstrates that cephalograms synthesized by our algorithms using either orthogonal projection or perspective projection have better image contrast than those by conventional RarCast methods \citep{moshiri2007accuracy,kumar2007comparison}. Therefore, soft-tissue cephalometric landmarks like subnasale, upper lip, lower lip and soft tissue pogonion can be identified easily.

The accuracy of landmarks in synthetic cephalograms using RayCast from 3D CBCT volumes has been validated in previous research \citep{farman2005dentomaxillofacial,farman2006development,moshiri2007accuracy}. Our Type I cephalogram synthesis method is an improved version of RayCast. Therefore, the accuracy of landmarks in our Type I synthetic cephalograms is guaranteed in principle. The improvement lies in image contrast based on the optical properties of conventional X-ray films and image resolution using SR techniques, making synthetic cephalograms closer to real conventional cephalograms. With the above premises, using the Type I synthetic cephalograms as the target of cephalogram synthesis from 2D projections has practical value.

%The accuracy of landmarks in synthetic cephalograms using RayCast from 3D CBCT volumes has been validated in previous research \citep{farman2005dentomaxillofacial,farman2006development,moshiri2007accuracy}. Our method for cephalogram synthesis from 3D CBCT volumes is an improved version of RayCast. Therefore, the accuracy of landmarks in our synthetic cephalograms is guaranteed in principle. Our method improves RayCast in terms of image contrast based on the optical properties of conventional X-ray films and image resolution using SR techniques. Fig.\,\ref{Fig:LandmarkLabelSynthesized} and Tab.\,\ref{Tab:landmarkDetection} display the high successful detection rates of landmarks in synthetic cephalograms using an automatic landmark detection algorithm learned from the existing database of conventional cephalograms. This indicates that the synthetic cephalograms share similar features with real conventional cephalograms. Therefore, our synthetic cephalograms from 3D volumes are close to real cephalograms in terms of landmark location, image contrast and image resolution. With this premise, using the synthesised cephalograms from 3D CBCT volumes as the target of cephalogram synthesis from 2D projections has practical value. 
% In Fig.\,\ref{Fig:sigmoidCurve}(b), the intensity transform between ray cast projections and conventional cephalograms formed by the black dots is an very coarse approximation due to the scarcity of dental CBCT volumes and their corresponding cephalograms. With matching pairs, more accurate intensity transform can be learned.

In our Type II cephalogram synthesis, pix2pixGAN is capable to improve image contrast and reduce the perspective deformation, as demonstrated in Fig.\,\ref{Fig:dualProjectionResults} and Fig.\,\ref{Fig:lineProfile}. Therefore, using synthetic cephalograms from 2D projections for cephalometric analysis is promising. But it is worth noting that some information, especially for low contrast high frequency structures, is missing or incorrect in the Type II synthetic cephalograms compared with Type I synthetic cephalograms. For example, in Fig.\,\ref{Fig:dualProjectionResults}(j) the cranial sutures indicated by the arrow are visualized. However, in the 2D projections in Figs.\,\ref{Fig:dualProjectionResults}(g)-(i), they are barely seen. As a result, they are not visible in the output of pix2pixGAN, no matter \modified{whether} one projection or dual projections \modified{are} used as the input. Another example is the circular region marked by $\text{F}_1$ in Fig.\,\ref{Fig:lineProfile}. Nevertheless, dominant structures are preserved, as highlighted by the positions of the major crests and troughs in our dual-projection output in Fig.\,\ref{Fig:lineProfile}. These structures guarantee the accuracy of manual landmark identification, as demonstrated by Fig.\,\ref{Fig:LandmarkLabelSynthesized} and Tab.\,\ref{Tab:landmarkDetection} where all the manual detection landmarks in Type I and Type II cephalograms are within the 4\,mm precision range.

With the existing database of conventional cephalograms, automatic cephalometric landmark detection algorithms are developed. In order to transfer these algorithms to synthetic cephalograms, the synthetic cephalograms should share as many features as possible with conventional cephalograms to get high detection accuracy. Due to the low image contrast in RayCast synthetic cephalograms, the SDRs are low, as displayed in Tab.\,\ref{Tab:landmarkDetection}. Image contrast is improved in MIP100 synthetic cephalograms. Therefore, the 4\,mm-SDR increases from 45.6\% to 81.5\%. With our proposed Type I synthesis method, the synthetic cephalograms are close to conventional cephalograms in terms of image contrast and resolution. Meanwhile, all anatomical structures including low intensity ones are contained in Type I synthetic cephalograms compared with MIP100 synthetic cephalograms. Therefore, the highest 2\,mm-SDR is achieved in Type I synthetic cephalograms. It demonstrates that the landmark detection model learned from the ISBI dataset is applicable for landmark detection in our Type I synthetic cephalograms. 

Compared with the synthetic cephalograms with the original sigmoid transform, those with our proposed modified sigmoid transform have only subtle difference in image contrast for the soft-tissues. However, such subtle difference substantially affects the automatic landmark detection, as shown in Tab.\,\ref{Tab:landmarkDetection}. It implies that the automatic landmark detection algorithm is very susceptible to image quality change. Therefore, it is a sensitive image quality indicator for synthetic cephalograms. For our Type II synthetic cephalograms, due to some inaccurate structures, the SDRs are still not high enough. For example, the mandible angles in Figs.\,\ref{Fig:LandmarkLabelSynthesized}(j)-(l) are more blurry than those in Figs.\,\ref{Fig:LandmarkLabelSynthesized}(g)-(i) respectively, causing difficulty for the automatic landmark detection algorithm in landmark identification.
Nevertheless, as shown in Tab.\,\ref{Tab:landmarkDetection}, the SDRs in our Type II synthetic cephalograms, which only require two CBCT projections for each synthesis, are still comparable to those in the Type I synthetic cephalograms with the original sigmoid transform and higher than those in the RayCast synthetic cephalograms.

\modified{For cephalogram synthesis from 3D CBCT volumes, because of the unsupervised setting using unpaired data, the learned CycleGAN model does not focus on our desired image contrast transform task only. Instead, it learns to synthesize other structures as well, including the positioner (Fig.\,\ref{Fig:cephalogramComparison}(d)) and other undesired anatomical structures (Fig.\,\ref{Fig:TypeILineProfiles}). Improving CycleGAN for cephalogram synthesis from 3D CBCT volumes is our future work. In this work, we apply the Type I synthetic cephalograms as the target of our Type II synthesis. Therefore, we choose our proposed modified sigmoid transform, which is a robust analytic method, for our Type I synthesis.
}

\modified{Pix2pixGAN is superior to CycleGAN in learning perspective deformation, as demonstrated by Figs.\,\ref{Fig:dualProjectionResults}-\ref{Fig:lineProfile}. CycleGAN achieves little success for tasks that require geometric changes, as reported by \citep{zhu2017unpaired}. Therefore, learning geometric change is one major limitation of CycleGAN. CycleGAN uses unpaired patches for training, which loses the pixel-to-pixel geometric relationship of Fig.\,\ref{Fig:dualProjection} in this work.
As illustrated in Fig.\,\ref{Fig:dualProjection}, another contribution of this manuscript is to apply two projection views for cephalogram synthesis instead of solely one projection view. Fig.\,\ref{Fig:dualProjection} indicates that learning the two-to-one mapping (from two cone-beam projections to one orthogonal-projection cephalogram) is easier than the one-to-one mapping (from one cone-beam projection to one orthogonal-projection cephalogram). For CycleGAN, indeed it has the ease of patch selection since unpaired patch-to-patch synthesis is possible. However, for the cycle consistency part, CycleGAN needs to learn the one-to-two mapping (from one orthogonal-projection cephalogram to two cone-beam projections), which is very difficult. }

In Type II, we divide each projection into four patches ($2 \times 2$) instead of more patches because of three reasons: 
\begin{itemize}
\item Section 3.2.3 tells that if the patches are selected quadrant-wisely ($2 \times 2$), the anatomical structures in the input cone-beam projection patch and those in the target patch are paired;

\item In our experiments, the stitching effect with $2 \times 2$ patches is already apparently visible in the result of 1-projection output, as displayed in Fig.\,10(q). Using our proposed 2-projection approach, the stitching effect is not so obvious in the result of 2-projection output. When selecting more patches, more patches are needed to stitch into one image. Then the stitching effect will become non-negligible and hence degrades image quality. 

\item The blurring effect caused by the large patch size in GANs will be mitigated by the SR step in our pipeline.
\end{itemize}
% Although soft tissues are visualized better in cephalograms synthesized by MIP, the likelihood of false findings is increased as well \citep{cattaneo2008comparison}.

\begin{figure}[t]
\centering
\begin{minipage}[b]{0.32\linewidth}
\centering
\begin{small}
Target
\end{small}
\end{minipage}
\begin{minipage}[b]{0.32\linewidth}
\centering
\begin{small}
Input
\end{small}
\end{minipage}
\begin{minipage}[b]{0.32\linewidth}
\centering
\begin{small}
pix2pixGAN
\end{small}
\end{minipage}

\begin{minipage}[b]{0.32\linewidth}
\subfigure[]{
\includegraphics[width = \linewidth]{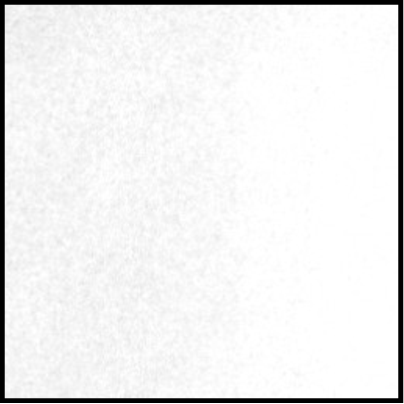}
}
\end{minipage}
\begin{minipage}[b]{0.32\linewidth}
\subfigure[]{
\includegraphics[width = \linewidth]{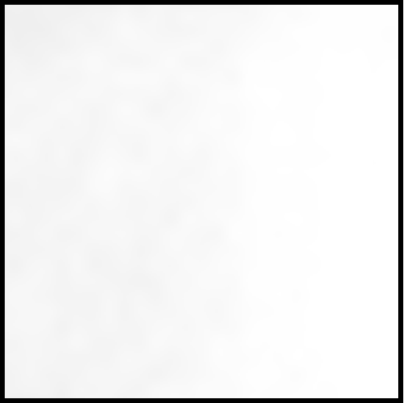}
}
\end{minipage}
\begin{minipage}[b]{0.32\linewidth}
\subfigure[3.63, 36.93, 0.290]{
\includegraphics[width = \linewidth]{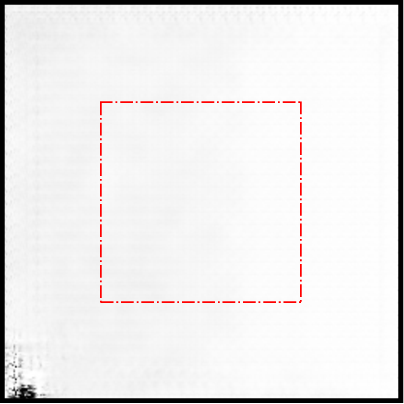}
}
\end{minipage}

\caption{\modified{An SR example where the SSIM value of the pix2pixGAN output is low. The RMSE (left), PSNR (mid), and SSIM (right) values are displayed in (c). Due to the bright/white color, a black box is added to show the patch boundary. The area inside the red dash box is valid after stitching.}}
\label{Fig:specialCaseSR}
\end{figure}

\begin{figure}[t]
\centering
\includegraphics[width = 0.31\linewidth]{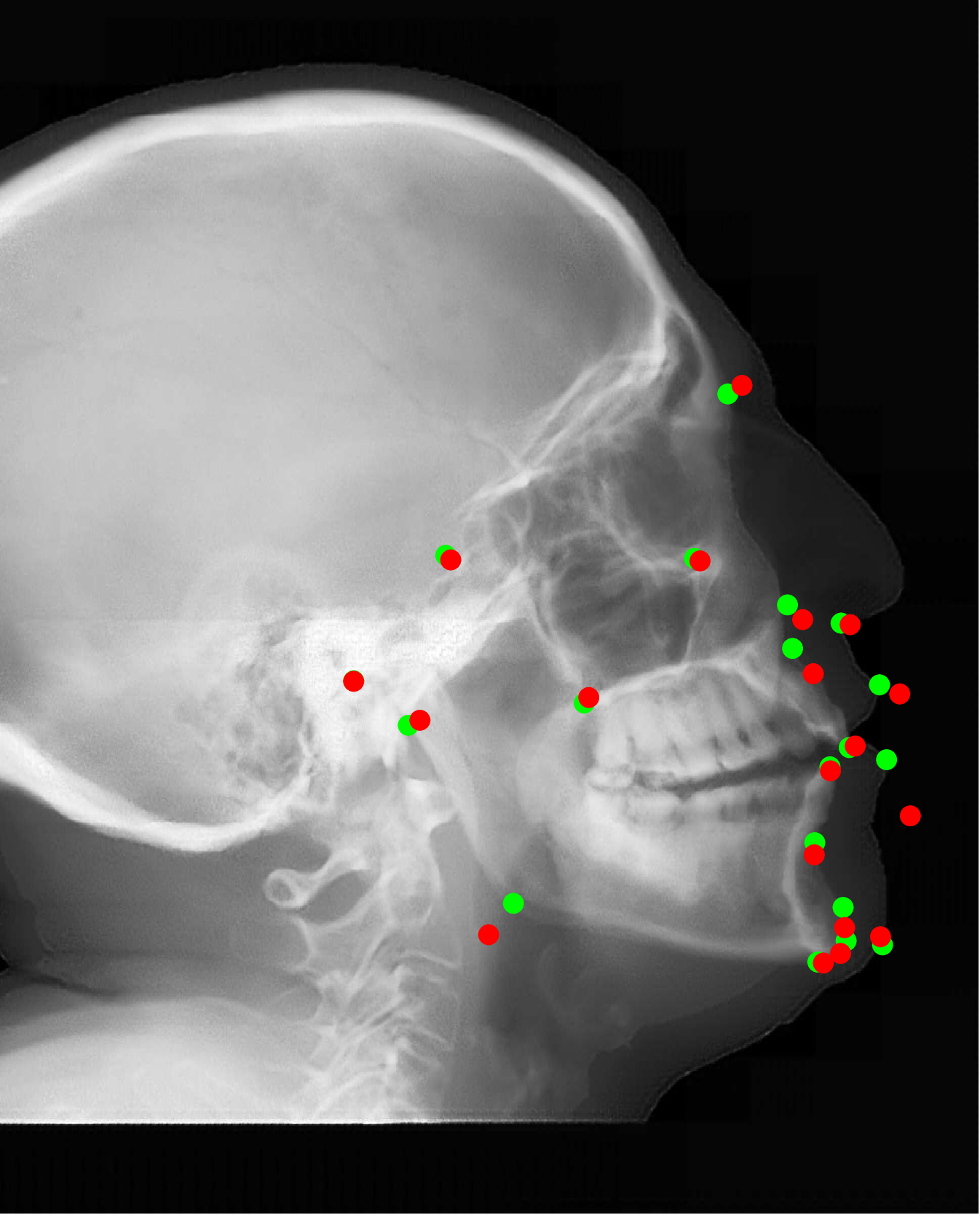}
\caption{\modified{One special case where the cephalomatric landmark detection accuracy is low due to the bad position of the patient head.  The green (reference) labels are manual detection landmark positions in Type I synthetic cephalograms, while the red labels are automated detection positions.}}
\label{Fig:specialCase}
\end{figure}

In our experiments, we have not observed any cases where the current methods fully fail. However, we do observe two special cases where the quality metrics have low values:

\begin{itemize}
\item For SR test, some results have very low SSIM values. An example is displayed in Fig.\,\ref{Fig:specialCaseSR}. The RMSE , PSNR , and SSIM values for Fig.\,\ref{Fig:specialCaseSR}(c) are 3.63, 36.93 and 0.290, respectively. The result patch has very low RMSE and high PSNR. However, the SSIM is very low, because the target patch has very low variation. As a consequence, the SSIM metric is sensitive to error. Because of such low SSIM cases, the average SSIM in Tab.\,\ref{Tab:ResultsOfSR} is not high. In our SR experiments, apparent incorrect pixels are all located near boundaries, as displayed in Fig.\,\ref{Fig:specialCaseSR}(c). Since overlapped patches are selected for SR, only the area inside the red box is valid after stitching, where no such apparent error exists. Therefore, such error is not a problem for our application. 

\item In one situation the cephalometric landmark detection has very low accuracy: when a patient cannot position the head well due to certain neck/spinal diseases, 5 landmarks among the 19 landmarks are beyond the 4\,mm precision range, as displayed in Fig.\,\ref{Fig:specialCase}. Especially, it fails to detect some evident landmarks in the Type I synthetic cephalogram like the Gonion landmark and the lower lip landmark.
\end{itemize}

\section{Conclusion And Outlook}
In this work, we have proposed a method to synthesize cephalograms from 3D CBCT volumes with improved image contrast based on the optical properties of conventional X-ray films and improved image resolution using SR techniques. We have also proposed a deep learning method to synthesize cephalograms directly from dual 2D X-ray projections for low dose purpose, which achieves higher accuracy compared with using one projection only. The accuracy of the synthesized landmarks is validated preliminary by manual landmark detection and our proposed automatic cephalomatric landmark detection method. 

In this work, proof-of-concept experiments have been carried out. In the future, clinical dental CBCT volumes/projections and their corresponding conventional 2D cephalograms are desired for further clinical verifications. One step further, with matching pairs of clinical data, an end-to-end pipeline can be set up, which allows to optimize image contrast and image resolution for optimal automatic landmark detection.

%\ 

%\noindent \textbf{Declaration of Competing Interest} 
%None.
%%Harvard
\bibliographystyle{model2-names}\biboptions{authoryear}
\bibliography{cephal}

%\section*{Supplementary Material}
%
%Supplementary material that may be helpful in the review process should
%be prepared and provided as a separate electronic file. That file can
%then be transformed into PDF format and submitted along with the
%manuscript and graphic files to the appropriate editorial office.

\end{document}